\documentclass[structabstract]{aa} 

\usepackage{txfonts}

\usepackage{natbib}
\usepackage[english]{babel}
\usepackage{graphicx}
\usepackage{verbatim}
\usepackage{multirow}
\usepackage{subfigure}
\usepackage{hyperref}
 \hypersetup{colorlinks=true,citecolor=blue}
\usepackage{longtable}
\usepackage{longtable,lscape}

\begin{document}

\title{Radio galaxies of the local universe}
\subtitle{All-sky catalog\thanks{The master catalog (described in Table 1) and the catalog listing all radio matches per galaxy, are only available in electronic form at the CDS via \url{http://cdsweb.u-strasbg.fr/cgi-bin/qcat?J/A+A/} or at \url{http://ragolu.science.ru.nl}.}, luminosity functions, and clustering}

\author{Sjoert van Velzen\inst{1} \and Heino Falcke\inst{1,2,3, 4} \and Pim Schellart\inst{1} \and Nils Nierstenh\"ofer\inst{5} \and Karl-Heinz Kampert\inst{5} }

\institute{
   IMAPP, Radboud University, P.O. Box 9010, 6500 GL Nijmegen, The Netherlands \email{s.vanvelzen@astro.ru.nl}
   \and ASTRON, Dwingeloo, The Netherlands 
   \and Max-Planck-Institut f\"ur Radioastronomie Bonn, Germany
   \and NIKHEF, Science Park, Amsterdam, Netherlands  
   \and Department of Physics, Bergische Universit\"at Wuppertal, Wuppertal, Germany
}

\date{Received April 11, 2012; accepted May 14, 2012}
  \abstract
  { 
    To understand the feedback of black holes on their environment
    or the acceleration of ultra-high energy cosmic rays in the
    present cosmic epoch, a comprehensive inventory of radio galaxies
    in the local universe is needed. This requires an all-sky catalog
    of radio-emitting galaxies, that hitherto has not been available.}
  { We present such an all-sky sample. Our catalog allows one to build
    volume-limited subsamples containing all low-power
    radio galaxies, similar to the prototypical low-power radio galaxies Cen~A or
    M87, within some hundred Mpc.}
 { We match radio emission from the NVSS and SUMSS surveys to galaxies of the 2MASS
   Redshift Survey (2MRS) using an image-level algorithm that properly treats the
   extended structure of radio sources. }
 { The bright master sample we present contains 575 radio-emitting
   galaxies with a flux greater than 213~mJy at 1.4~GHz. Over 30\% of
   the galaxies in our catalog are not contained in
   existing large-area extra-galactic radio samples. 
   We compute the optical and radio luminosity functions and
   the fraction of radio galaxies as a function of galaxy luminosity. 
   94\% of the radio galaxies within $z=0.03$
   are of Hubble type E/S0. The local galaxy density in a sphere of
   2~Mpc centered on the radio galaxies is 1.7 times higher than
   around non-radio galaxies of the same luminosity and morphology, which is a
   statistically significant enhancement ($>3\sigma$).}
 { Our sample presents the deepest all-sky catalog of low-power radio
   galaxies. The observed enhancement of the galaxy density around 
   radio galaxies suggests a causal
   relation between external galaxy properties, such as environment or
   merger history, and the formation of powerful jets in the present
   universe. Since the enhancement is observed with
   respect to galaxies of the same luminosity and Hubble type, it is not
   primarily driven by black hole mass. Our automated matching
   procedure is found to select radio-emitting galaxies with high
   efficiency (99\%) and purity (91\%), which is key for future
   processing of deeper, larger samples.}\keywords{}
  \maketitle

\section{Introduction}\label{sec:intro}
The final episode in the history of black hole accretion and galaxy formation takes place in our cosmic backyard, the local universe. A large sample of nearby radio galaxies may thus be considered an important anchor point for theories of black hole growth and downsizing  \citep[e.g.,][]{Alexander12} and can be used to study feedback from radio jets \citep{Fabian03,Mathews03} on the environments of their host galaxies.

Nearby radio galaxies allow one to study a regime of fainter jets and thus, assuming standard jet-disk coupling \citep{Rawlings91, Falcke95I}, lower rates of accretion onto the black hole. At these low rates, the mode of accretion \citep{Narayan95} may switch to a `jet-dominated mode' or `radio mode' \citep*{Falcke04}, similar to the `low-hard state' of X-ray binaries \citep*{Fender04}, implying that radio surveys are an ideal tool to find active black holes in this regime. Large samples of radio-loud active galatic nuclei (AGN) provide increasing evidence that two different modes of accretion onto super-massive black holes indeed exist \citep{Ghisellini01, Koerding06, Ghisellini11,Wu11, Best12}.

The first catalogs of extra-galactic radio sources were constructed by gathering a (heterogeneous) set of optical follow-up observations \citep{Schmidt68, Veron74, Kuehr81,Laing83,Jones92}. With the advent of wide-field optical surveys and deep radio surveys with high angular resolution, it has become possible to match cataloged sources at both wavelengths, and thus systematically construct  large catalogs of radio sources with distance information \citep{Condon02,Ivezic02,Best05a,Sadler07,Mauch07,Kimball08,Donoso09,Brown11}. Such catalogs have played a key role in the study of star formation, AGN, and the interplay between them. 

While at intermediate to large redshifts ($z \gtrsim 0.1$) one can survey a limited area of the sky to obtain a representative slice of the extra-galactic volume, studies of the local universe require full-sky coverage (i.e., 4$\pi$ solid angle) to map the anisotropy of the matter distribution. None of the modern catalogs of active black holes meet this requirement.

With the recent release of the 2MASS redshift survey \citep[2MRS; ][]{Huchra12}, currently the deepest all-sky redshift survey, it has become possible to construct an extra-galactic radio catalog that covers 90\% of the volume of the local universe. Indeed we made this our goal: obtain a complete, all-sky catalog of galaxies that emit at radio wavelengths. We have constructed this catalog in the most systematic way possible to ensure that the selections effects are well-defined (opposed to compiling a list of known radio sources from the literature). Since the radio galaxies of our catalog are a subset of normal galaxies and, by construction, the sample is the largest of its kind, it present a powerful tool for a statistical study of the relation between black hole activity and galaxy environment. 

Our primary motivation for building an all-sky radio catalog is to obtain a volume-limited sample of galaxies that could be powerful enough to accelerate ultra-high energy cosmic rays (UHECRs, charged particles with an energy in excess of $~5\times 10^{20}$~eV) and study the magnetic field and energetics of these sources. For this study, full-sky coverage is key because interactions with photons of the cosmic microwave background limit the distance an UHECR can travel to about 100~Mpc, the so-called, GZK horizon \citep{Greisen66,ZatsepinKuzmin66}; if the sources of UHECRs are rare ($<10^{-5}~{\rm Mpc}^{-3}$), the full-sky has to be searched to obtain a sizable sample.

A good candidate source of UHECRs is Cen~A: this radio galaxy may just be powerful enough to accelerate protons up to the ultra-high energy scale \citep[for a recent review see][]{Biermann12}. 
Since Cen~A is the nearest and one of the best studied radio galaxies, it provides a good anchor for the rest of the catalog; the flux of Cen~A at 100~Mpc ($\sim 1$~Jy at 1~GHz and $K=11$) is comfortably within the limits of existing surveys. Our main goal can thus be summarized in one sentence: \emph{find all radio galaxies within 100~Mpc that are as a luminous as Cen~A}. Since Cen~A is relatively faint at optical wavelengths, this requirement actually allows us to find most typical radio galaxies within 200~Mpc.  

In this paper, we focus on the construction and properties of the radio-bright sample (a larger sample, obtained by lowering the radio flux limit, will be presented in a future publication). In section \ref{sec:cat} we discuss our matching algorithm and we compare our sample to existing radio catalogs. In the third section we present the number counts (sec. \ref{sec:volsample}), the cross-correlation of radio galaxies with the local matter distribution (sec. \ref{sec:cc}), and the luminosity functions (sec. \ref{sec:lumfunc}). We close with a discussion (sec. \ref{sec:discussion}). In the second paper in this series, we will discuss the magnetic fields, jet power, and energy injection of the radio galaxies with particular emphasis of its relevance for UHECRs.

\section{Catalog construction}\label{sec:cat}
As explained in the introduction, we wish to obtain a flux-limited, all-sky catalog of extra-galactic radio sources, from which a volume-limited sample can be derived. In this section we discuss the construction of this catalog. First, we will describe the input (sec \ref{sec:input}), followed by a detailed discussion of the matching algorithm (sec. \ref{sec:matching}). In section \ref{sec:columns} we present the columns our catalog. We assess the completeness in section \ref{sec:completeness} and probability for random matches in section \ref{sec:randommatch}. A comparison to existing extra-galactic radio catalogs is presented in section \ref{sec:othercat}. Finally, we briefly discuss some newly identified radio-emitting galaxies in section \ref{sec:notes}.

\subsection{Input}\label{sec:input}
In the following paragraphs we discuss the input for our catalog: the 2MASS Redshift Survey (2MRS), the NRAO VLA Sky Survey (NVSS), and the Sydney University Molonglo Sky Survey (SUMSS). 

\begin{figure}
\centering 
\includegraphics[trim=5mm 2mm 2mm 5mm, clip, width=.48\textwidth]{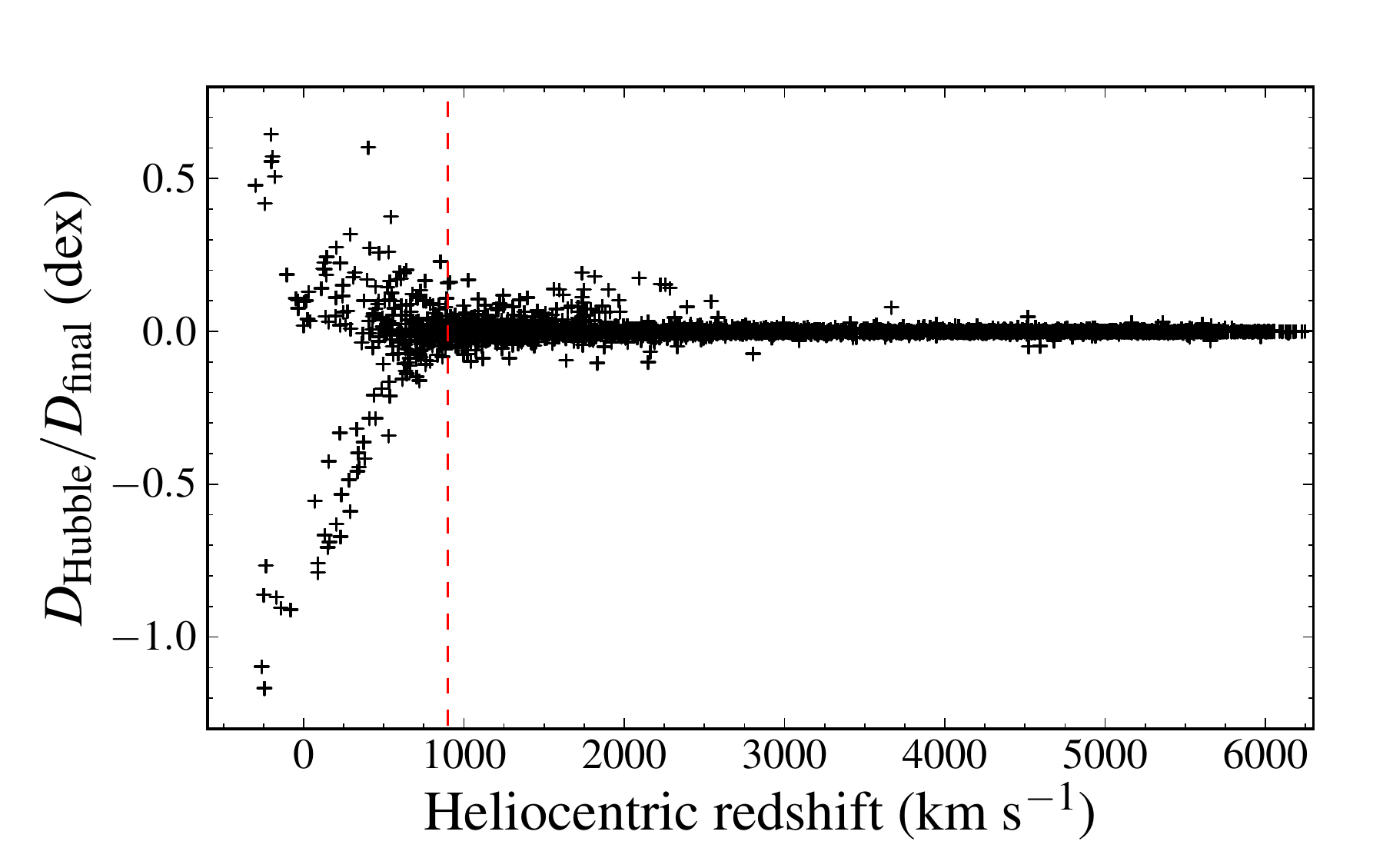}
\caption{The difference between the Hubble distance, after our correction for peculiar velocity ($D_{\rm Hubble}$), and the final distance ($D_{\rm final}$). The latter is the inverse-variance weighted mean of $D_{\rm Hubble}$ and median redshift-independent distance from NED-D. The fractional uncertainty on the Hubble distance decreases with redshift, hence the convergence to zero difference. The dashed line indicates the redshift cut that is applied in the computation of the luminosity functions and clustering (sec. \ref{sec:ana}).}\label{fig:NED-D}
\end{figure}

\subsubsection{Redshift survey}\label{sec:2MRS}
To be able to select a volume-limited sample of radio sources, we need a measurement of the distance to each source. Currently, the deepest all-sky redshift survey is the 2MRS \citep[][]{Huchra12}, which covers essentially the entire extra-galactic sky accessible at optical wavelengths (91\% of the full-sky). The targets of this survey are galaxies from the 2~Micron All-Sky Survey \citep[2MASS,][]{Skrutskie06}, contained in the extended source catalog \citep{Jarrett00, Jarrett04}, selected with the following requirements: 
\begin{itemize}
\item $K_s<11.75$ and detected in the $H$-band
\item $E(B-V)<1$
\item $|b|>5^{\circ}$ for $30^{\circ}<l<330^{\circ} $; $|b|>8^{\circ}$ otherwise.
\end{itemize}
Where $E(B-V)$ is the extinction from the maps of \citet{schlegel98}; $l$ and $b$ are the Galactic longitude and latitude, respectively. $K_s$ is the isophotal magnitude (measured in an elliptical aperture defined at the 20~mag per square arcsecond isophote). The $K$-band is centered at $\lambda\approx2.2\,\mu m$, the near infrared (NIR) part of the electromagnetic spectrum.

The 2MRS cuts select 44,599 galaxies from 2MASS; the current catalog contains redshifts for 97.6\% of these. The median redshift is 0.028, 90\% of the sample is contained within $z<0.052$. We applied no cuts on the 2MRS catalog.

Some care has to be taken when converting redshifts of nearby-galaxies to distances using the Hubble law. We adopt the approach used by \citet{Blanton05} for the construction of the NYU Value-Added Galaxy Catalog. First, we shift to the Local Group barycenter using the heliocentric velocity determination of \citet{Yahil77}. In this frame we estimate the most likely distance and its uncertainly using the model of the local velocity field of \citet{Willick97} based on the IRAS 1.2 Jy redshift survey \citep{Fisher95}. We adopt $H_0=72\,{\rm  km}\,{\rm s}^{-1}$ and $\Omega_m=0.3$, $\Omega_\Lambda=0.7$ to convert the peculiar-velocity corrected redshifts to a distance. 

For galaxies with a radial velocity (measured in Local Group barycenter) below $6,000 \, {\rm km}\,{\rm s}^{-1}$ we attempt to improve the Hubble law distance using published redshift-independent distances (e.g., Cepheids, Supernova Type Ia, Tully-Fisher relation) as listed in NED-D\footnote{\url{http://ned.ipac.caltech.edu/Library/Distances/} (v5.1) compiled by I. Steer and B. F. Madore.}. For 3351 galaxies (8\%) we compute the inverse variance weighted mean distance modulus from our peculiar-velocity corrected Hubble distance and the median redshift-independent distance modulus.  The corrections to the luminosity distances derived solely from the Hubble law are modest, $-0.0006\pm 0.07$~dex in the mean. Corrections larger than 0.5~dex are observed only for galaxies with a Local Group barycenter radial velocity smaller than $187~{\rm km}\,{\rm s}^{-1}$ (Fig. \ref{fig:NED-D}). For the radio-emitting galaxies in our final sample (section \ref{sec:mfinal}), the coverage of NED-D is 100\% below $250~{\rm km}\,{\rm s}^{-1}$, so the uncertainty on the radio luminosity due to peculiar velocities is not significant. 

\begin{figure*}[t]
\centering
\includegraphics[trim=12mm 2mm 180mm 4mm,  width=.47\textwidth]{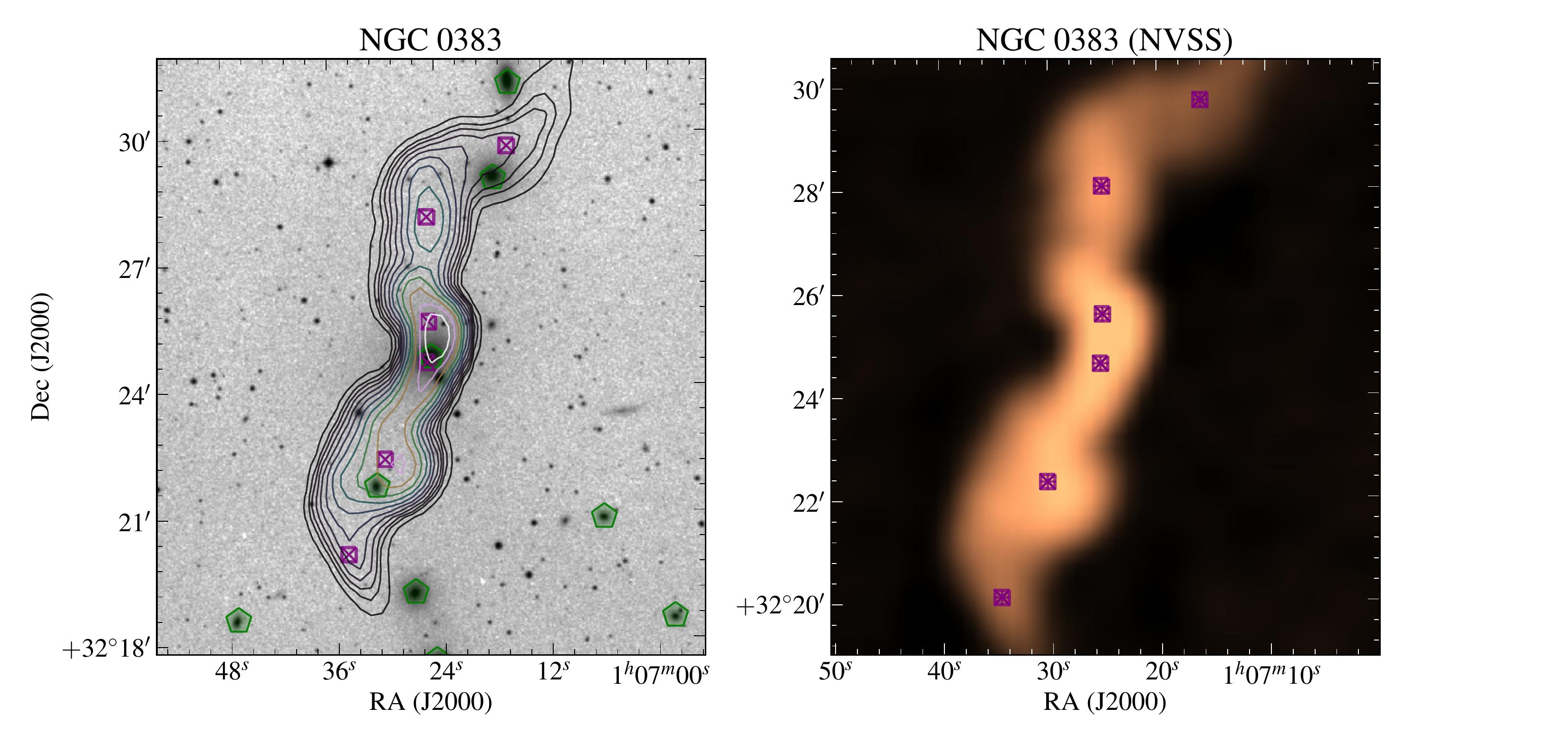}
\includegraphics[trim=12mm 2mm 180mm 4mm, clip, width=.47\textwidth]{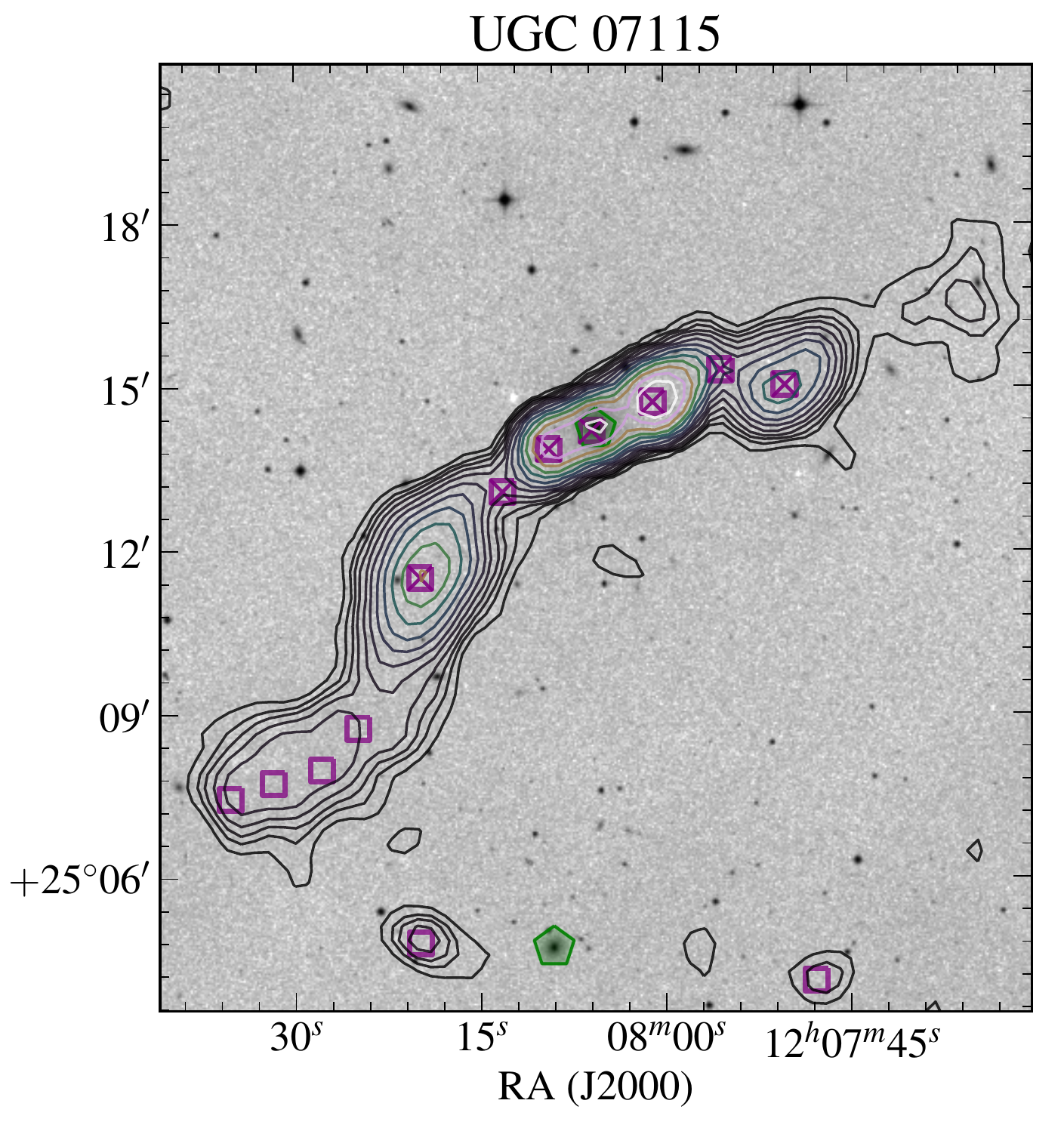}
\caption{Two examples to illustrate the challenges that arise when matching optical images with a resolution of $\sim 2"$ to radio images with a resolution of $45"$. The contours show the radio emission from NVSS; 2MRS galaxies are labeled with green pentagons. The entries of the NVSS catalog are shown by purple squares, we see that both jets are decomposed into multiple elliptical Gaussians (6 for NGC~383 and 11 for UGC~7115). The radio entries matched to this galaxy by the angular distance requirement (Eq. \ref{eq:creq}) are labelled with crosses. For UGC~07115, the remaining four jet components are included after applying a friends-of-friends algorithm (sec. \ref{sec:cmatch}).}\label{fig:0383catmatch}
\end{figure*}

\subsubsection{Radio catalogs}\label{sec:radio_cat}
No single radio survey covers the full sky, we thus have to combine multiple surveys. To avoid sky-dependent selection effects, these surveys have to be similar in depth, frequency and resolution. With these requirements only one set of radio surveys remains\footnote{Other radio surveys such as FIRST \citep{Becker95} or AT20G \citep{Murphy10}, could be considered more suitable for matching to optical catalogs of galaxies. However this set of surveys does not meet our requirement of covering the entire sky at a similar frequency.}: NVSS \citep[][]{Condon98} and SUMSS \citep[][]{Bock99,Mauch03}, covering the Northern and Southern Hemisphere, respectively. 

The NVSS catalog is derived from images obtained with the VLA in D and DnC configurations at 1.4~GHz and covers the sky north of declination $-40^{\circ}$ (75\% of the full-sky). The size of the point-spread-function (PSF) of these images is $45''$~FWHM and the astrometric accuracy is better than $1"$ for $F_\nu>15$~mJy. The typical rms brightness fluctuations in the NVSS images are 0.45 mJy/beam and the completeness limit is about 2.5 mJy. 

The SUMSS catalog is constructed from data obtained with MOST \citep[Molonglo Observatory Synthesis Telescope; ][]{Mills81, Robertson91} at 843 MHz. The resolution of the SUMSS images is $43"$~FWHM. The typical astrometric accuracy in right ascension is $2-3"$ at 10 mJy and $1"$ above 20 mJy; position uncertainties in declination ($\delta$) are typically a factor of $\rm{csc}(|\delta|)$ higher than in right ascension. The rms noise level of the SUMSS images is $\sim 1$ mJy/beam, the completeness limit is about 10 mJy. The survey covers the sky south of declination $-30^{\circ}$ with $|b|>10^{\circ}$. The sky below $|b|=10^{\circ}$ that is not in the NVSS footprint is covered by MGPS-2 \citep[The second epoch Molonglo Galactic Plane Survey; ][]{Murphy07} which is the Galactic counterpart to SUMSS. The MGPS-2 catalog, however, currently only contains compact sources which makes it unsuited for our matching algorithm. We therefore removed the $|b|<10$ region for $\delta<-40$, reducing the area covered by our catalog from 91\% to 88\%. 

It is important to realize that both the SUMSS and NVSS catalog are constructed by fitting elliptical Gaussians to the radio maps. Each entry in the catalog is a separate Gaussian with a measured major/minor axis and integrated flux. Extended radio sources are often broken up into multiple Gaussians, hence a single astrophysical object is not necessarily represented by a single entry in the radio catalogs.

\subsubsection{Flux limit}\label{sec:fluxlims}
As explained in the introduction, our primarily goal to obtain a volume limited sample of radio galaxies as bright as Cen~A.  We use the flux of Cen~A at the distance that contains 95\% of the galaxies in 2MRS to find a conservative value for the radio flux limit. Using $F_{1400} = 1330$~Jy for all the radio emission of Cen~A \citep{Cooper65} at 1.4~GHz and a spectral index\footnote{We define the spectral index, $\alpha$, by $F_\nu \propto \nu^{\alpha}$.} $\alpha=-0.6$, we obtain the following flux limits: $F_{1400}>213,\, F_{843} >289$~mJy for NVSS, SUMSS. We apply these limits to the sum of the integrated radio flux of all Gaussians that are matched to a galaxy. Extending these flux limits to the completeness limits of the radio surveys ($\sim 10$~mJy) will be the subject of future work. 

\subsection{Matching}\label{sec:matching}
The next step is to match the radio surveys to the redshift survey to find radio-emitting galaxies. Our aim is to automate the matching as much as possible. This reduces (or at least parametrizes) human bias and substitutes expensive man-hours with cheap cpu-time. Since this is our first attempt at fully automated cross-wavelength matching, we will also inspect all matches manually.

Two challenges have to be met: (i) the FWHM of the PSF of the radio images is over an order of magnitude greater than the NIR images, (ii) many radio sources are resolved and will appear in the radio catalogs as multiple entries which can be offset from the galaxy by several arcminutes (e.g., giant radio galaxies). Our approach to this problem is to proceed in two steps: first we match on the catalog-level (sec. \ref{sec:cmatch}), then we assess these matches at the image-level (sec. \ref{sec:imrej}). We designed the first step to be ``generous'', i.e., all potential matches should be found, at the cost of a large background of false identifications. At the image-level this background is often trivially rejected. 

In a nutshell, our matching pipeline is based on the assumption that the galaxy is source or origin of the radio emission, but we allow this emission to be displaced and even disconnected from the galaxy; it is optimized to recover radio emission from both starforming galaxies, and systems with more complicated or asymmetric morphologies such as FR~II galaxies or ``head-tail'' sources.

\subsubsection{Catalog-level matching}\label{sec:cmatch}
We use the size of the radio source and the angular distance between this source and the galaxy to define the first matching criterion: 
\begin{equation}\label{eq:creq}
   d_i < N_{\rm lim}\times  {\rm FWHM}_i  \quad {\rm OR} \quad d_i< d_{\rm lim} \quad ,
\end{equation}
where $d_i$ is angular distance between the $i$th radio entry and the galaxy center and ${\rm FWHM}_i$ is the deconvolved major axis of the radio Gaussian. We adopted the following cuts: $N_{\rm lim}=3$ and $d_{\rm lim}=2'$. In section \ref{sec:completeness} we verify that these settings are indeed generous enough. 

For the coordinate matching we use \verb k3match , a new\footnote{k3match is available under GNU General Public License at \url{pschella.github.com/k3match}} and efficient implementation of 3 dimensional binary tree search. It can find matches between two sets of points on a sphere in $O(N\log(N))$ time as opposed to the $O(N^2)$ time needed for a brute force search (P. Schellart et al. 2012 in prep).

In some cases, the requirement set by Eq. \ref{eq:creq} will miss a small fraction of the radio emission (e.g., UGC~7115 in Fig. \ref{fig:0383catmatch}: the start of the jet is matched to the galaxy, but the rest of the radio emission, extending 6' further out, is not). This problem is solved by extending the matches using a friend-of-friend algorithm. We add a new radio entry to the group of matches if $d_{ij}< \max(N_{\rm lim}\times{\rm FWHM}_i, d_{\rm lim})$, with $d_{ij}$ the angular distance between the current radio match $i$ and the potential new entry $j$ (i.e., a link length equivalent to Eq. \ref{eq:creq}), and we repeat this procedure until no new matches are found. This approach successfully recovers the entire structure of nearly all extended radio sources (as demonstrated by the images in Appendix \ref{sec:atlas}).

For each galaxy, the total radio emission is simply given by the sum of the integrated flux of all the components that are matched to the galaxy. Using Eq. \ref{eq:creq} we obtain 1273 galaxies with a total radio flux above our flux limit (section \ref{sec:fluxlims}),  with a total of 8452 matches between galaxies and entries in the radio catalogs. The generous criteria used at this stage are required to match galaxies with displaced radio counterparts, but they inevitably yield false matches; we discuss the rejection of these false matches in the following section. 

\begin{figure*}[t!]
\centering
\includegraphics[trim=10mm 5mm 5mm 5mm,  width=.3\textwidth]{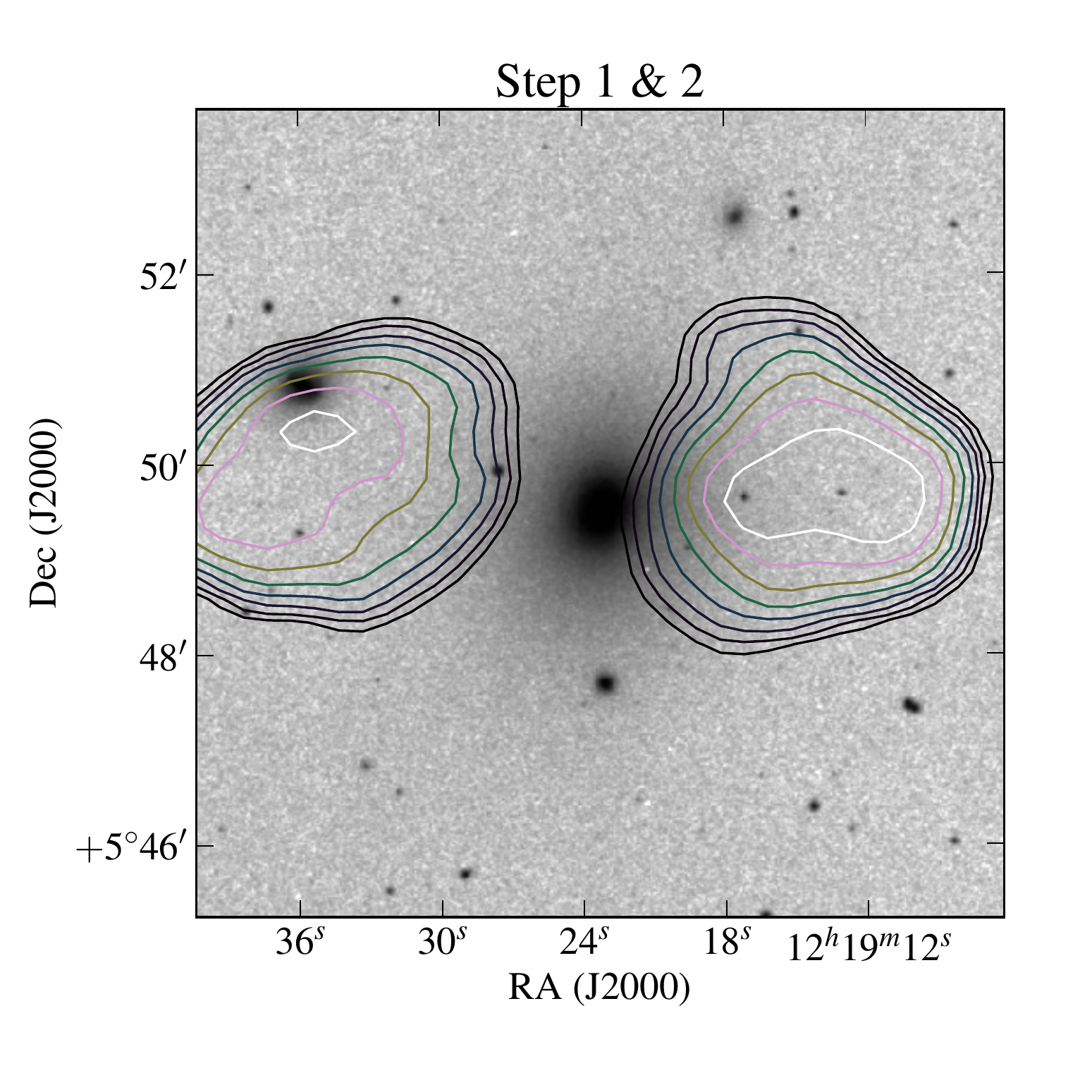}
\includegraphics[trim=10mm 5mm 5mm 5mm, clip, width=.3\textwidth]{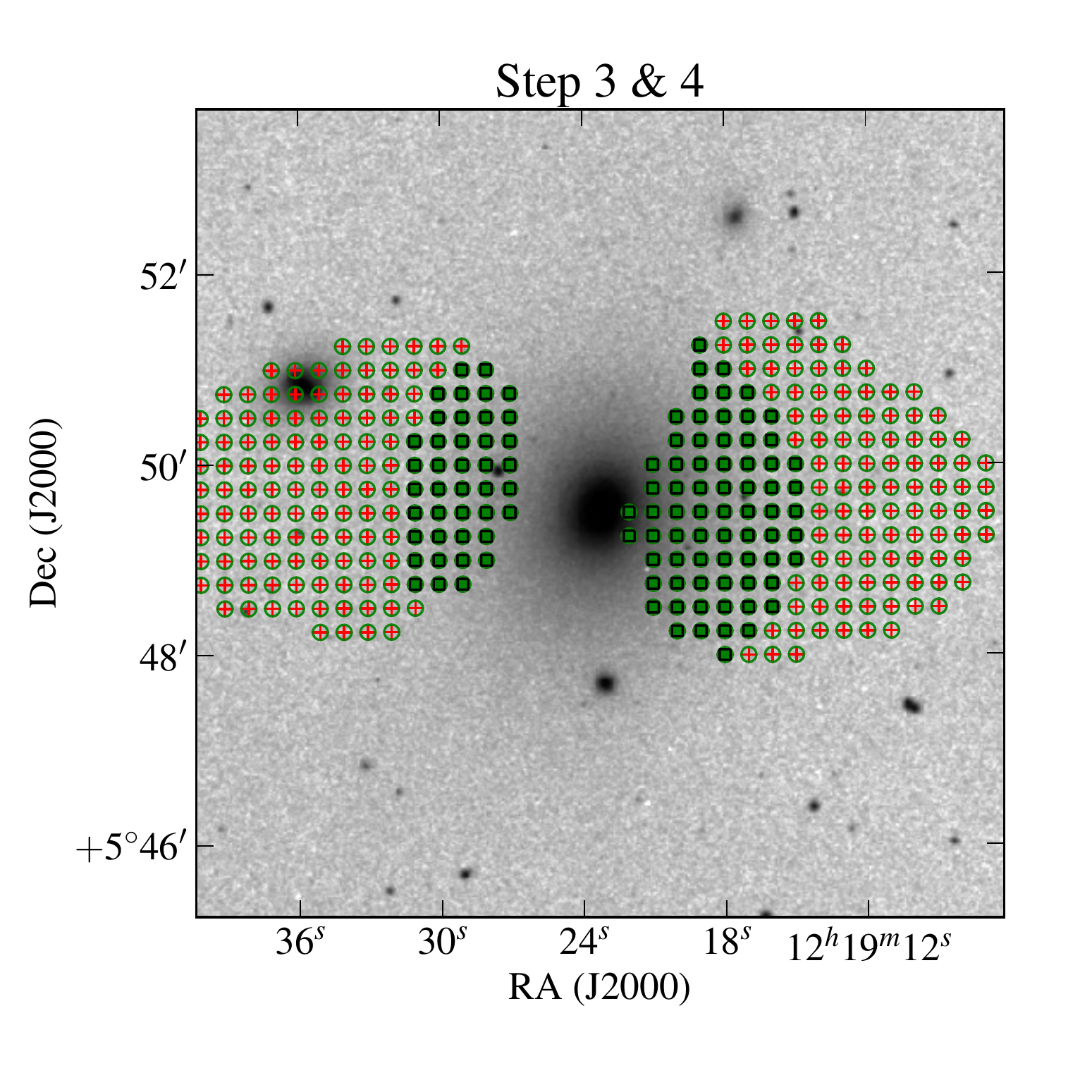}
\includegraphics[trim=10mm 5mm 5mm 5mm, clip, width=.3\textwidth]{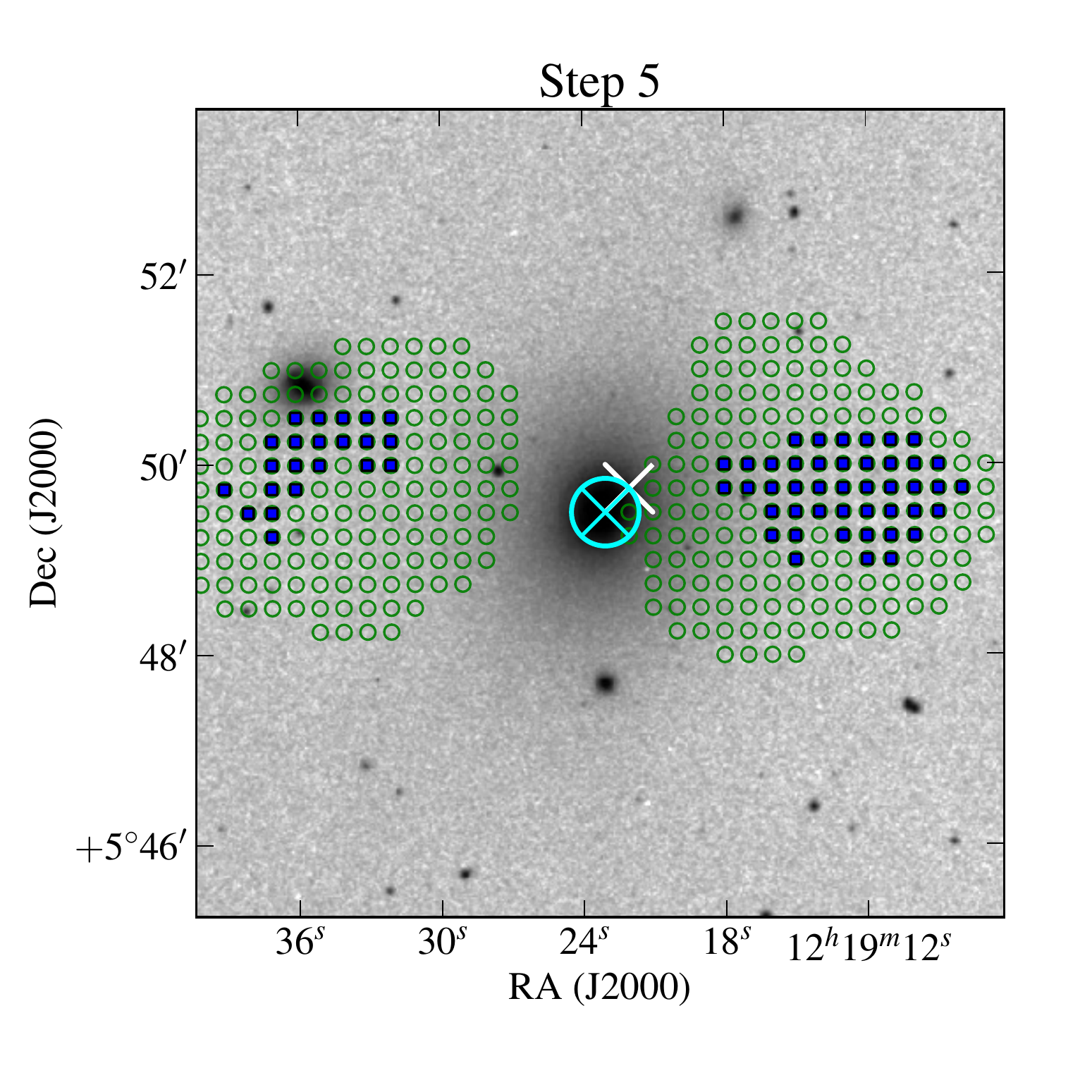} \\ 
\includegraphics[trim=10mm 5mm 5mm 5mm,  width=.3\textwidth]{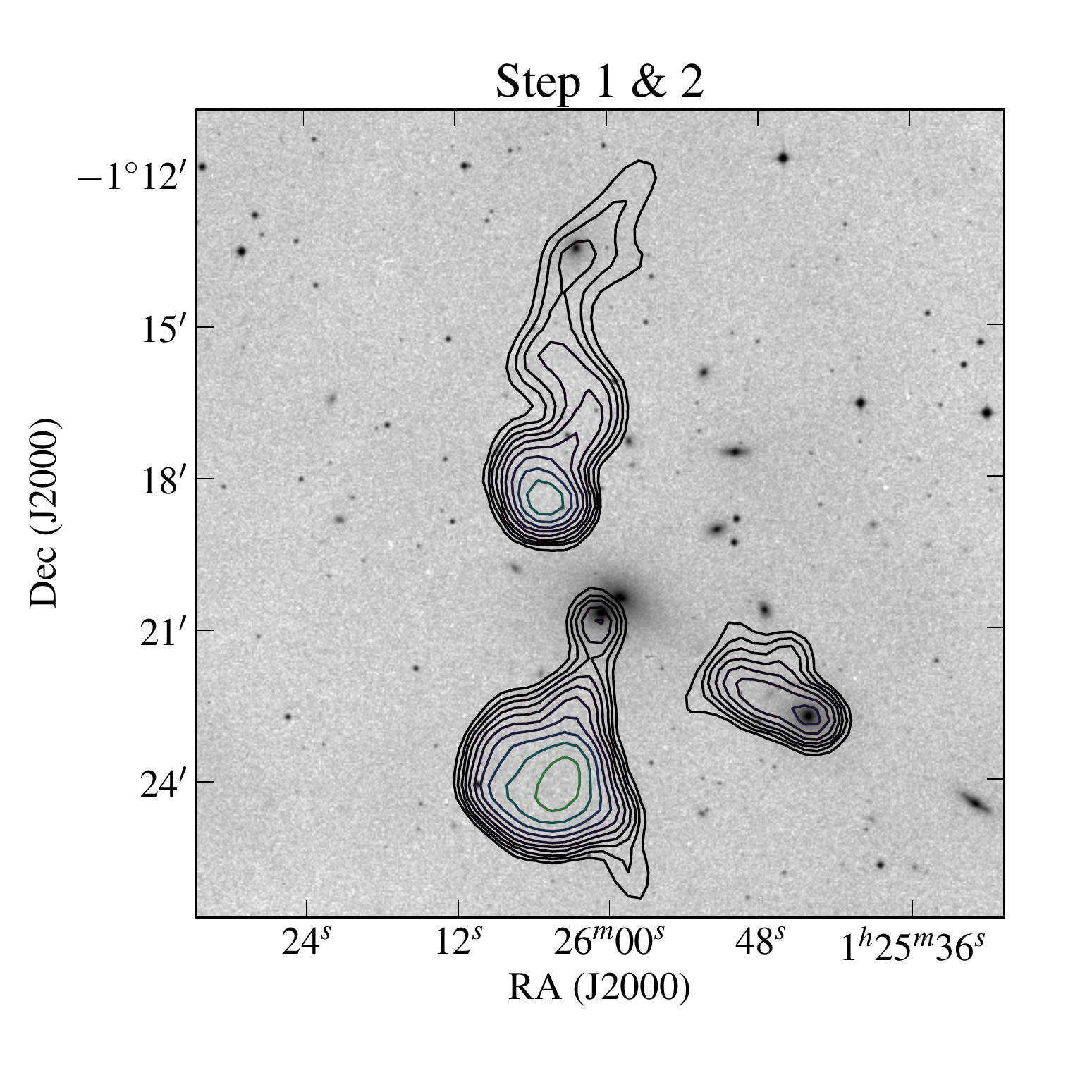}
\includegraphics[trim=10mm 5mm 5mm 5mm, clip, width=.3\textwidth]{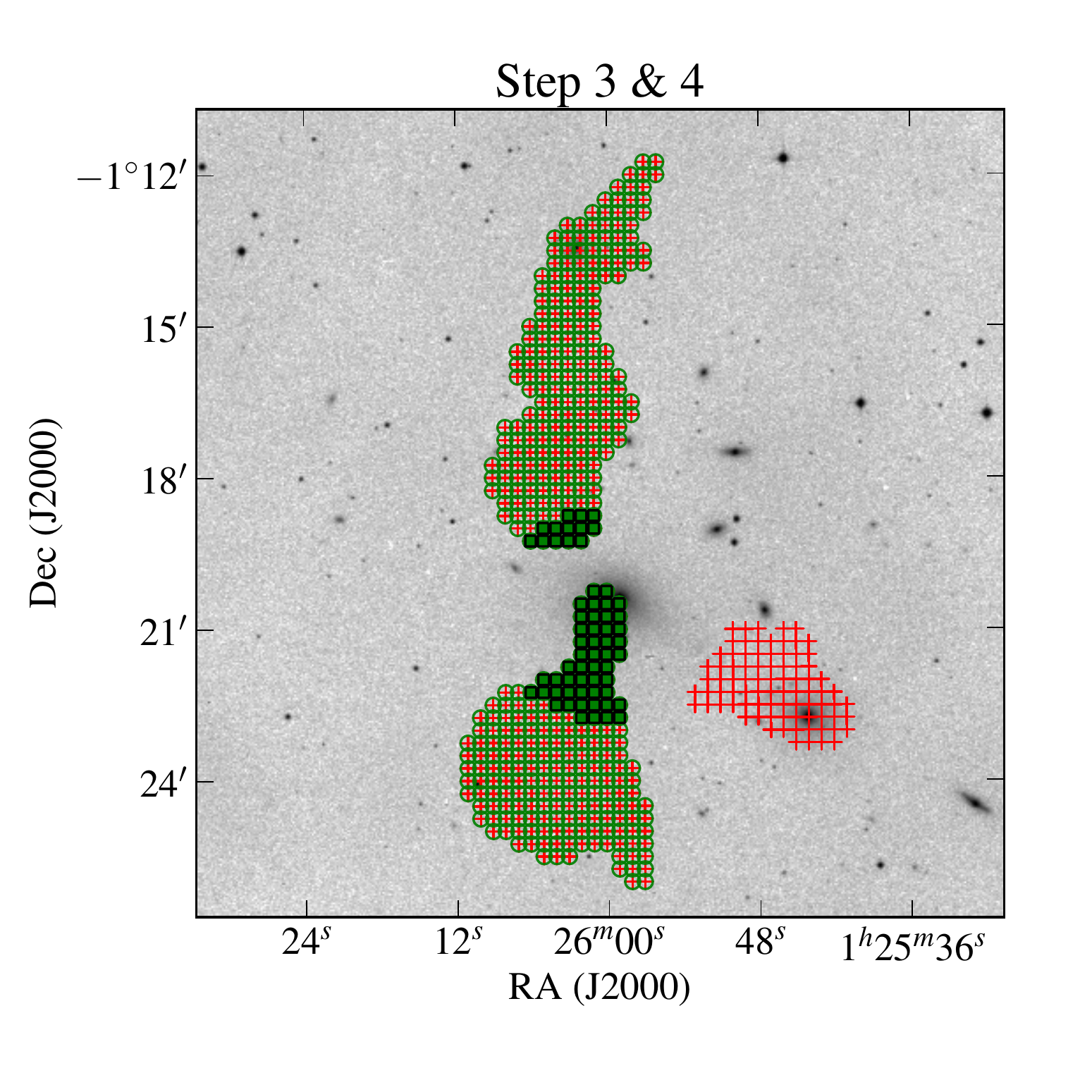}
\includegraphics[trim=10mm 5mm 5mm 5mm, clip, width=.3\textwidth]{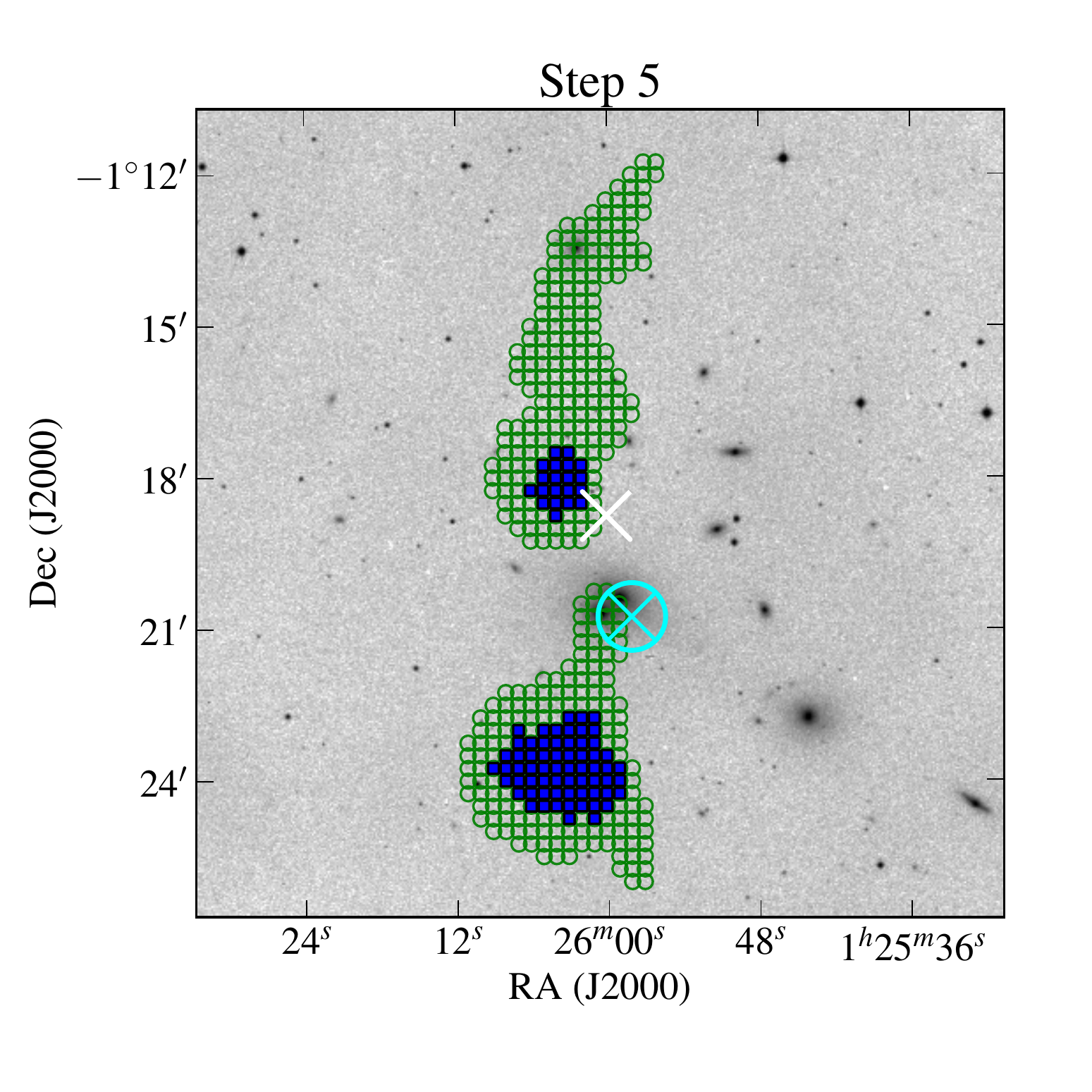}
\caption{The image-level matching algorithm at work (top: UGC~7360, bottom: NGC~0547). In step 1 \& 2 we determine the size of the frame and draw contours. In step 3 we find the pixels above the value of the outer contour (labelled with red plusses) and in step 4 we link these pixels to the optical center of the galaxy (green squares). In step 5, we find the pixels that are connected to the group obtained in the previous step (shown by green circles), these define the radio emission matched to this galaxy. Finally we measure the geometrical center (white cross) and the flux-weighted center (cyan circled cross). For the latter, we use only the pixels that contain 80\% of the total flux of the group (labeled with blue squares).}\label{fig:centers_example}
\end{figure*}

\subsubsection{Image-level rejection}\label{sec:imrej}
We now proceed to remove false matches from the set of 1273 galaxies from the catalog-level matching (section \ref{sec:cmatch}) using the information contained in the pixels of the radio images. We stress that in nearly all cases these false matches are trivially identified (e.g., the left panel of Fig. \ref{fig:0383catmatch}, in this field all galaxies are matched to the radio emission, but the source of this emission is undoubtedly the central galaxy, NGC~0383). The only challenge is to parametrize the judgement of a trained astronomer\footnote{This is a classical problem of Computer Vision; it may likely be solved more efficiently using tools from this field.}. Again our approach is to use the galaxy as a starting point to identify (disconnected) radio emission as originating from a single source (e.g., IC~2722 in Fig. \ref{fig:Newly2}).

Our algorithm consists of the following steps (in Fig. \ref{fig:centers_example} we show two examples).
\begin{enumerate}
\item Make a cut-out of the radio image centered at the coordinates of the galaxy. The width of this image ($I$) is set by the maximum angular distance between the galaxy and the radio sources matched to this galaxy.
\item \label{step:cont} Draw contours on this sub image. The level of the lowest contour is given by $\max(\sigma(I) /7, 3\times \sigma(I_{50}) )$, where $\sigma(I)$ is the standard deviation of the pixels of the image obtained in the previous step and $I_{50}$ is subset of this image containing the pixels below the median of the image.
\item \label{step:g1} Find all pixels contained by lowest contours that are within $\max(\mathrm{FWHM}_i,\, 30")$ of the location of the galaxy, where $\mathrm{FWHM}_i$ is the deconvolved major axis of the radio sources matched to this galaxy. If no pixels are found, we reject this galaxy. Most galaxies are rejected at this step since they are not connected to the radio contours (e.g., left panel of Fig. \ref{fig:0383catmatch}). Radio sources that consist of two lobes (FR~II morphologies) are not rejected because the galaxy connects the two lobes (e.g., UGC~7360 in Fig. \ref{fig:centers_example}).
\item \label{step:g2} Find all pixels within the lowest contour that are connected to pixels of the previous step. Any radio matches outside this group are rejected. This step removes unrelated radio emission that has been matched at the catalog-level (e.g., the head-tail source that was initially matched to NGC~0547, Fig. \ref{fig:centers_example}). 
\item \label{step:c1} If more than one elliptical Gaussian from the radio catalogs is matched to a galaxy, we try to estimate the center of the radio emission.
\begin{figure*}[t!]
\centering
{\includegraphics[trim=11mm 2mm 8mm 5mm, clip, width=.48\textwidth]{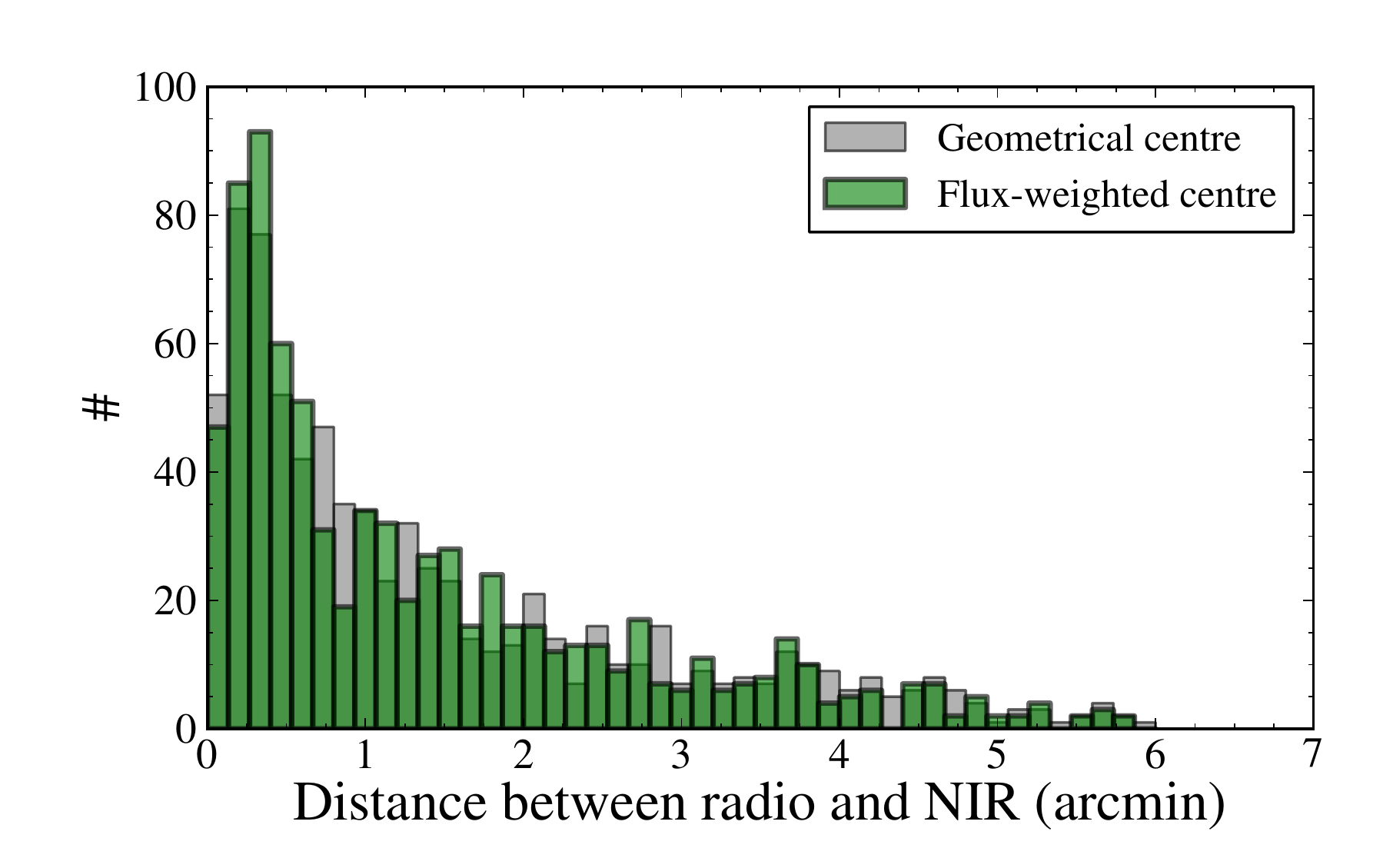}} \quad
{\includegraphics[trim=11mm 2mm 8mm 5mm, clip, width=.48\textwidth]{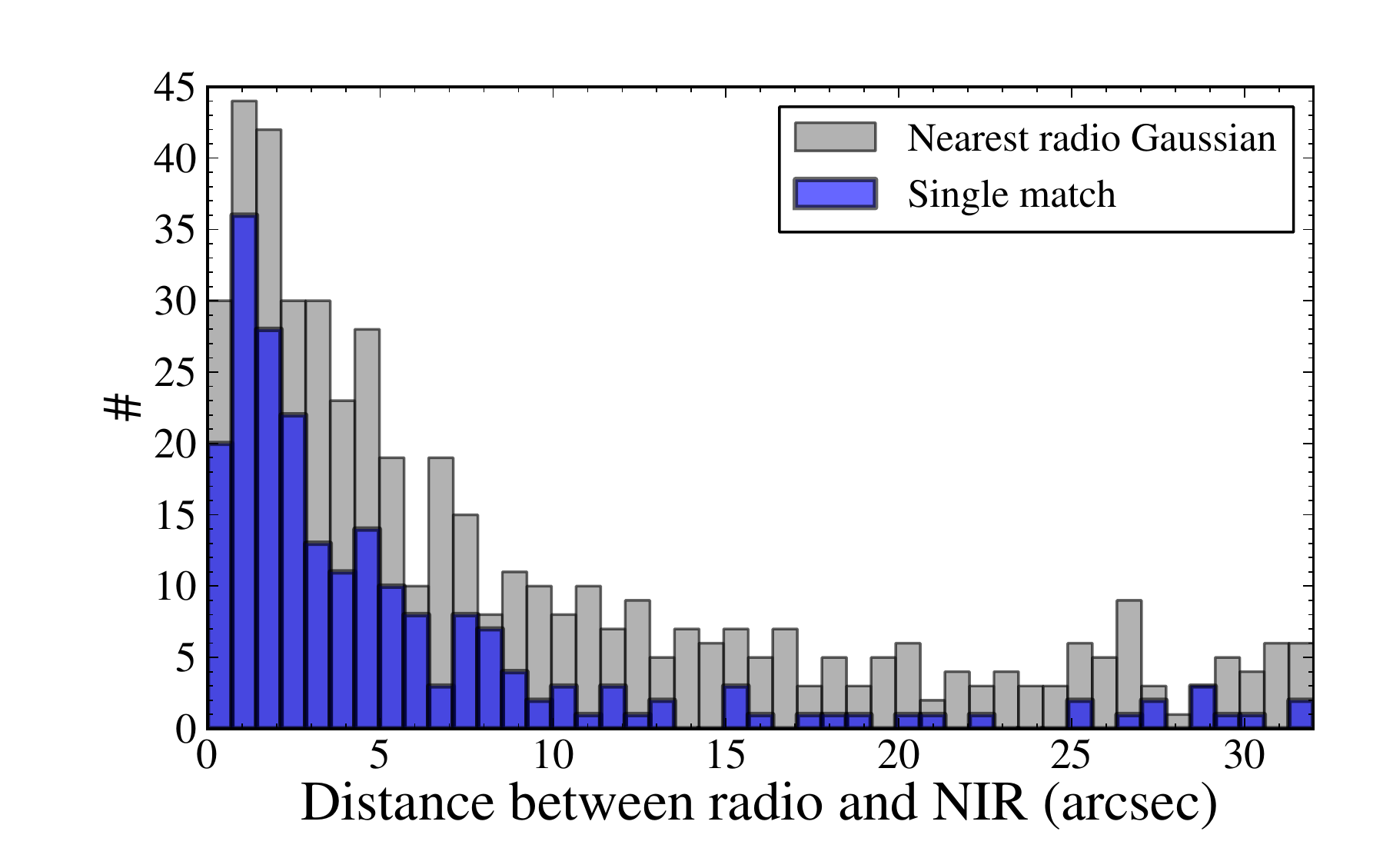}}
\caption{Histogram of angular distance between the NIR center of the galaxy (i.e., 2MRS coordinates) and different measure of the center for the radio emission for the candidate matches from the catalog cuts (Eq. \ref{eq:creq}). Left: the geometrical center (Eq. \ref{eq:geocen}) and the flux-weighted center (Eq. \ref{eq:fluxcen}). Right: the minimum angular distance between the galaxy and any of the matched radio sources for galaxies with a single match (thick blue line) and multiple matches (thin grey line).}\label{fig:hdist}
\end{figure*}
We compute the geometrical center by minimizing
\begin{eqnarray}\label{eq:geocen}
  d_{i} =\sum_{j} d_{ij} \quad,
\end{eqnarray}
where $d_{ij}$ is distance between pixel $i$ and $j$ and the sum runs over all pixels in the group (i.e., pixels obtained in step \ref{step:g2}). We also compute a flux-weighted center by minimizing
\begin{eqnarray}\label{eq:fluxcen}
    d_{i} =\max(d_{ij} w_{ij}) \quad,
 \end{eqnarray}
where the weight $w_{ij}$ is unity for the pixels that contain 80\% of the total flux in the group and zero otherwise. 
The geometrical method (Eq. \ref{eq:geocen}) yields a robust measurement of the center for radio sources that have at least one symmetry axis that intersects the location of the galaxy, while the flux-weighted center gives a better estimate for radio sources with two bright spots at equal distance to the galaxy. 
\end{enumerate}  
Step \ref{step:g1} rejects 412 galaxies that are not close enough to the outer contour of the radio emission.  After removing unrelated radio emission in Step \ref{step:g2}, we al left with 769 galaxies above the radio flux limit. 

To define our final sample, we use a cut on the angular distance of the galaxy to the geometrical and the flux-weighted center. Since these measures of the center are not alway appropriate (e.g., for head-tail sources), we also consider the minimum angular distance between the galaxy and the radio matches as well as the minimum angular distance between the galaxy and the second-highest contour (if the galaxy is within this contour we set this distance to zero). In Fig. \ref{fig:hdist} we show the distribution of these distances. Based on these distributions, we adopt the following criteria for galaxies with multiple radio matches:
\begin{eqnarray}
d&<& \left\{
  \begin{array}{l l}
    1.2'  & \quad \mbox{to geometrical center} \\
    1.2' & \quad \mbox{to flux-weighted center} \\
    15"  &\quad \mbox{to second highest contour}  \\
    8"  &\quad \mbox{to nearest radio catalog entry } \label{eq:dcute}
      \end{array} 
    \right.   
\end{eqnarray} 
Galaxies with multiple radio matches should pass \emph{at least one} of these angular distance criteria.
For galaxies with a single radio match, the flux-weighted or geometrical center should not be used. Instead, we require at least one of the following two criteria:
\begin{eqnarray}
  d&<&  \left\{
    \begin{array}{l l}
      8" &\quad  \mbox{to  radio  catalog entry}\\
      0" & \quad  \mbox{to second highest contour} 
    \end{array} 
  \right.  \label{eq:dcutp}
\end{eqnarray}
The second of these two requirements implies that the coordinates of the galaxy are \emph{within} the second highest radio contour. For point sources this requirement is superfluous (the galaxy always lies within the highest contour). For extended radio emission, however, the galaxy can be within the second highest contour while the distance to the radio catalog entry is $>8"$.

The image-level cuts (Eqs. \ref{eq:dcute} \& \ref{eq:dcutp}) reject 119 galaxies; 78 sources are flagged for manual inspection because more than one galaxy is matched to the same radio source. Besides these 78 ambiguous matches, we will also inspect the rejected and accepted matches to measure how well our image-level algorithm is preforming.

\subsubsection{Manual inspection \& classification}\label{sec:mfinal}
The sample obtained after the catalog-level matching (Eq. \ref{eq:creq}) is small enough to inspect all matches. We stress that, at this point, we do not add new sources to the sample, the only goal is to identify spurious or ambiguous results from the automatic matching pipeline. 

For 51 galaxies the radio emission matched to the galaxy by the image-level algorithm (sec \ref{sec:imrej}) was rejected after human inspection, hence the purity of the sample after our fully automated image-level matching procedure is 91\%. Some of these initially accepted matches are very hard to classify or reject automatically (e.g., blended sources or, very rarely, Galactic emission), but most are due to artifacts in the radio images. We found that in only six cases, our image-level cuts (Eqs. \ref{eq:dcutp} \& \ref{eq:dcute}) rejected a galaxy while at inspection we found this to be genuine match, implying an efficiency of 99\%. 

The sources above the flux limit (sec \ref{sec:fluxlims}) that remain after manual inspection comprise the final sample: 575 radio-emitting galaxies. Using the morphology of the radio emission, we classify these into four categories, the number of sources in each class is given in brackets.
\begin{itemize}
\item \emph{Point Sources (97)}: a single, unresolved radio source.
\item \emph{Starforming galaxies (52)}: extended radio emission that coincides with the extended NIR emission from the galaxy. 
\item \emph{Jets \& Lobes (407)}: resolved radio emission beyond the NIR image of the galaxy that appears to originate from the center of galaxy. We did not attempt to subdivide this sample using the \citet{Fanaroff74} classification scheme because most sources with an FR~II morphology would not be classified as such based on their radio luminosity. 
\item \emph{Unknown (19)}: if none of the above classes apply. These sources are \emph{not} removed from our sample because they formally pass our image-level cuts (Eqs. \ref{eq:dcutp} \& \ref{eq:dcute}). Yet their morphology suggests that the radio emission is not associated with the host galaxies. Users of the catalog may decide to remove the sources in this class.
\end{itemize}
We stress that these classes are based on morphology only, they are not complete (i.e., some starforming galaxies or jets are classified as point sources because the radio emission is not resolved).

\begin{table*}[t]
\caption{Columns of the master catalog}\label{tab:master}
\centering
\begin{tabular}{l c p{290pt} c c}
\hline \hline
  Column name & Units & Description & Source\tablefootmark{*} \\
\hline
 \verb 2MASX & & Target name of the object from 2MRS database & 2MRS \\ 
  \verb NED_id & & NED id of galaxy  &  NED \\ 
  \verb ra  & deg (J2000) & Right Ascension of galaxy & 2MRS \\ 
  \verb dec  & deg (J2000) & Declination of galaxy & 2MRS \\ 
  \verb l & deg  & Galactic longitude & 2MRS \\ 
  \verb b & deg  & Galactic latitude & 2MRS \\ 
  \verb Kmag & mag (AB) & $K$-band isophotal magnitude, corrected for Galactic extinction & 2MRS \\
  \verb Kmag_err & mag (AB) & uncertainty on \verb K  & 2MRS \\
  \verb z &   & Heliocentric redshift  & 2MRS\\
  \verb zdist &  &  Peculiar-velocity corrected redshift  & \ref{sec:2MRS} \\
  \verb zdist_err  &  & Uncertainty on \verb z_dist & \ref{sec:2MRS} \\
  \verb D & Mpc & Inverse-variance weighted mean of distance from NED-D and peculiar velocity corrected Hubble distance ($h=0.72$) & \ref{sec:2MRS} \\ 
  \verb D_err & Mpc & Uncertainty on the former row  & \ref{sec:2MRS} \\ 
  \verb gal_type & & Galaxy morphological type code: -9 to 9, encodes Hubble sequence, 98 if galaxy has never been examined \citep[][Table A8]{Huchra12} & 2MRS\\
  \verb n_nvss & \# &  Number of radio Gaussians matched to galaxy (after manual inspection) & NVSS, sec. \ref{sec:mfinal} \\
  \verb n_sumss & \# &  Number of radio Gaussians matched to galaxy (after manual inspection) &  SUMSS, sec. \ref{sec:mfinal} \\ 
  \verb F1400 & mJy & Sum of integrated flux at 1.4~GHz of all radio matches (zero if \verb n_nvss=0 ) & NVSS \\ 
  \verb F1400_err & mJy & Uncertainty on the former row & NVSS \\ 
  \verb F843 & mJy & Sum of integrated flux at 843~MHz of all radio matches (zeros if \verb n_sumss=0 ) & SUMSS \\ 
  \verb F843_err & mJy & Uncertainty on the former row  & SUMSS \\ 
  \verb Fsyn & mJy &  Flux at 1.1~GHz, obtained from 1.4~GHz or 843~MHz flux using $\alpha=-0.6$ & NVSS or SUMSS\tablefootmark{a} \\ 
 \verb Lsyn & ${\rm erg}\,{\rm s}^{-1}$ & $\nu L_\nu$ at 1.1~GHz & NVSS or SUMSS\tablefootmark{a}\\ 
  \verb sum_ma & deg & Sum of deconvolved FWHM of the major axis of all matched radio Gaussians & NVSS or SUMSS\tablefootmark{a} \\ 
  \verb max_ma & deg & Largest deconvolved major axis FWHM of all matched radio sources & NVSS or SUMSS\tablefootmark{a} \\ 
  \verb lim_ma  & bool & Limit flag on major axis (=0 if source is resolved) & NVSS or SUMSS\tablefootmark{b}  \\
  \verb max_dist_to_gal & deg & Maximum angular distance between the galaxy and the radio matches& NVSS or SUMSS \\
  \verb min_dist_to_gal & deg & Minimum distance between the galaxy and the radio matches& NVSS or SUMSS\\
  \verb contour_dist & deg & Angular distance to the second-highest radio contour (zero if galaxy is within contour) & sec. \ref{sec:imrej} \ \\
  \verb geo_cen_ra & deg & RA of the geometrical center &  Eq. \ref{eq:geocen}\tablefootmark{c} \\
  \verb geo_cen_dec & deg & Decl. of the geometrical center &  Eq. \ref{eq:geocen}\tablefootmark{c} \\
  \verb flux_cen_ra & deg & RA of the flux-weighted center & Eq. \ref{eq:fluxcen}\tablefootmark{c} \\
  \verb flux_cen_dec & deg & Decl. of the flux-weighted center &  Eq. \ref{eq:fluxcen}\tablefootmark{c} \\
  \verb class & & Classification based on morphology: \verb p =\emph{Point Sources}, \verb g =\emph{Starforming galaxies,} \verb j =\emph{Jets \& Lobes}, \verb u =\emph{Unknown} & sec. \ref{sec:mfinal} \\
\hline
\end{tabular}
\tablefoot{\tablefootmark{*}{We lists what catalog provided the input for each parameter of our master catalog, we refer to a section or equation in this paper for parameters that not trivially obtained from the original catalog.} In the overlap region ($-40<\mathrm{Decl.}<-30$) entries from both SUMSS and NVSS are matched to the same galaxy. How this information is combined depends on the parameter in question: \tablefootmark{a}{when both NVSS and SUMSS matches are available, we the pick the largest value.} \tablefootmark{b}{when both NVSS and SUMSS data is available, we raise the major axis limit flag if the source is not resolved in both catalogs.}  \tablefootmark{c}{when both NVSS and SUMSS matches are available, we use the mean.}  }
\end{table*}

\subsection{Catalog columns}\label{sec:columns}
We combine the properties of the 575 galaxies of our final sample into a single master catalog; the columns of this catalog are described in Table \ref{tab:master}.  In Table~\ref{tab:near} we list the 50 most luminous sources within 120~Mpc. A selection of the full catalog is shown in Appendix \ref{sec:cats}. Since for some applications it is useful to have acces to the individual radio matches per galaxy, we also provide a basic catalog, listing all matched radio Gaussians per galaxy.

\subsection{Completeness}\label{sec:completeness}
\begin{figure}[t]
\centering
\includegraphics[trim=12mm 2mm 8mm 10mm, clip, width=.48\textwidth]{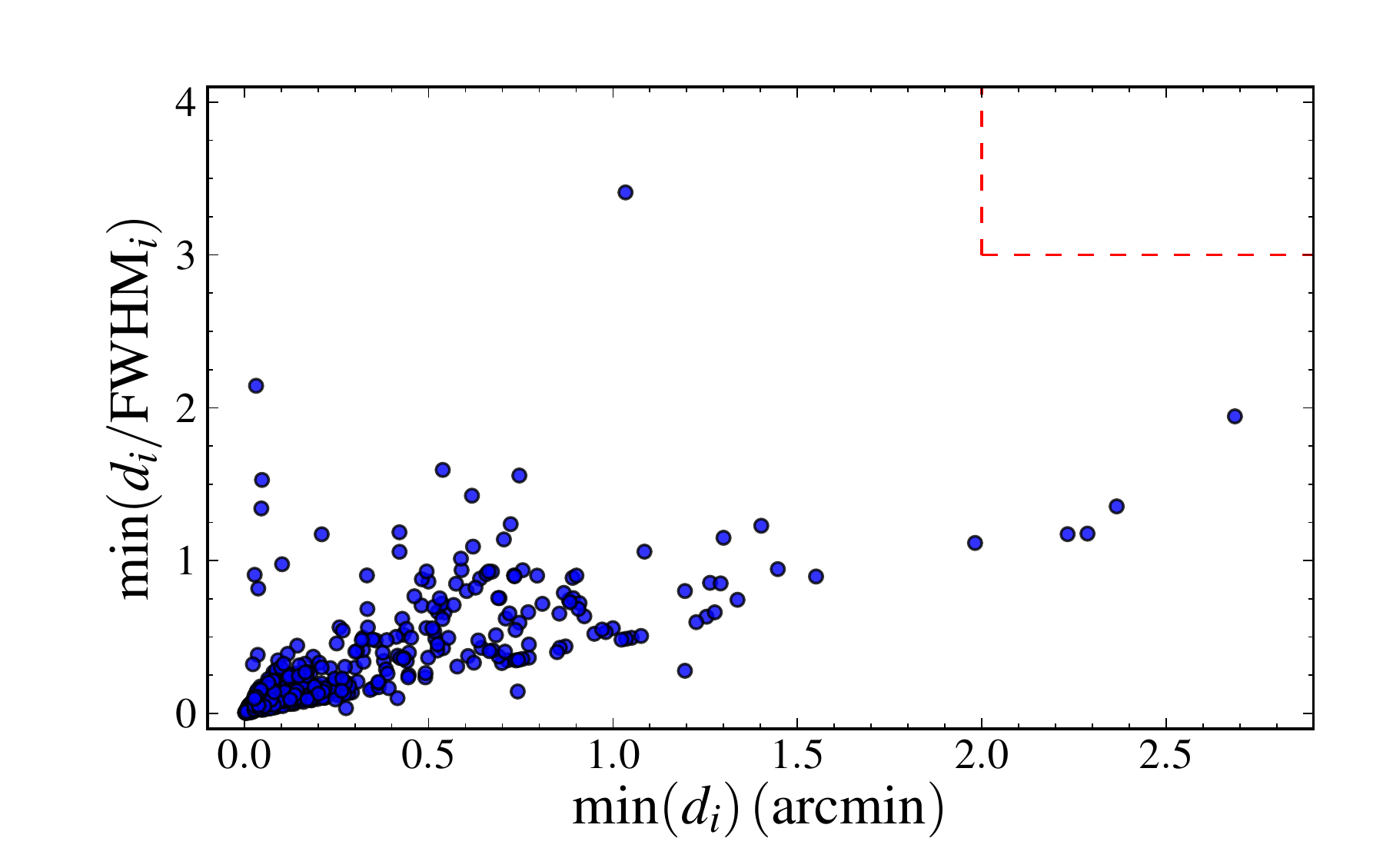}
\caption{The minium angular distance between the radio matches and center of the galaxy ($d_i$) and the minimum of this distance weighted by the major axis ($d_i/ {\rm FWHM}_i$) for the 575 galaxies of our master catalog. These parameters were used to select the sample of potential radio emitting galaxies (Eq. \ref{eq:creq}). We see that our cuts, $d_i<\max(2', 3\times {\rm FWHM}_i)$ (dashed red lines), are not limiting the number of galaxies in the final sample.}\label{fig:finaldist}
\end{figure}

The 575 galaxies of our master catalog are a subset of the potential matches obtained from the catalog parameters (sec. \ref{sec:cmatch}). We can check if the latter sample was large enough by comparing the minium angular distance between the radio Gaussians and the galaxies to the cuts that we have adopted. In Fig. \ref{fig:finaldist} we see that the catalog selection is not limiting the size of the final sample, because all galaxies are well within one of the two requirements placed by Eq. \ref{eq:creq}. 

We also confirm that the small difference in frequency between NVSS (1.4~GHz) and SUMSS (843~MHz) does not lead to a noticeable selection effect: the number of NVSS sources over the number of SUMSS sources is 3.65, which is very close to fraction of the area (within the 2MRS footprint) probed by these surveys, 3.74.

Since we manually inspected all matches obtained from the catalog-level selection (sec. \ref{sec:cmatch}), the completeness of our master catalog is primarily determined by the completeness of the input catalogs. As discussed in section \ref{sec:input}, the completeness of NVSS and SUMSS is nearing 100\% at our flux limits and the completeness of 2MRS is 97.6\%. However, at the radio flux limit of our catalog, the systematic uncertainty on the total flux of sources with a complex radio morphology slightly decreases the completeness; a few sources may be missing from our final sample because our estimate of the total flux is below the true value of the total flux. This is not a serious limitation since most complex sources have been described in the literature which allows us to improve the estimate of their total flux (see section \ref{sec:othercat}). 
\begin{table} 
 \caption{\label{tab:near}The 50 most luminous radio sources within 125 Mpc}
 \begin{tabular}{p{110pt} c c c}
\hline 
NED name & $\nu L_\nu$ & $D$ & Class \\ & (${\rm erg}\,{\rm s}^{-1}$) & (Mpc) &  \\
\hline \hline 
ESO 075- G 041 & $41.80$ & $120$ & j\\
NGC 1275 & $41.37$ & $79$ & p\\
2MASX J17131541-2502266 & $41.26$ & $122$ & j\\
UGC 01841 NOTES01 & $41.00$ & $98$ & j\\
NGC 1128 NED02 & $40.97$ & $97$ & j\\
MESSIER 087 & $40.92$ & $18$ & j\\
NGC 5532 & $40.92$ & $105$ & j\\
NGC 3862 & $40.90$ & $91$ & j\\
Fornax A & $40.87$ & $21$ & j\\
NGC 7236 & $40.87$ & $116$ & j\\
IC 4296 & $40.84$ & $51$ & j\\
2MASX J12543570-1234070 & $40.79$ & $69$ & j\\
CGCG 114-025 & $40.77$ & $70$ & j\\
PKS 0718-34 & $40.77$ & $119$ & j\\
UGC 11465 & $40.76$ & $106$ & j\\
NGC 0315 & $40.73$ & $70$ & j\\
NGC 6048 & $40.63$ & $114$ & j\\
NGC 0383 & $40.63$ & $72$ & j\\
NGC 6251 & $40.60$ & $110$ & j\\
ESO 460- G 004 & $40.59$ & $105$ & j\\
NGC 0547 & $40.59$ & $77$ & j\\
NGC 5090 & $40.45$ & $47$ & j\\
UGC 02755 & $40.45$ & $108$ & j\\
UGC 12064 & $40.42$ & $74$ & j\\
Cen A & $40.37$ & $3.6$ & j\\
2MASX J17124278-2435477 & $40.36$ & $103$ & j\\
2MASX J01532586+7115067 & $40.35$ & $102$ & j\\
NGC 1044 NED01 & $40.31$ & $91$ & j\\
2MASX J17204089-0111573 & $40.29$ & $124$ & u\\
CGCG 403-019 & $40.27$ & $119$ & j\\
UGC 7360 & $40.26$ & $32$ & j\\
NGC 1167 & $40.23$ & $74$ & p\\
NGC 5127 & $40.22$ & $71$ & j\\
UGC 02783 & $40.12$ & $91$ & j\\
ESO 443- G 024 & $40.11$ & $73$ & j\\
NGC 6328 & $40.11$ & $59$ & p\\
UGC 11294 NOTES01 & $40.07$ & $67$ & j\\
UGC 07115 & $40.06$ & $96$ & j\\
NGC 0741 & $40.06$ & $80$ & j\\
IC 0383 & $40.04$ & $117$ & j\\
2MASX J17354374-0720527 & $40.02$ & $104$ & j\\
NGC 4696 & $40.00$ & $42$ & p\\
NGC 3343 & $40.00$ & $92$ & j\\
NGC 6466 & $39.97$ & $100$ & j\\
NGC 7075 & $39.96$ & $75$ & j\\
UGC 12127 & $39.96$ & $123$ & j\\
NGC 2329 & $39.95$ & $85$ & j\\
2MASX J05444416+1648501 & $39.95$ & $85$ & j\\
NGC 5141 & $39.95$ & $78$ & j\\
IC 4374 & $39.95$ & $90$ & p\\
 \end{tabular}
 \end{table} 

\subsection{Probability of random matches}\label{sec:randommatch}
To estimate the contamination due to background radio sources we shuffle the R.A. of the 2MRS galaxy catalog to obtain a randomized, uniform sample. The catalog-level cuts (Eq. \ref{eq:creq}) yield 233 galaxies with a total of 1332 matches to the radio catalogs. Our image-level algorithm rejects 218 of these, leaving only 15 galaxies. After checking this sample for imaging artifacts, we are left with 13 sources that genuinely passed the image-level cuts (Eqs. \ref{eq:dcute} \& \ref{eq:dcutp}). Two of the 13 matches with the randomized 2MRS catalog are radio sources that have already been unambiguously identified as local radio-emitting galaxies using the original 2MRS data, hence these should not be included in the estimate of the number of background matches. 

All of the matches of the shuffled galaxy sample are to extended radio sources; most are background double sources whose geometrical center is within 1.2~arcminutes of the galaxy. As expected, they are classified into the \emph{Unknown} morphological class (sec. \ref{sec:mfinal}). To conclude, the expected number of background matches for extended radio sources in our catalog is $\approx 11$, or  2\% contamination. This number is of the same order as the number of galaxies in our catalog that have an unclassifiable radio morphology, confirming that most of these are very likely to be random matches.

\subsection{Comparison to other catalogs}\label{sec:othercat}
We now compare our sample of 575 radio-emitting galaxies with some well-known catalogues of extra-galactic radio sources.

 First, we consider the \citet{Kuehr81} catalog, which is essentially complete for sources with $F_{5000}>1$~Jy at 5~GHz and covers roughly the whole sky excluding $|b|<10^{\circ}$.  Three radio galaxies with $z<0.05$ from this catalog are not found in our final sample: NGC~1265 ($K=12.4$, $z=0.025$), ESO~252-GA018 ($K=12.0$, $z=0.034$), and NGC~7385 ($K=13.3$, $z=0.026$). The $K$-band flux of these galaxies is \emph{below} the flux limit of 2MRS ($K<11.75$), hence by construction they are not included in our final sample. We shall not add these three galaxies to our final sample by hand because this would dramatically complicate the selection function ---they would need to be excluded to obtain a well-defined volume-limited sample.
We can also test the completeness of the \citet{Kuehr81} catalog using our list of final sources. We limit our sample using the completeness limit of the Kuehr catalog $F_{5000}>1$~Jy, converted to NVSS and SUMSS using $\alpha=-0.6$ and restrict to at $|b|>10$, to find 16 radio-emitting galaxies that are not listed in \citet{Kuehr81}. 

Next, we compare our catalog to that of \citet*{Condon02}, obtained by a careful manual selection of UGC galaxies that are detected in NVSS.  We restrict this catalog using our $K$-band and radio flux limits to 131 sources and find that five of these are not contained in our final sample. All five are nearby ($z<0.0017$) starforming galaxies whose very extended, low surface brightness radio emission has not been fully included in the NVSS catalog, causing them to fall (just) below our flux limit. For the main science goal of this work (``find all local radio galaxies''), these five galaxies are irrelevant. 

We also compare the total flux measured at 1.4~GHz by \citet{Condon02} by a re-analysis of the NVSS images to our estimate using the sum of the fitted elliptical Gaussian listed in the NVSS catalogs. The flux they report is on average $0.04 \pm 0.13$~dex, higher than our work. Six radio galaxies show an offset larger than 0.3~dex: NGC~0315, NGC~891, IC~342 NGC~5127, NGC~6946, and NGC~7236.
For NGC~0315, this is due a hotspot in the jet that has been missed by the friends-of-friends algorithm. The missing flux for the other sources is due to their complex radio morphology that is not entirely captured by the superposition of elliptical Gaussians. For these six cases, we use the total flux reported by \citet{Condon02} as final value for the flux in our catalog, this reduces the the flux offset of the other 125 sources that are common to both catalogs to $0.010 \pm 0.06$~dex.

We should also compare our catalog to that of \citet*{Jones92}, 193 southern extra-galactic radio sources with a flux density greater than 0.4~Jy and a size greater than 0.5'. After restricting this catalog to galaxies detected in 2MRS, we find three sources that are not in our final sample: WKK~4452,  ESO~137-G006 and ESO~137-G00. All three, however, are detected below $|b|<10$, which is outside the footprint of SUMSS (and thus outside the footprint of our catalog). In a future release, we will complete our coverage of the southern sky below $|b|=10$ using the MGPS-2 images and these radio galaxies will be included. 

We find a mean ratio between the 843~MHz flux measured by \citet{Jones92} and this work of $0.015\pm0.17$~dex. Only Fornax~A showed an offset greater than 0.3~dex, which is due the complex morphology of this source.  For this galaxy, we use the total flux measured by \citet{Jones92} as final value for the flux in our catalog (this reduces the flux offset to $-0.014\pm0.1$~dex). We also manually adjusted our catalog flux of Cen~A to the total flux (i.e., inner and outer lobes) as measured by \citet{Cooper65}. 

Of the 575 radio sources in our final sample, 209 are \emph{not} contained in the union of the three catalogs described above, plus the catalog of radio sources in the 6dFGS survey \citep{Mauch07} and the collection of known extra-galatic radio sources and AGN of \citet{Veron74,VCV06}.

\subsection{Newly found radio galaxies}\label{sec:notes} 
In the previous paragraph we found that over 30\% of the galaxies in our sample are not contained in existing large-area samples of extra-galactic radio sources or AGN. Some of these have been identified as such by surveys that cover a smaller area of the sky, but in many cases this work is the first to classify these galaxies as radio emitting. Below we discuss some notable examples of these genuinely new identifications. We also list some sources that have been identified as (candidate) extra-galactic sources (e.g., based on their radio morphology or radio spectrum) by earlier work, but for which this work is the first to provide a redshift of the galaxy. The NED name of the galaxy can be used to find the images of these sources in Appendix \ref{sec:atlas}.
\begin{itemize}
  \item J00112171+5231437 blended point source or head-tail source, member of a cluster that contains a radio relic \citep{2011A&A...528A..38V}.  
  \item NGC 0349 jet over 5' long,  non-detection in the CRATES catalog of flat-spectrum radio sources  \citep{2007ApJS..171...61H}. 
  \item J03204016+2727485 complex geometry, multiple hotspots?
  \item J03212595+1806093 large ($5'$) radio galaxy, shows both lobe emission and strong central emission. 
  \item NGC 1477 complex morphology, strong emission from galaxy center and faint lobes, contained in the CRATES catalog \citep{2007ApJS..171...61H}. 
  \item J05444416+1648501 extended jet emission, observed but not detected at VLBI scales: VERA 22~GHz flux $<0.1$~Jy \citep{2007AJ....133.2487P}. 
  \item J06120351-3257472 large source with a triple morphology (detected in both NVSS and SUMSS), galaxy is member of a rich cluster \citep{1980ApJS...42..565D}.
  \item J07331844-3654533 two large lobes ($\sim 10^2$~kpc), detected in both NVSS and MGPS-2.     
   \item J11032753-4657471 large surface brightness difference between the two lobes. 
  \item IC 2722  ``relaxed FR~II'', looks like central engine has turned off. 
  \item ESO 505- G 014 spectacular narrow and long (10') jet. 
  \item J13595556-1056266 clearly extended jet emission, detected by Chandra \citep{2010ApJS..189...37E}.
  \item J16103572-0511173 jet morphology, but galaxy is not exactly at the center of the radio emission.
  \item J16390277-6505079 complex morphology, seems to show a hot spot displaced from current jet axis. Galaxy is located in the Great Attractor region \citep{2006MNRAS.369.1131R}.
  \item J17050125-2445099 two lobes and core emission, in Ophiuchus region \citep{2000MNRAS.316..326H}.
  \item J17124278-2435477 complex morphology, in Ophiuchus region \citep{2000MNRAS.316..326H}.
  \item J17131541-2502266 two lobes, with $\nu L_\nu=2\times 10^{41}\,{\rm erg}\,{\rm s}^{-1}$, this is one of brightest radio sources within 125~Mpc. It was identified as a potential double source (with unknown redshift) by \citet*{1982PASAu...4..278S}. 
  \item J19264081+4123284 head-tail source.
  \item J19300192-1509191 resolved radio emission, most likely central jet emission. Detected by ROSAT \citep{1999A&A...349..389V}, radio, X-ray and optical (without redshift) matched by \citet{2000ApJS..129..547B}.
  \item J20430830-5059054 resolved radio emission, morphological class not clear.
  \item J21492796-6429194 slightly resolved, identified as radio source in the direction of the Abell cluster (i.e., without redshift) by \citet{1990MNRAS.247..387R}.
  \item J22380121+4107363 two non-axisymmetric lobes.
  \item MCG -05-55-032 wide-angle morphology, classified as a point source in the CRATES catalog \citep{2007ApJS..171...61H}. 
\end{itemize}

\section{Analysis}\label{sec:ana}

The 575 radio-emitting galaxies we identified  in the previous section are a subset of a much larger flux-limited sample of ``normal'' galaxies, which allows for detailed comparison of both classes. In section we will work with two samples of radio galaxies. We define sample~A, the volume-limited sample that is used to measure the Hubble type abundance (sec. \ref{sec:volsample}) and clustering (sec. \ref{sec:cc}) of radio galaxies. And we define sample~B, which is limited only by the radio luminosity of the galaxy, to compute the luminosity functions (sec. \ref{sec:lumfunc}). A detailed analysis of the magnetic field and jet power of the radio galaxies will be presented in an accompanying paper (van Velzen et al. 2012).

\begin{figure}[t]
\centering
{\includegraphics[trim=12mm 12mm 8mm 10mm, clip, width=.48\textwidth]{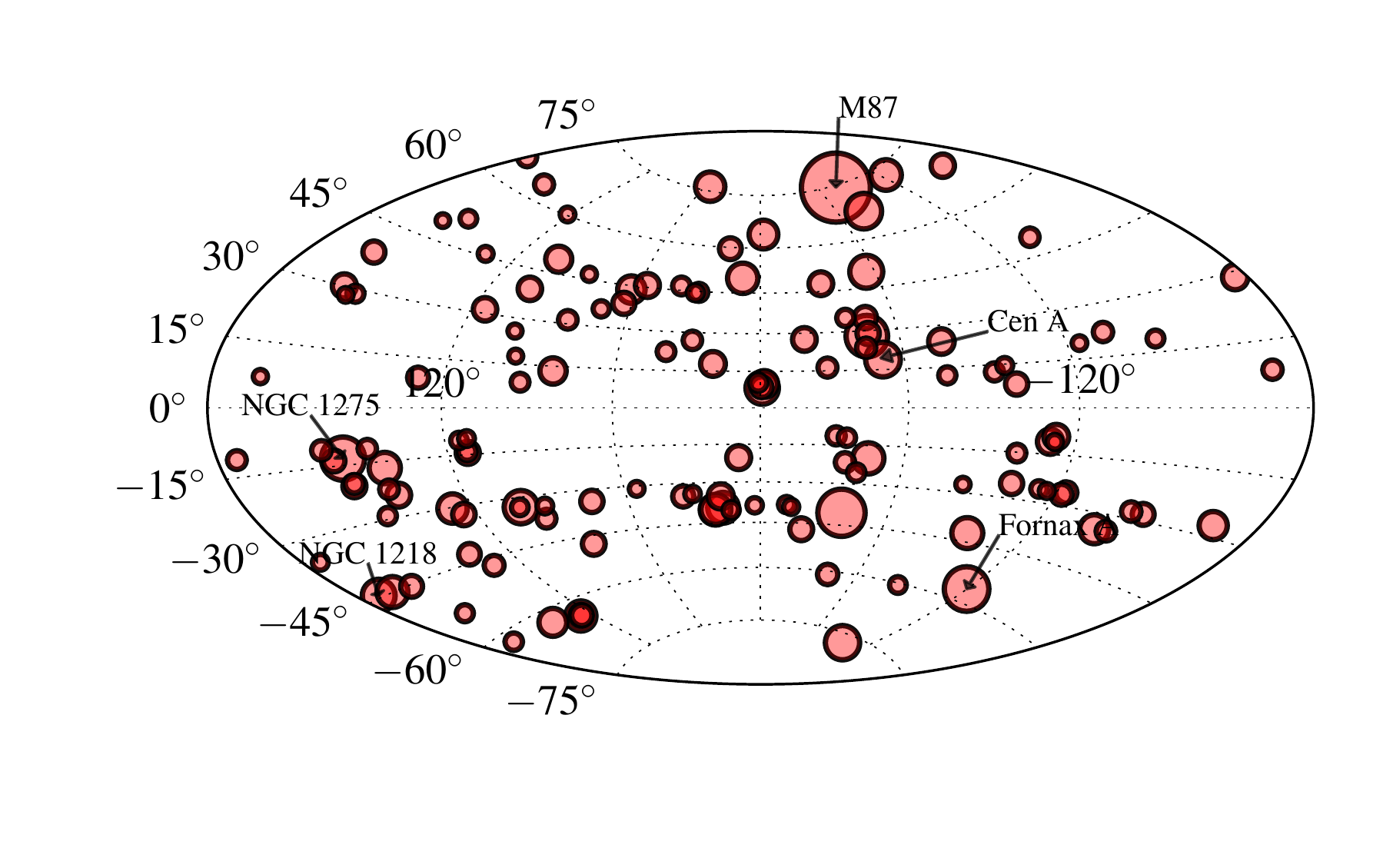}} \caption{Map in Galactic coordinates with the radio galaxies of the volume-limited sample ($z<0.03$). The area of the circles is proportional to the radio flux of the source. The location of some famous sources is indicated (M87 and NGC~1275 are the brightest members of the Virgo and Perseus cluster, respectively).}\label{fig:lb_sky}
\end{figure}

\subsection{Number counts in a volume-limited sample}\label{sec:volsample}
Our master catalog is limited by both the flux limit of the redshift survey and the radio flux limit we imposed. To obtain a volume-limited sample, we thus have to make a cut on both the $K$-band luminosity and the radio luminosity. 
We shall restrict our sample to $z=0.03$, or comoving distance of 130~Mpc; the corresponding minimum $K$-band luminosity is $M_{K}=-23.78$, this limit is just high enough to select all galaxies with a $K$-band luminosity equal to Cen~A ($M_{K}=-23.8$). Applying this distance and NIR luminosity cut to 2MRS yields 10964 galaxies; our master catalog is reduced from 575 to 153 galaxies. 
The flux limit applied to the radio catalogs was $F_{1400}>213$~mJy, $F_{843} >289$~mJy for NVSS, SUMSS, respectively. Hence for $z<0.03$, our catalog contains all sources with a radio luminosity greater than $\nu L_\nu = 5.8\times 10^{39}\,{\rm erg}\,{\rm s}^{-1}$ at 1.4~GHz and $\nu L_\nu = 4.8\times 10^{39}\,{\rm erg}\,{\rm s}^{-1}$ at 843~MHz (or $L_{1400}>4.2 \times 10^{23}\,{\rm W}\,{\rm Hz}^{-1}$, $L_{843}>5.6 \times 10^{23}\,{\rm W}\,{\rm Hz}^{-1}$). Applying this radio luminosity requirement selects 461 galaxies of which 74 are below $z=0.03$, all of these also obey the cut in $K$-band luminosity. None of these radio-emitting galaxies are morphologically classified as \emph{Starforming}. In the following we shall refer to this set of 74 sources within $z=0.03$ as sample~A: the volume-limited sample of radio galaxies. We list the properties of the galaxies in sample~A in Table \ref{tab:A}.

If we restrict sample~A to galaxies with a radio luminosity that is greater or equal than Cen~A ($\nu L_\nu = 2.6 \times 10^{40}\,{\rm erg}\,{\rm s}^{-1}$ at 1~GHz), only 24 radio galaxies within $z=0.03$ remain.

\begin{figure}[t]
\centering
{\includegraphics[trim=5mm 12mm 8mm 5mm, clip, width=.48\textwidth]{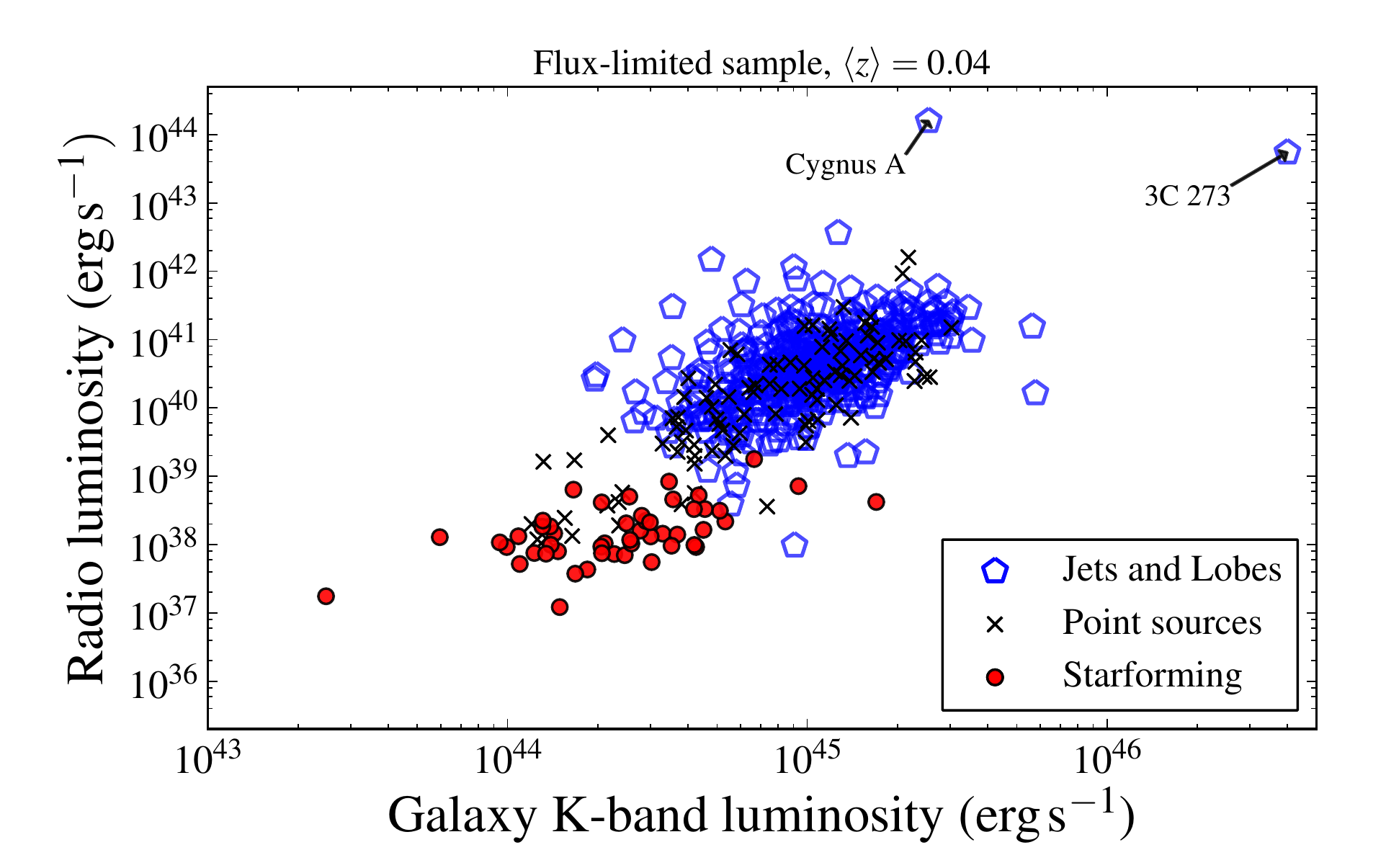}} \\[6pt]
{\includegraphics[trim=5mm 2mm 8mm 5mm, clip, width=.48\textwidth]{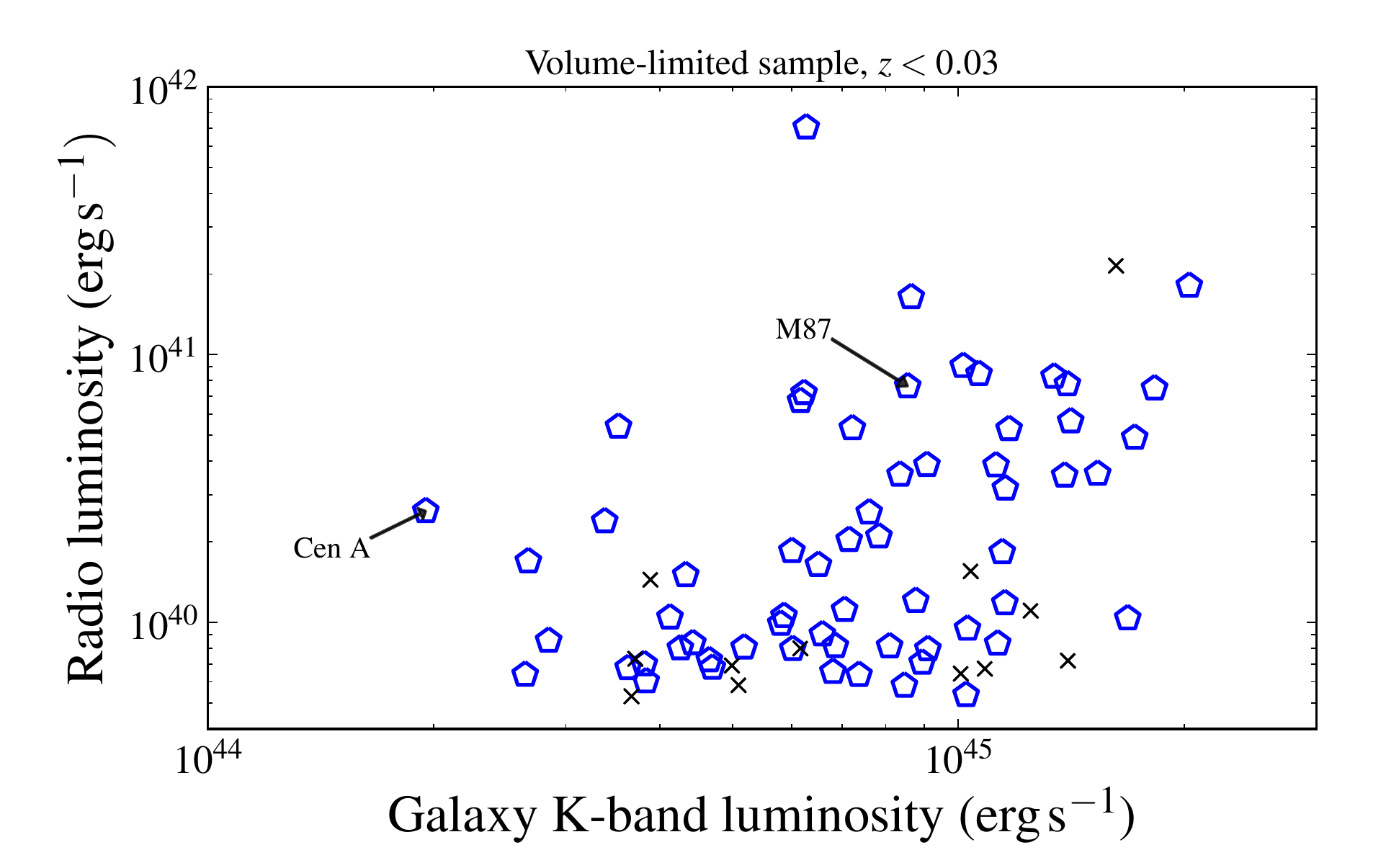}}
\caption{The $K$-band and 1~GHz luminosity ($\nu L_\nu$) for all extra-galactic radio sources in our sample (top) and a volume-limited sample (bottom). Most sources that are morphologically classified as AGN (i.e., the \emph{Jets and Lobes} class) are more luminous than $\nu L_\nu = 10^{39}~{\rm erg}\,{\rm s}^{-1}$. Our volume-limited sample shows that the galaxy NIR luminosity and radio luminosity are not strongly correlated.}\label{fig:K-radio}
\end{figure}

\begin{figure}[t]
\includegraphics[trim=5mm 2mm 8mm 5mm,  width=.48\textwidth]{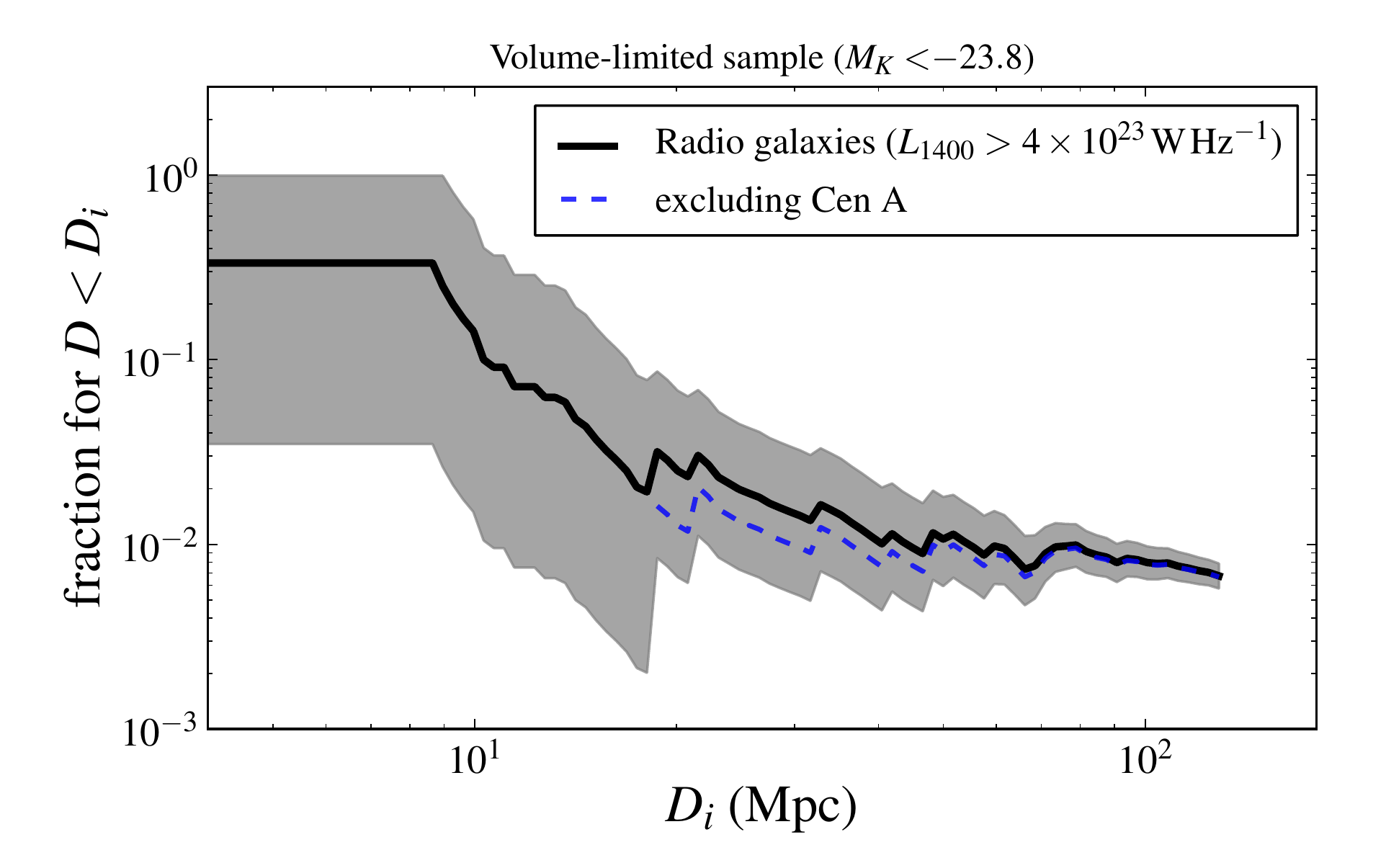}\caption{The fraction of radio galaxies in the volume-limited sample (sample~A, defined in sec. \ref{sec:volsample}) as a function of distance to the local group barycenter. The grey area shows the 90\%~CL, based on Poisson statistics for the number of observed radio galaxies. Since our sample probes over 90\% of the volume, this uncertainty reflects the cosmic variance, rather than the statistical significance of this observation. Cen~A remains the only radio galaxy up to 16~Mpc, where M~87 joins the ranks (at this distance, the number of normal galaxies in the volume-limited sample is 62).}\label{fig:radio-fraction}
\end{figure}

In Fig. \ref{fig:lb_sky} we show a map in Galactic coordinates with the location of the 74 radio galaxies of sample~A. The radio and $K$-band luminosity of both the full flux-limited sample and volume-limited sample are shown in Fig. \ref{fig:K-radio}. The 74 radio galaxies comprise 0.65\% of all galaxies in the volume-limited sample; in Fig. \ref{fig:radio-fraction} we show this fraction as a function of distance to the barycenter of the Local Group.

Using the 2MRS morphological galaxy classification for sources with $K<11.25$, we find the following Hubble type fractions for our volume-limited sample of radio galaxies: 47\% Ellipticals, 47\% S0~galaxies, and 3\% Spirals. For the $\sim 10^4$ normal galaxies within $z=0.03$ the E, S0 and S fractions are 14\%, 31\%, and 54\%; for massive normal galaxies with $M_{K}<-25$, these fraction are 26\%, 40\%, and 33\%. To allow a direct comparison to the galaxy morphology distribution of radio galaxies, we compute the Hubble type fractions of non-radio galaxies in bins of absolute $K$-band magnitude and we compute the mean fraction weighted by the number of radio galaxies in each luminosity bin. We thus find the Hubble type fractions for non-active galaxies drawn from the observed mass distribution of radio galaxies: 27\%, 38\%, and 34\% (E, S0, and S).

\subsection{Clustering of radio galaxies}\label{sec:cc}
\begin{figure}
\centering
\includegraphics[trim=10mm 15mm 10mm 12mm, clip, width=.48\textwidth]{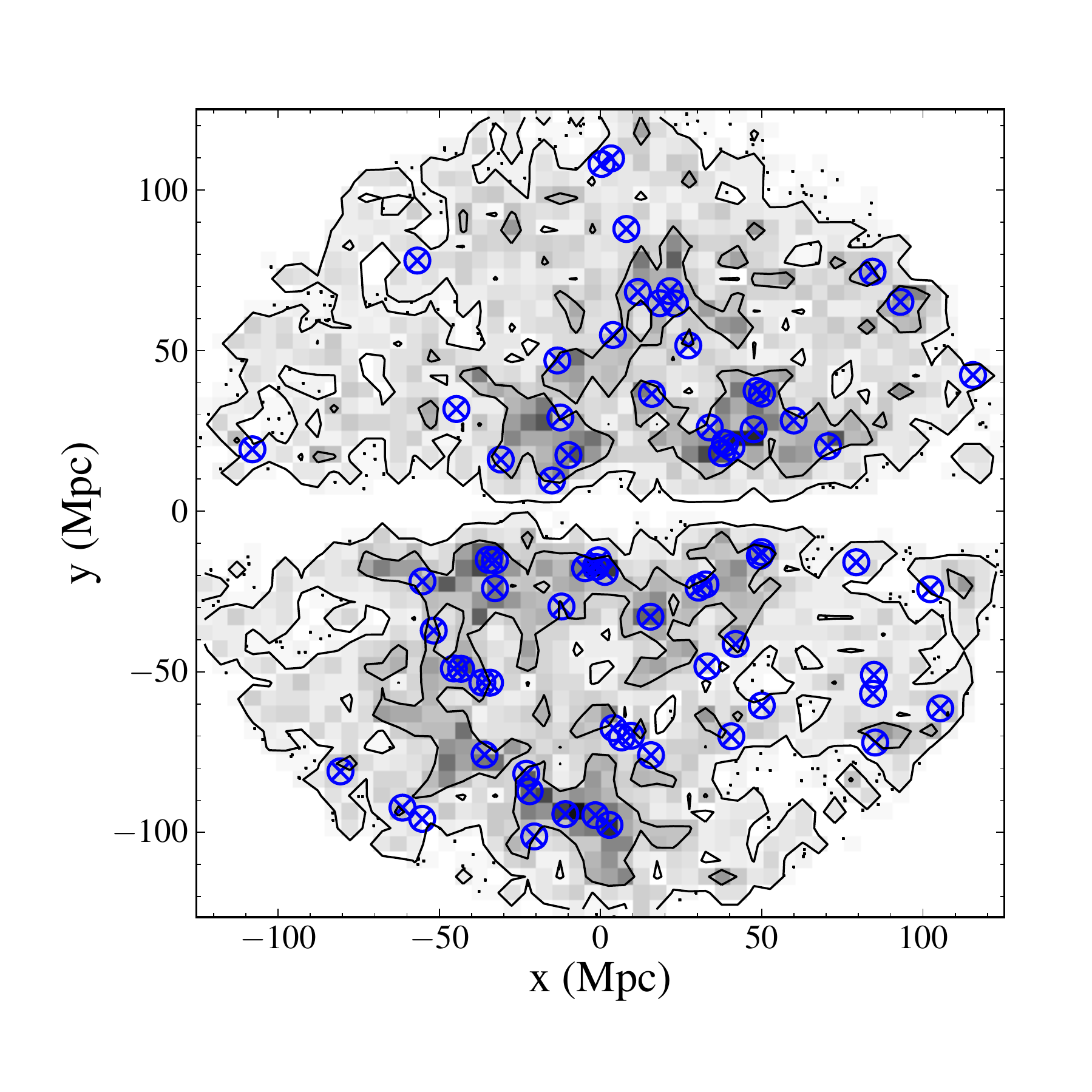}
\caption{Projected density of the volume-limited sample of $\sim 10^4$ galaxies ($z<0.03$) along $b=l=0$ in Galactic coordinates. The contour encompass 95\% and 50\% of the sample. The lack for sources around $y=0$ is due to the disk of the Milky Way. One clearly sees that the radio galaxies in our volume-limited sample (blue circled crosses) reside in regions of high galaxy density.}\label{fig:topview}
\end{figure}

\begin{figure}[t]
\includegraphics[trim=5mm 17mm 6mm 5mm, clip, width=.48\textwidth]{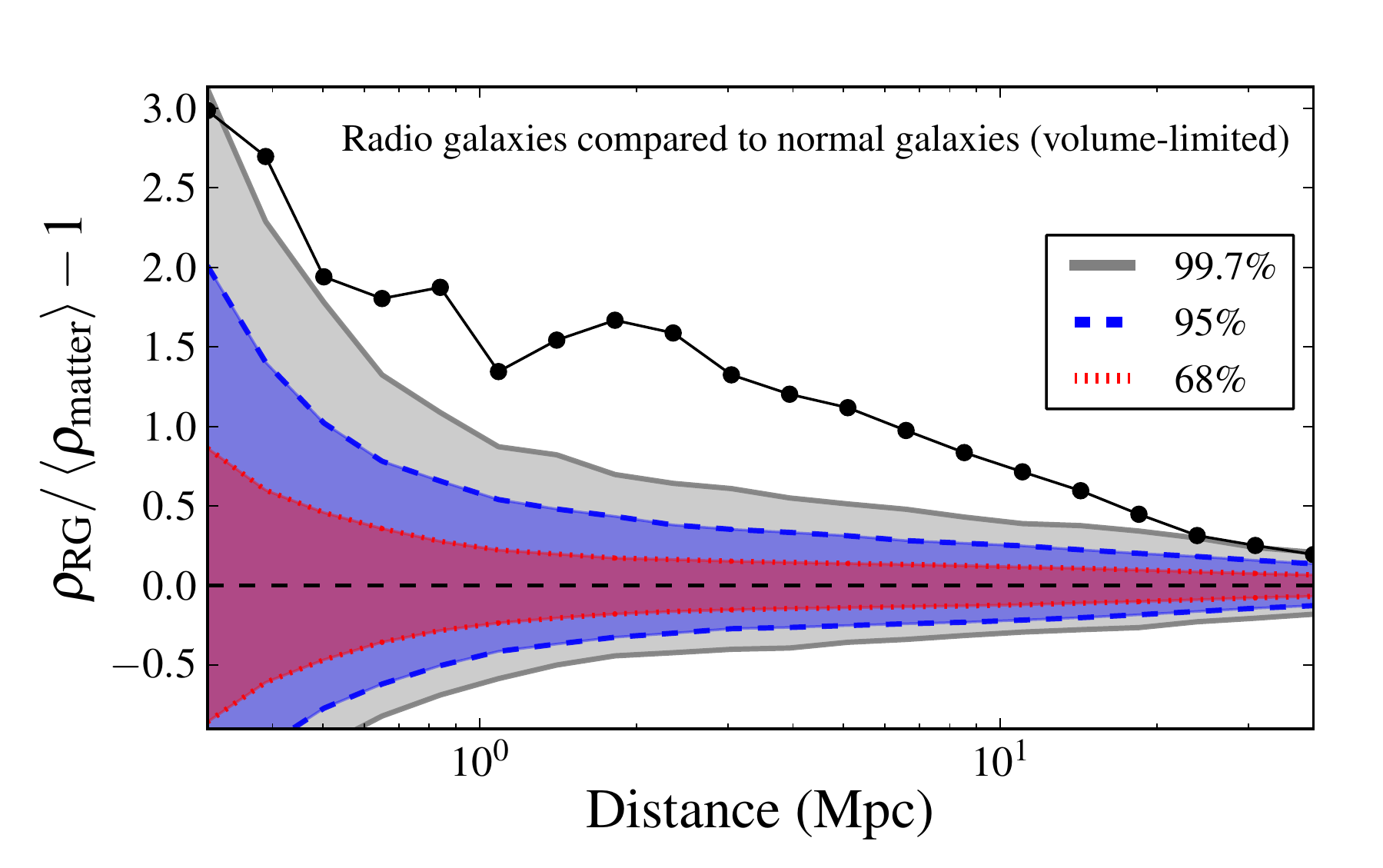}
\includegraphics[trim=5mm 17mm 6mm 5mm, clip, width=.48\textwidth]{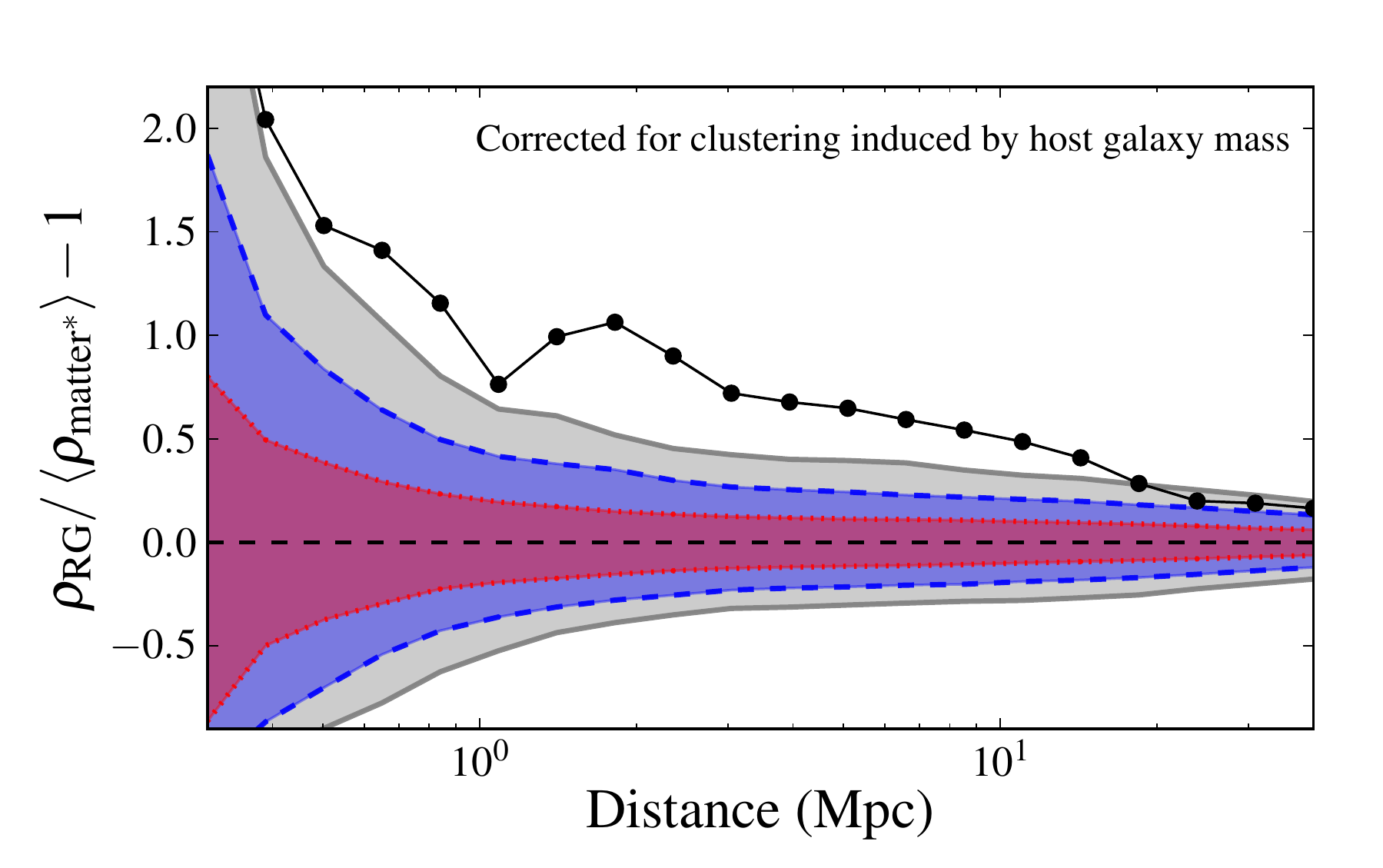}
\includegraphics[trim=5mm 2mm 6mm 5mm, clip, width=.48\textwidth]{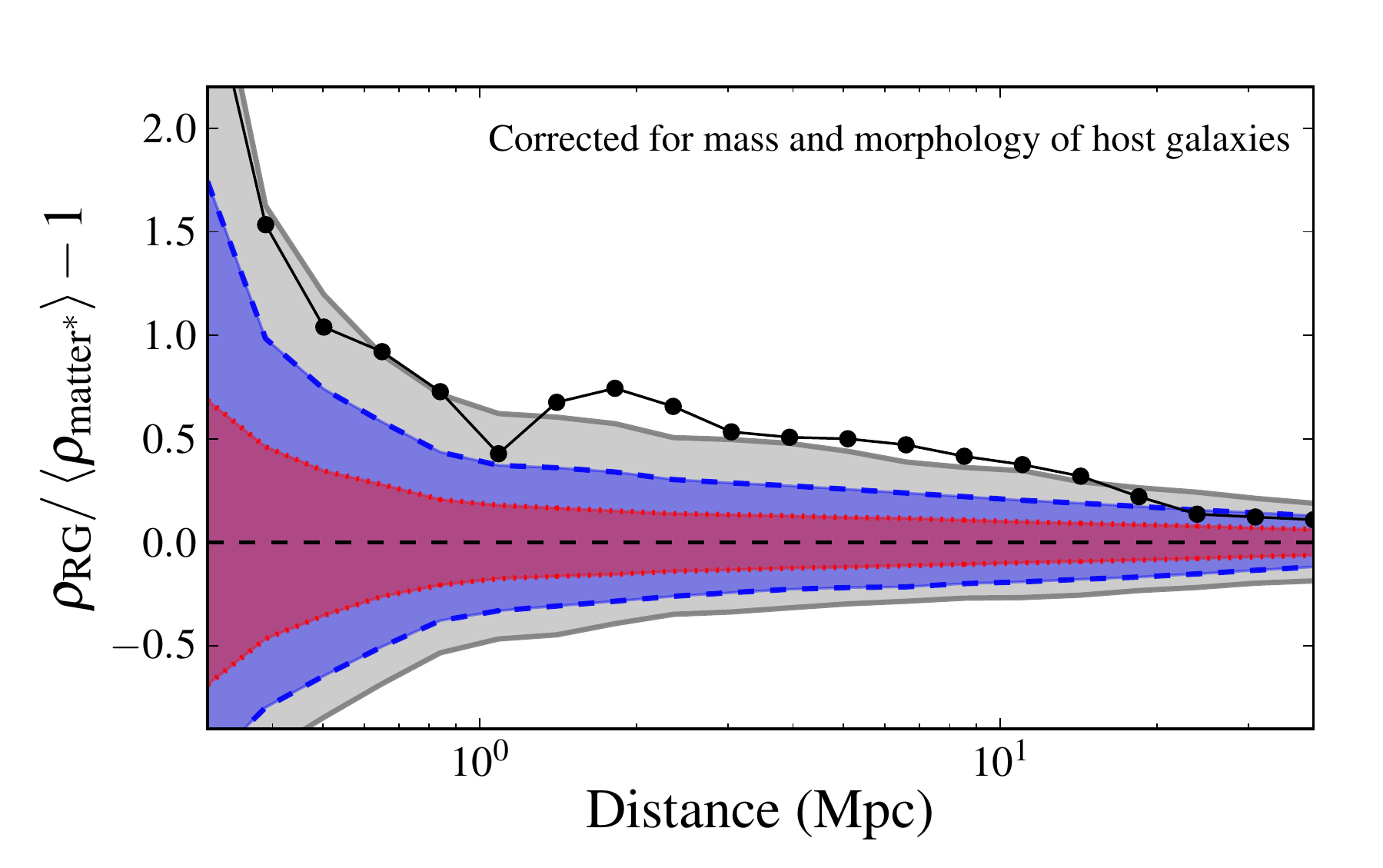}
\caption{The fractional density excess of radio galaxies in the volume-limited sample (defined in sec. \ref{sec:volsample}). Top: the excess with respect to a volume-limited sample of non-radio galaxies (i.e., $z<0.03$ and $M_{K}<-23.78$). Middle: the excess with respect to normal galaxies drawn from the $K$-band luminosity function of radio galaxies. Bottom: the fractional density excess corrected for clustering induced by the morphological type of the host galaxies \emph{and} their luminosity. Even after correcting for the clustering of their massive and predominantly early-type hosts, radio-loud jets are observed to occur in regins of high galaxy density.}\label{fig:cc3d}
\end{figure}

Our sample of radio galaxies is a subset of an all-sky redshift survey, allowing us to study the clustering of radio galaxies with respect to the matter distribution. We use the volume-limited sample of radio galaxies within $z=0.03$ (sample~A, defined in section \ref{sec:volsample}) and we restrict to $z>0.003$, yielding 73

 radio galaxies. In Fig. \ref{fig:topview} we show a ``top view'' of this volume-limited sample; more than 50\% of the radio galaxies are within the contour that encompasses half of the normal galaxy sample, an obvious sign of enhanced clustering. We quantify this clustering by counting the number of pairs between radio galaxies and normal 2MRS galaxies as a function of comoving distance, $\rho_{\rm RG}(d)$, i.e., the total galaxy count within a comoving distance $d$ of each radio galaxy. We compare this to the number of pairs for random subsets of 73 non-radio galaxies of the volume-limited sample, $\rho_{\rm matter}(d)$. In the top panel of Fig. \ref{fig:cc3d} we show $\rho_{\rm RG}/\left<\rho_{\rm matter}\right>-1$: a highly significant excess of clustering with respect to the normal galaxy distribution is clear.

Clustering is well-known to correlate with galaxy mass and morphology \citep[e.g.,][]{Dressler80,Bamford09} and the majority of the radio galaxies in our sample have massive, early-type hosts. Hence galaxy luminosity and morphology are confounding factors in the analysis of radio galaxy clustering. One can remove the clustering due to host mass by computing the density excess with respect to the radio-loud galaxy mass distribution. We thus measure $\left<\rho_{\rm matter^*}\right>$ by drawing random subsets of non-radio galaxies from the observed $K$-band luminosity distribution of radio galaxies. We show the result in the middle panel of Fig. \ref{fig:cc3d}; while the amplitude of the excess decreases, it remains highly significant. By forcing the random subsets of non-radio galaxies to have, on average, the same Hubble type (E, S0, or S) abundance as the radio galaxies we also correct the density excess for clustering due to morphology. We find that the galaxy number density around our sample of radio galaxies remains enhanced by a factor of 1.7 at 2~Mpc (Fig. \ref{fig:cc3d} bottom panel); the probability that this enhancement is observed for a set of non-radio galaxies with the same luminosity and Hubble type distribution as the radio galaxies is less than $0.3\%$. 

The brightest galaxy of a cluster is often found to be radio-loud \citep*{Matthews64,Burns90} and this will contribute to the observed density excess of radio galaxies. We therefore preform a simple test to find the brightest cluster galaxies and estimate this contribution. First, we define a cluster or group as a collection of more than 5 galaxies from the volume-limited sample ($M_{K}<23.8$) that can be connected by 3~Mpc strings (i.e., a friends-of-friends cluster finding algorithm with a comoving link length of 3~Mpc). Of the 73 radio galaxies from sample~A that we use here, 32 are in such groups. Of these, 18 are the brightest member of their group (or cluster). After removing these galaxies, the enhanced clustering (corrected for mass and morphology of the host galaxies) of the remaining 55 radio galaxies falls between the $2\sigma$-level and the $3\sigma$-level for all distances we consider; at 2~Mpc, the fractional density excess is 50\%.

\subsection{Luminosity function}\label{sec:lumfunc}
\begin{figure*}[t]
\centering
\includegraphics[trim=5mm 2mm 8mm 5mm, clip, width=.48\textwidth]{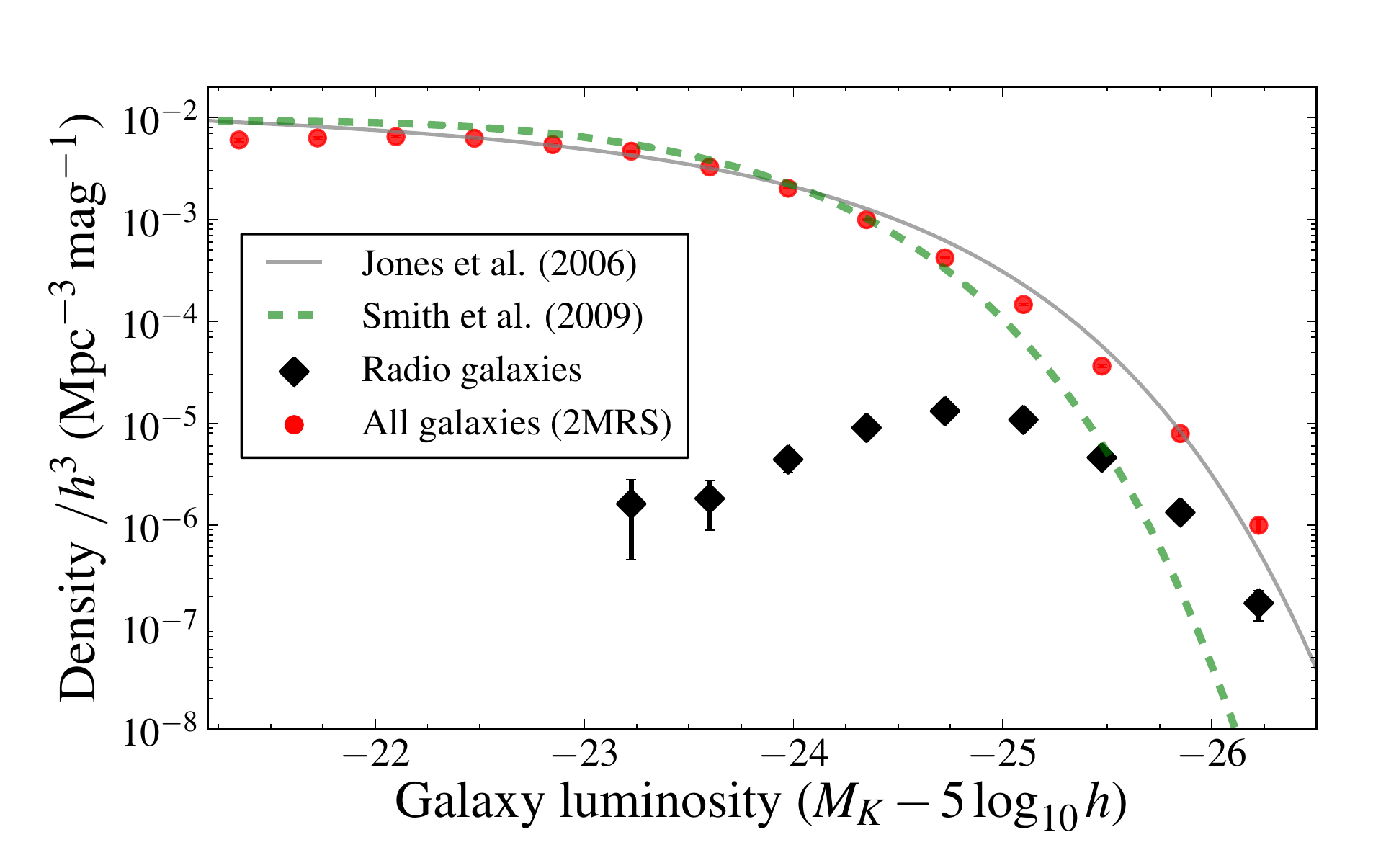} \quad
\includegraphics[trim=5mm 2mm 8mm 5mm, clip, width=.48\textwidth]{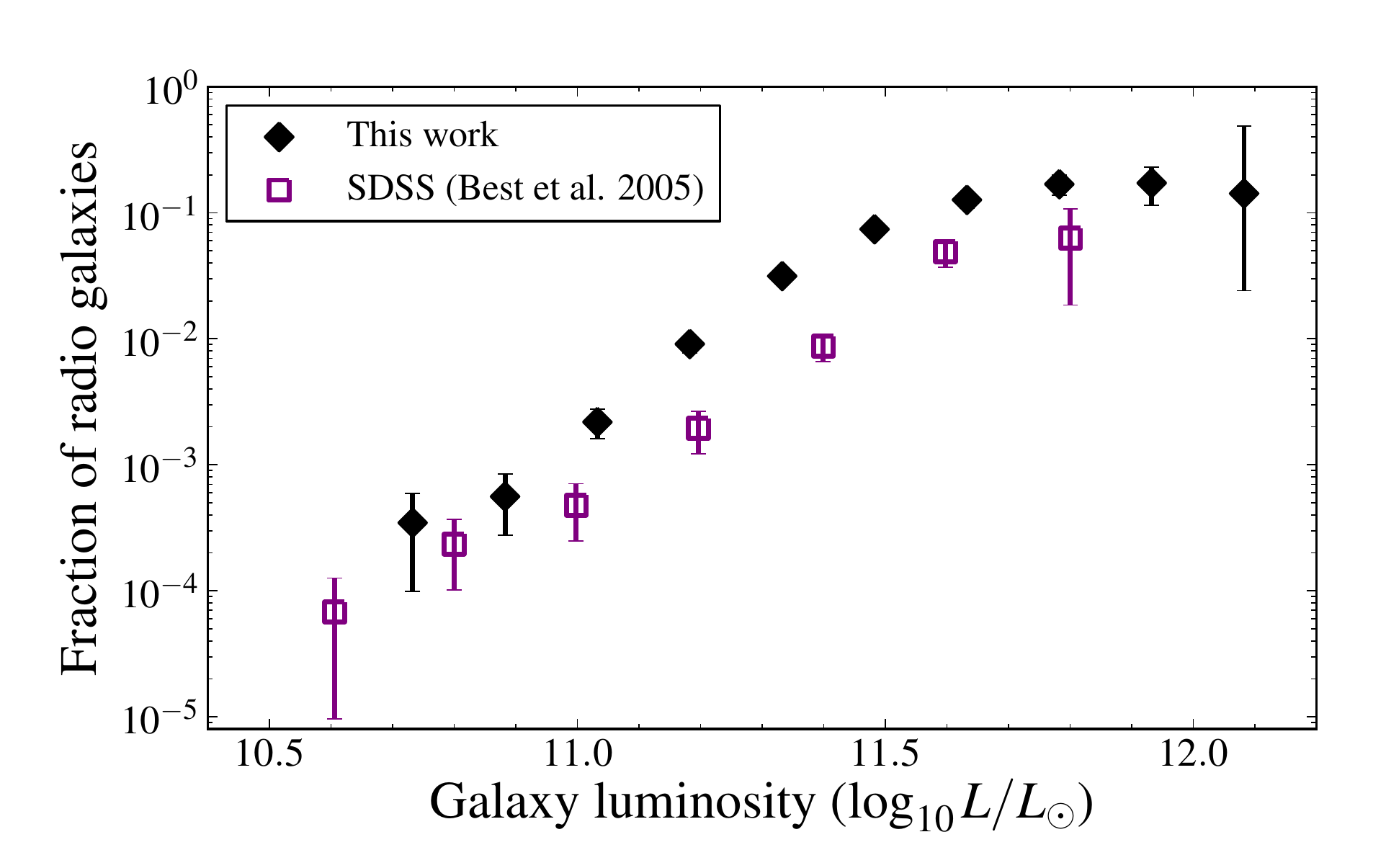}
\caption{Left: the $K$-band luminosity function for powerful radio galaxies (sample~B, $L>10^{24}\,{\rm W}\,{\rm Hz}^{-1}$) and normal galaxies. We also show Schechter function fits to galaxies from the UKIDSS Large Area Survey \citep{Smith09} and the 6dF Galaxy Survey \citep{Jones06}. The number density of powerful radio galaxies peaks at an absolute $K$-band magnitude of $-25.5$ (using $h=0.72$). Right: the  fraction of radio galaxies as a function of galaxy luminosity. We see at rapid increase from zero for $L<5\times10^{10}~L_\odot$ to $\sim 20$\% for the brightest galaxies. We also show the fraction of radio-loud AGN (with the same minimum radio luminosity) in SDSS \citep{Best05b}, converted to the $K$-band luminosity using a mean mass-to-light ratio $M/L_K=0.8$ \citep{Bell03}.}\label{fig:K-lumfunc}
\end{figure*}

\begin{figure*}[t]
\centering
\includegraphics[trim=5mm 2mm 8mm 5mm, clip, width=.48\textwidth]{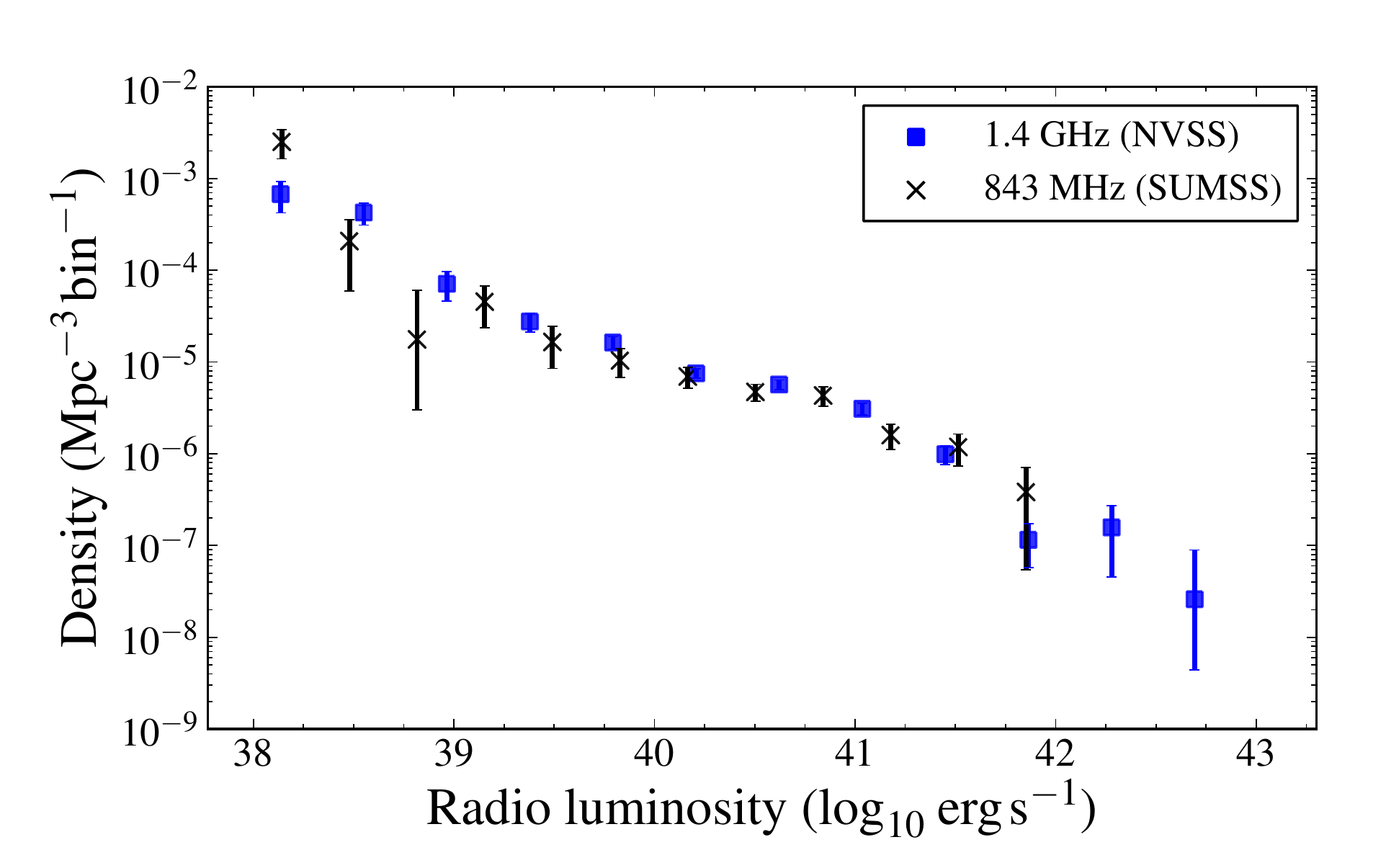} \quad
\includegraphics[trim=5mm 2mm 8mm 5mm, clip, width=.48\textwidth]{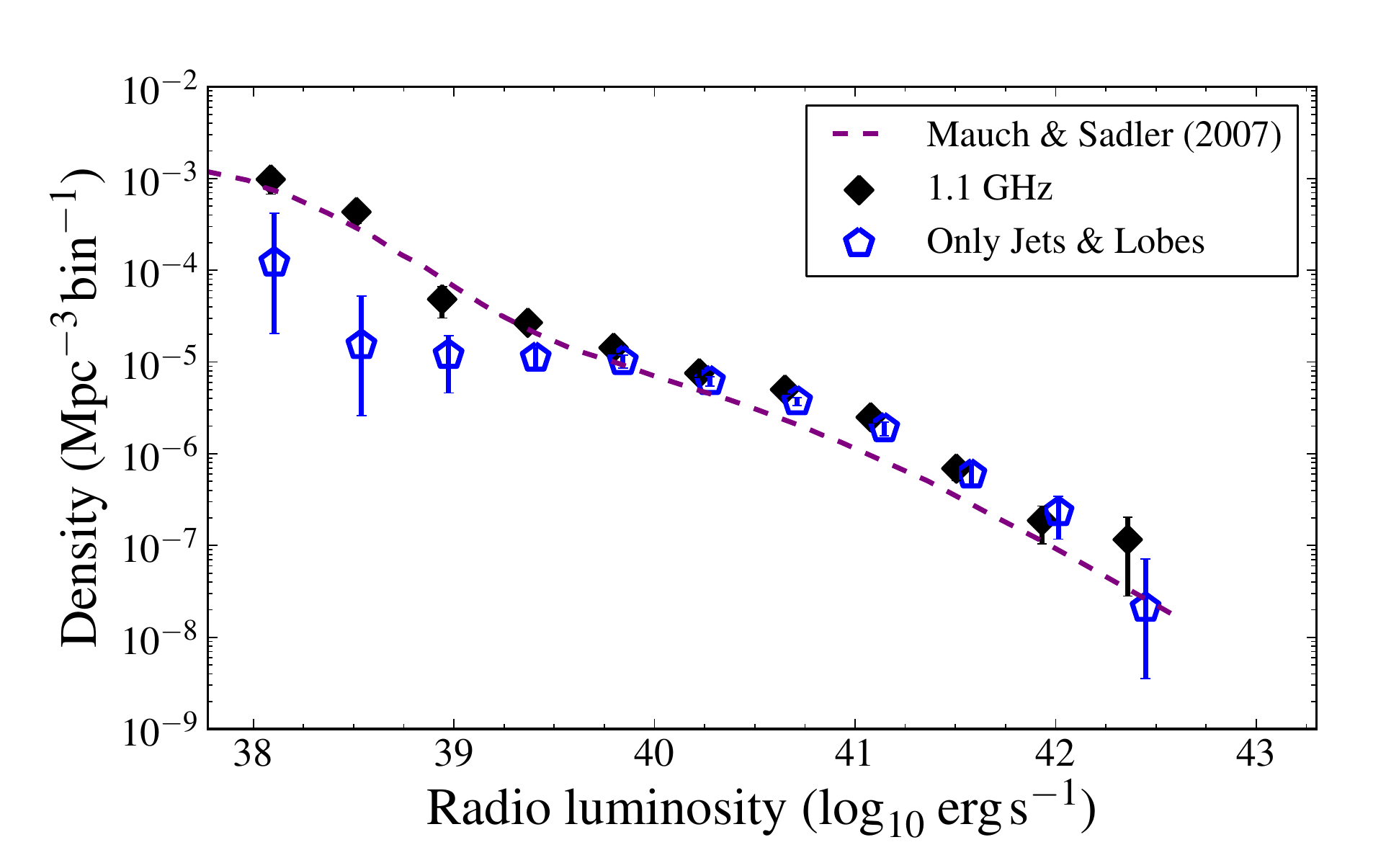}
\caption{The radio luminosity ($\nu L_\nu$) function of extra-galactic radio sources at different frequencies. Right: 1.4~GHz from NVSS ($\delta>-40$) and 843~MHz from SUMSS ($\delta<-30$, $|b|>10$). Left: the 1.1~GHz sample includes sources from both surveys by scaling to this frequency using the mean spectral index of $\alpha=-0.6$. For $\nu L_\nu > 10^{39}~{\rm erg}\,{\rm s}^{-1}$, the luminosity function for radio galaxies with a jet or lobe morphology (open diamonds) is similar to full sample, while below this scale the contribution from starforming galaxies becomes clear.}\label{fig:radio-lumfunc}
\end{figure*}

In this section, we use the $1/V_{\rm max}$ method \citep[e.g.,][]{Schmidt68}, to obtain the number density of radio galaxies and normal galaxies as a function of luminosity. The $1/V_{\rm max}$ method allows one to compute this density without applying a hard distance limit (such as the $z<0.03$ limit that was used to define sample~A in section \ref{sec:volsample}). In this section, we define sample~B: a set of powerful radio galaxies with $L_{1400}>10^{24}\,{\rm W}\,{\rm Hz}^{-1}$. As before, we also require a minimum redshift $z>0.003$ and exclude the sources with unknown morphology, because these are most likely random matches. The luminosity and minimum redshift cut have been picked to aid a comparison to previous work. These cuts leave 527 radio-emitting galaxies, with a median redshift of 0.042, and 401 powerful radio galaxies with a median redshift of 0.052. 

We find $\left<V/V_{\rm max}\right>=0.494\pm 0.012$ and $\left<V/V_{\rm max}\right>=0.496 \pm 0.014$  for all radio-emitting galaxies and the powerful radio galaxies, respectively,  which shows that density evolution with redshift or distance-dependent selection effects are not a significant influence on the derived radio source densities. 

In Fig. \ref{fig:K-lumfunc} we show the $K$-band luminosity function. One clearly sees that the powerful radio galaxies reside in the most luminous hosts and their density peaks at $M_{K} \approx -25.5$. The fraction of radio galaxies is a strong function of galaxy luminosity. No radio galaxies with $L_{1400}>10^{24}\,{\rm W}\,{\rm Hz}^{-1}$ are found for $M_{K}>-23.7$ and the fraction rises rapidly to an asymptotic value of 0.25 for the most massive galaxies. 
We also compute the luminosity function of normal galaxies which agrees reasonably well with Schechter functions that have been derived form the UKIDSS Large Area Survey \citep{Smith09} and the 6dF Galaxy Survey \citep{Jones06} (except at the bright end, which is a known issue with Schechter functions).

Recalling that the radio-emitting galaxies in our catalog originate from NVSS (at 1.4~GHz) and/or SUMSS (at 843~MHz), we first compute the luminosity function for each survey separately (Fig. \ref{fig:radio-lumfunc}, left panel). We find that, as expected, the luminosity functions from NVSS and SUMSS are nearly identical. To increase the statistical power, we combine both surveys by scaling the flux to 1.1~GHz using the mean spectral index of $\alpha=-0.6$  (Fig. \ref{fig:radio-lumfunc}, right panel).  We find that the luminosity function of radio-emitting galaxies with a jet or lobe morphology agrees well with the results of the entire sample, except below $\nu L_\nu = 10^{39}~{\rm erg}~{\rm s}^{-1}$. Below this luminosity, the contribution of starforming galaxies starts to dominate \citep[e.g.,][]{Condon02}. At $10^{41}\,{\rm erg}\,{\rm s}^{-1}$ our estimate of the number density is a factor 2--3 above the 1.4~GHz luminosity function derived form 2661 radio-loud AGN at $0.003 < z < 0.3$ \citep{Mauch07}. Below, we suggest an explanation for this difference.

\section{Conclusions \& Discussion}\label{sec:discussion}
Our conclusions can be summarized as follows.
\begin{enumerate}
\item A new extra-galactic radio catalog covering 88\% of the celestial sphere has been constructed. The catalog contains all galaxies brighter than $K=11.75$ that show radio emission at $F_{1400}>213$~mJy or $F_{843} >289$~mJy; it contains 575 sources.
Over 30\% of these radio-emitting galaxies are not contained in existing large-area catalogs.
\item Our matching algorithm identifies extra-galactic radio sources with 99\% efficiency and 91\% purity. All matches have been inspected manually to remove imaging artifacts and blended sources. The contamination due to background radio sources in the final sample is $2$\%; the \emph{Unknown} morphological class contains 19 potential random matches.
\item For a volume-limited sample within $z=0.03$ (130~Mpc), we find 27 galaxies with a radio and $K$-band luminosity greater or equal to Cen~A, which is 0.2\% of all 2MRS galaxies within the same volume. 
\item No obvious correlation between galaxy luminosity and total radio power is observed (Fig. \ref{fig:K-radio}).
\item The $K$-band luminosity function of radio galaxies peaks at $M_{K}=-25.5$ or a luminous mass of $\sim 10^{11}~M_\odot$ (Fig. \ref{fig:K-lumfunc}). 
\item Of radio galaxies within $z=0.03$, 94\% are of the E/S0 Hubble type; this higher than the fraction of bright ($M_{K}<-25$) normal galaxies that are of this morphological class (66\%) or the E/S0 fraction of non-active galaxies scaled to the same $K$-band luminosity distribution as the radio galaxies (65\%).
\item The fraction of radio galaxies as a function of $K$-band luminosity rises asymptotically to 20\% (Fig. \ref{fig:K-lumfunc}).
\item We have computed the fractional density excess of radio galaxies in a volume-limited sample as a function of distance from the radio galaxies. Radio galaxies are significantly more clustered than normal galaxies; in a sphere of 2~Mpc centered on the radio galaxies, the galaxy density is 2.7 times greater than around normal galaxies (Fig. \ref{fig:cc3d} top panel).  
\item After correcting the latter result for the extra clustering induced by the mass and Hubble type of the radio galaxies, the density excess remains significant ($>3\sigma$): at 2~Mpc the density around radio galaxies is 1.7 times higher than around non-radio galaxies with the same mass and galaxy morphology distribution (Fig. \ref{fig:cc3d} bottom panel). 
\end{enumerate}

The observation that radio-loud AGN are found in the most massive galaxies is well-known. Indeed, the median absolute $K$-band magnitude of radio galaxies in the NVSS-6dFGS sample \citep{Mauch07} of $M_{K}=-25.4$ is in excellent agreement with our results. The radio luminosity where the density of AGN overtakes that of starforming galaxies ($\nu L_\nu = 10^{39}~{\rm erg}\,{\rm s}^{-1}$) agrees with radio luminosity functions from other surveys at similar redshift \citep{Condon02,Mauch07}.

The decreasing fraction of radio galaxies with distance (Fig. \ref{fig:radio-fraction}) appears at odds with the cosmological trend of increasing AGN activity to $z\sim2$ \citep[e.g.,][]{Hopkins07}, yet most observers in other galaxies would find the same. Because radio galaxies are found in high density regions and the average astronomer lives in a galaxy inside a cluster, the local fraction of radio galaxies is typically observed to be higher than the cosmic average. 

For a standard mass-to-light ratio in the $K$-band, $M/L_K\sim 1$, we find that the fraction of galaxies with a radio luminosity greater than $L=10^{24}~{\rm W}~{\rm Hz}^{-1}$ is a factor 5 larger than the fraction of AGN with the same minimum radio luminosity in SDSS \citep{Best05b,Best07}. We also find  a factor 2--3 higher number density of radio sources at  $\nu L_\nu = 10^{41}~{\rm erg}\,{\rm s}^{-1}$ (or $10^{25}~{\rm W}\,{\rm Hz}^{-1}$) compared to the NVSS-6dFGS sample \citep{Mauch07}. If we restrict our sample to sources detected beyond $200$~Mpc (leaving 237 radio galaxies), while keeping $V_{\rm max}$ of each source fixed, these two discrepancies disappear. Hence a large-scale local overdensity of radio galaxies or a distance dependent selection effect are the most likely explanations for the larger radio galaxy density derived from our sample. We note that \citet{Mauch07} find $\left<V/V_{\rm max}\right>=0.532 \pm 0.006$ for their radio-loud AGN; this significant offset from 0.5 implies evolution (or selection bias) is present in their sample.

The conclusion of \citet{Ledlow95} that the number of radio galaxies detected in Abell clusters simply scales with the number of galaxies surveyed, appears incompatible with our detection of enhanced clustering of radio galaxies with respect to normal galaxies. Our clustering detection, however, is in agreement with more recent work at intermediate redshift ($z \gtrsim 0.1$). Using the SDSS galaxy sample, \citet{Kauffmann08} found a factor $\approx 2$ increase for the galaxy counts in a projected radius of 300~kpc around radio-loud AGN compared to radio-quiet AGN; \citet{Donoso10} found that the projected cross-correlation at 2~Mpc between radio-loud AGN and luminous red galaxies (LRG) is 25\% stronger than the auto-correlation of the radio-quiet LRG. \citet{Tasse08b} compared the environment of the host galaxies of powerful radio sources ($L_{1400} \gtrsim 10^{24}~{\rm W}\,{\rm Hz}^{-1}$) in the  XMM-LSS field \citep{Tasse08} to normal galaxies of the same mass and redshift. For $M>10^{10.5}~M_\odot$ their photometric-redshift-based density estimator finds an overdensity at 450~kpc, close to the distance where the enhanced clustering of radio galaxies in our sample exceeds the $3\sigma$-level. 

The enhanced density around radio galaxies suggests a causal relation between galaxy environment and jet power. Since this excess is measured with respect to galaxies of the same mass and morphological type, the mechanism behind this relation is not primarily driven by black hole mass. 
The gas cooling out of the hot atmospheres of the host galaxies is often suggested as a potential mechanism to turn on radio-loud jets \citep[][]{Burns90, Best05b}. 
 An interesting future application is to compare the luminosity of the core of the jet, which will be measured by a VLBA 8~GHz survey of 2MASS galaxies \citep{Condon11}, to the large scale radio emission contained in our catalog. This will allow a study of jet power as a function of black hole mass --- the latter can be estimated using the luminosity and Hubble type of the galaxy \citep[e.g.,][]{Caramete11b}.

The automated image-level matching presented here could be considered an improvement to the manual classification that has been used to construct similar samples in the past. For future studies of existing extra-galactic catalogs (e.g., the 2MASS Extended Source Catalog) as well as upcoming radio surveys (e.g., LOFAR, ASKAP), automated cross-wavelength identification will be key to manage the large number of sources.

\begin{acknowledgements}
SvV would like to thank M.~R. Blanton, E. K\"ording, R. Plotkin, H. R\"ottgering, and R. van~Weeren for useful discussions. In addition, we would like to thank the referee for the swift reply and the useful comments. HF acknowledges funding from the European Research Council (ERC) Advanced Grant. KHK acknowledges funding from the German Ministry of Education and Research (BMBF).

This research has made use of the NASA/IPAC Extragalactic Database (NED) which is operated by the Jet Propulsion Laboratory, California Institute of Technology, under contract with the National Aeronautics and Space Administration. This research has also made use of the SIMBAD database, operated at CDS, Strasbourg, France. The Two Micron All Sky Survey is a joint project of the University of Massachusetts and the Infrared Processing and Analysis Center/California Institute of Technology, funded by the National Aeronautics and Space Administration and the National Science Foundation. The Digitized Sky Survey was produced at the Space Telescope Science Institute under U.S. Government grant NAG~W-2166. The images of these surveys are based on photographic data obtained using the Oschin Schmidt Telescope on Palomar Mountain and the UK Schmidt Telescope. The plates were processed into the present compressed digital form with the permission of these institutions. 

The density contours in Fig. \ref{fig:topview} were created with \verb bovy_plot . All plots with celestial coordinates were produced with \verb APLpy , an open-source plotting package for Python hosted at \url{http://aplpy.github.com}.
\end{acknowledgements}

\bibliography{general_desk,ADS_desk}


\begin{appendix} 

\section{Example images}\label{sec:atlas}
Here we show images of examples of newly identified radio-emitting galaxies discussed in section \ref{sec:notes}. For the images of all 575 radio-emitting galaxies in our sample we refer to the online catalog, \url{http://ragolu.science.ru.nl/hcat.html}.
The following legend applies to all images:
\begin{itemize}
\item Green pentagram: 2MRS galaxy.
\item Green cross ($\times$): center of the frame (i.e., galaxy in question).
\item Purple square ($\square$): SUMSS/NVSS catalog source.
\item Purple cross ($\times$): matched catalog source by image-level algorithm.
\item Purple plus ($+$): manually accepted match.
\item Large magenta cross: geometrical center of radio emission.
\item Large red circled cross: flux-weighted center of radio emission.
\end{itemize}

\begin{figure*}[ht]
{\includegraphics[trim=10mm 5mm 39mm 5mm, clip, width=.47\textwidth]{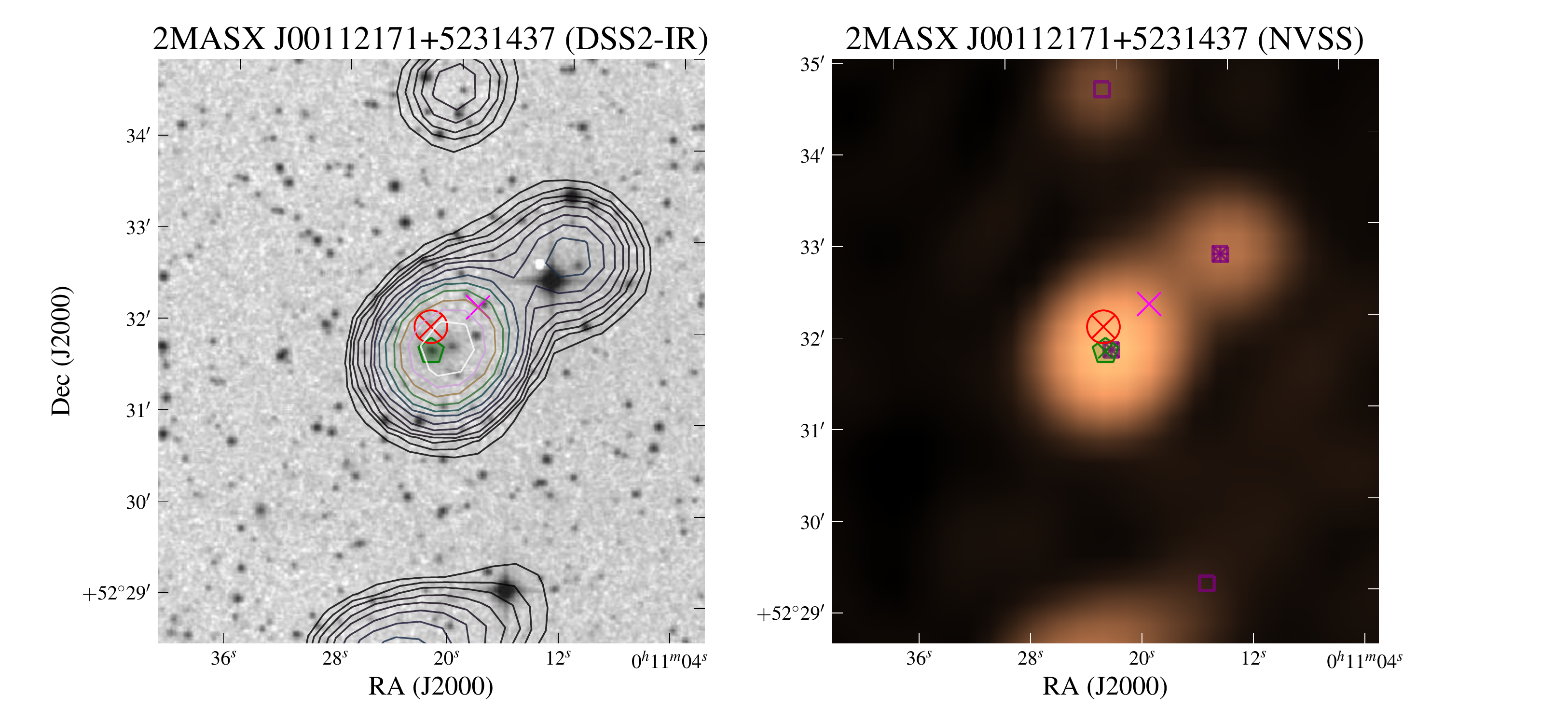}} \qquad
{\includegraphics[trim=10mm 5mm 39mm 5mm, clip, width=.47\textwidth]{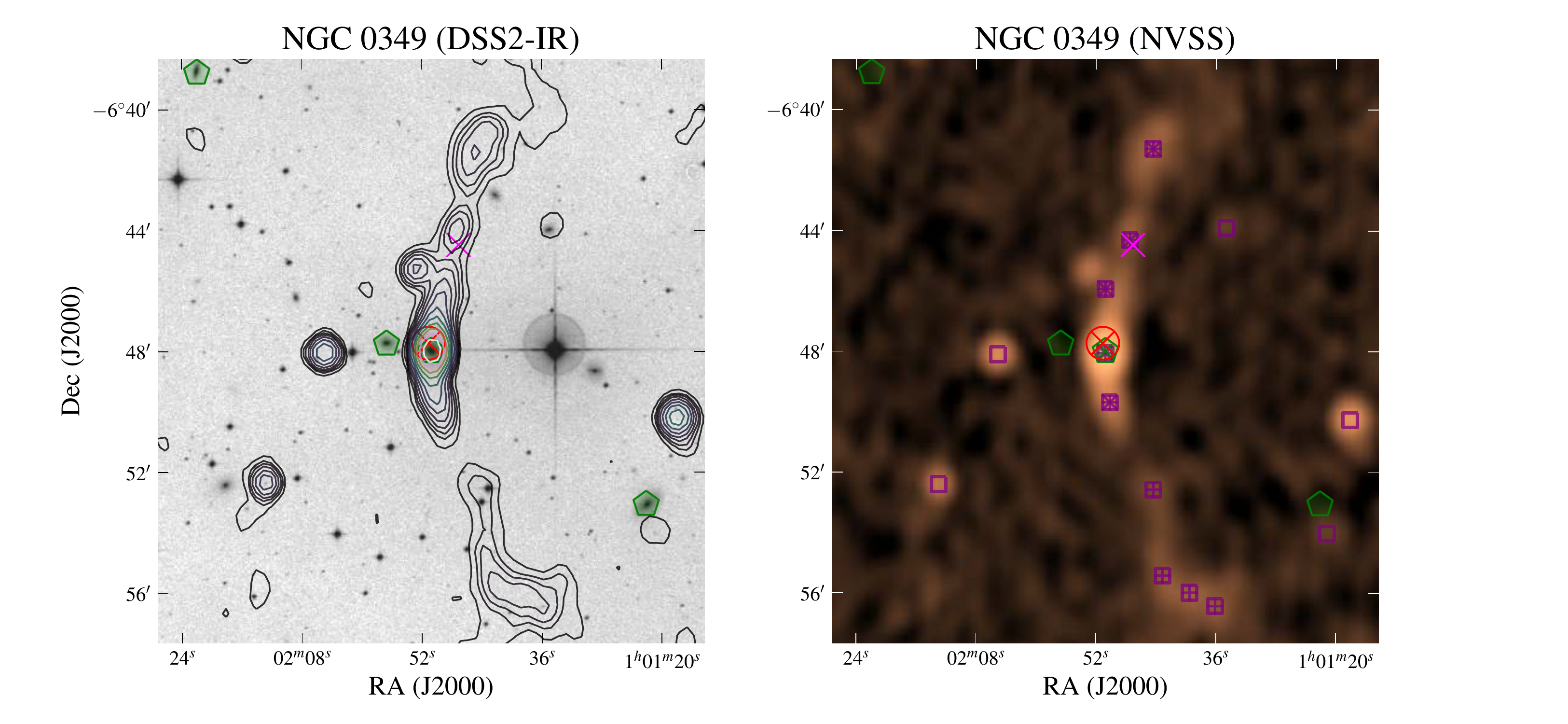}} \\[5pt]
{\includegraphics[trim=10mm 5mm 39mm 5mm, clip, width=.47\textwidth]{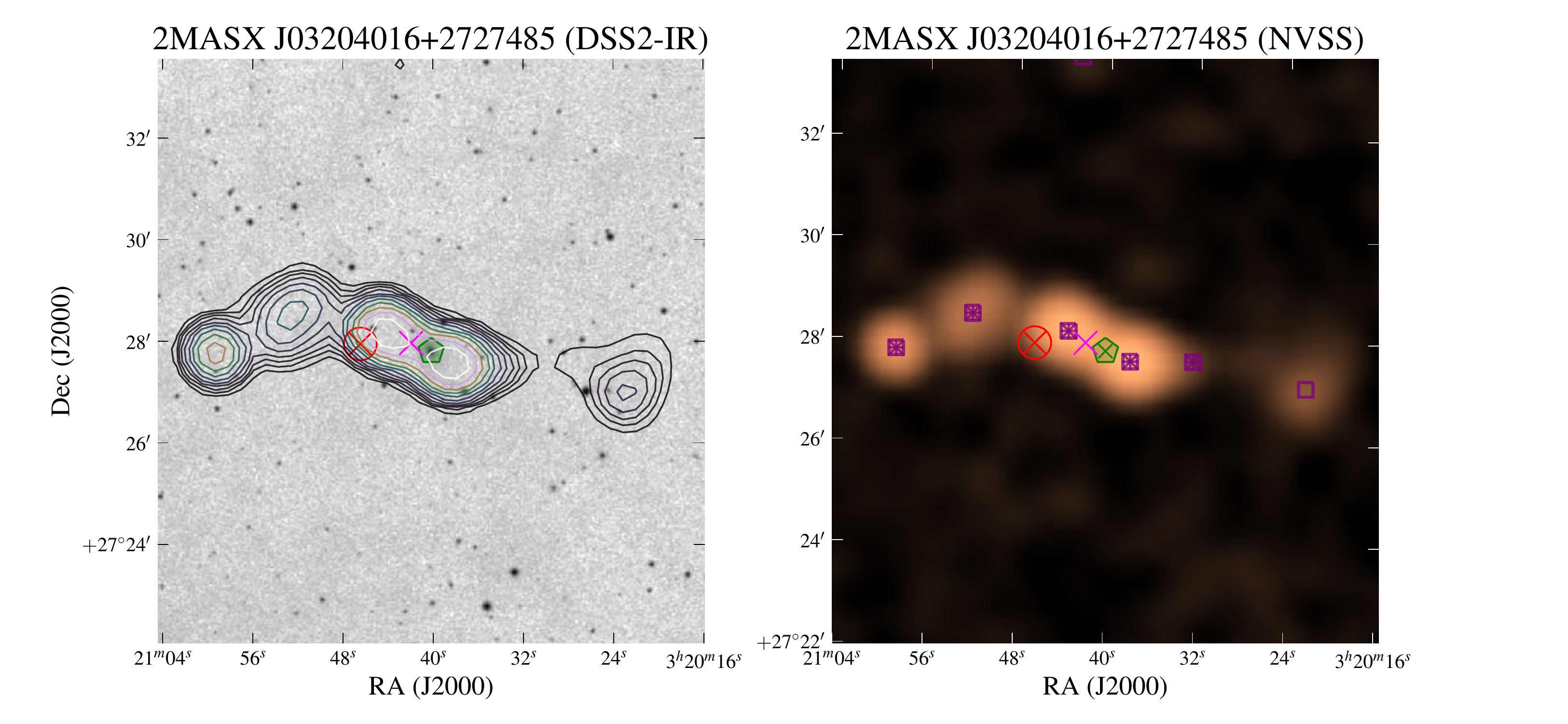}} \qquad
{\includegraphics[trim=10mm 5mm 39mm 5mm, clip, width=.47\textwidth]{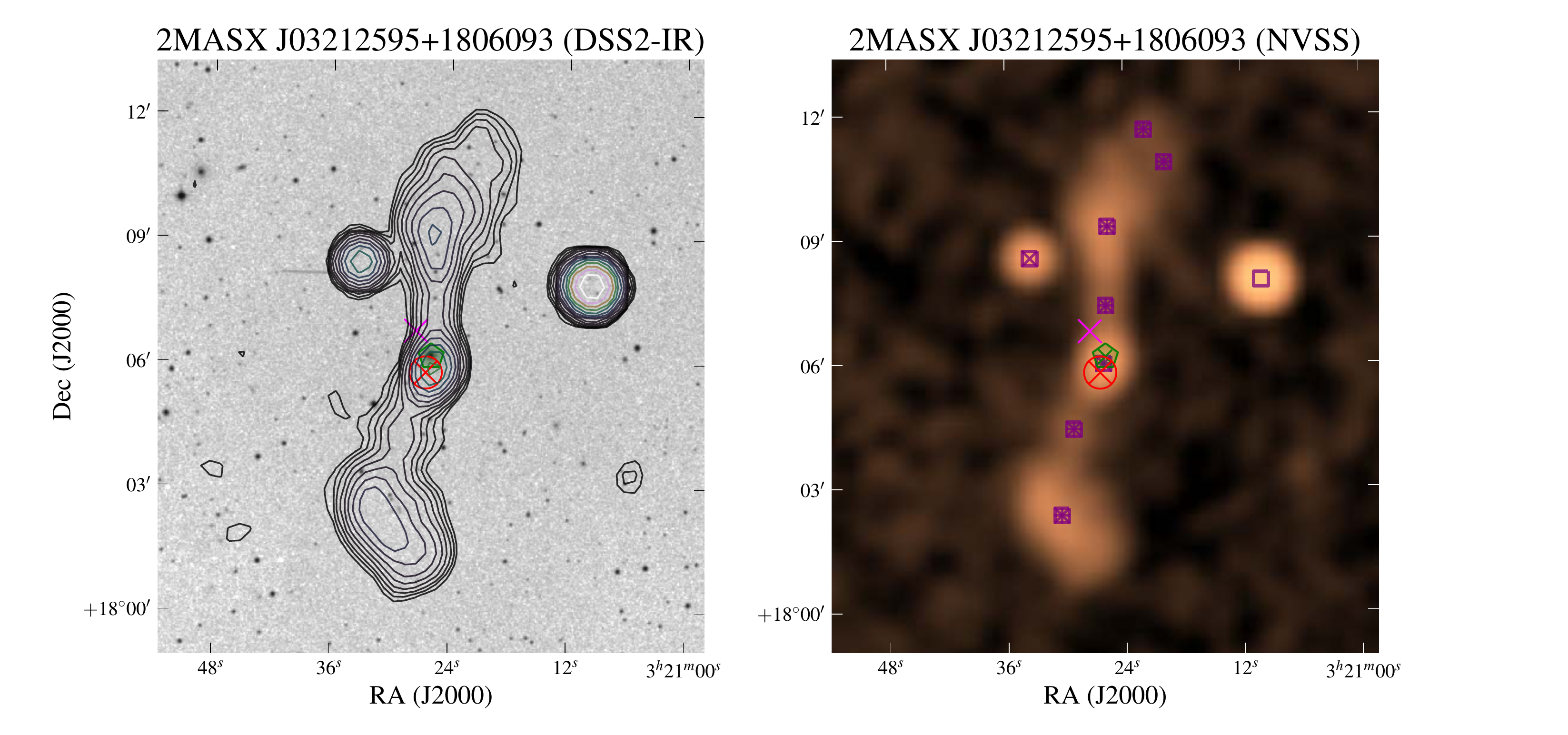}} \\[5pt]
{\includegraphics[trim=10mm 5mm 39mm 5mm, clip, width=.47\textwidth]{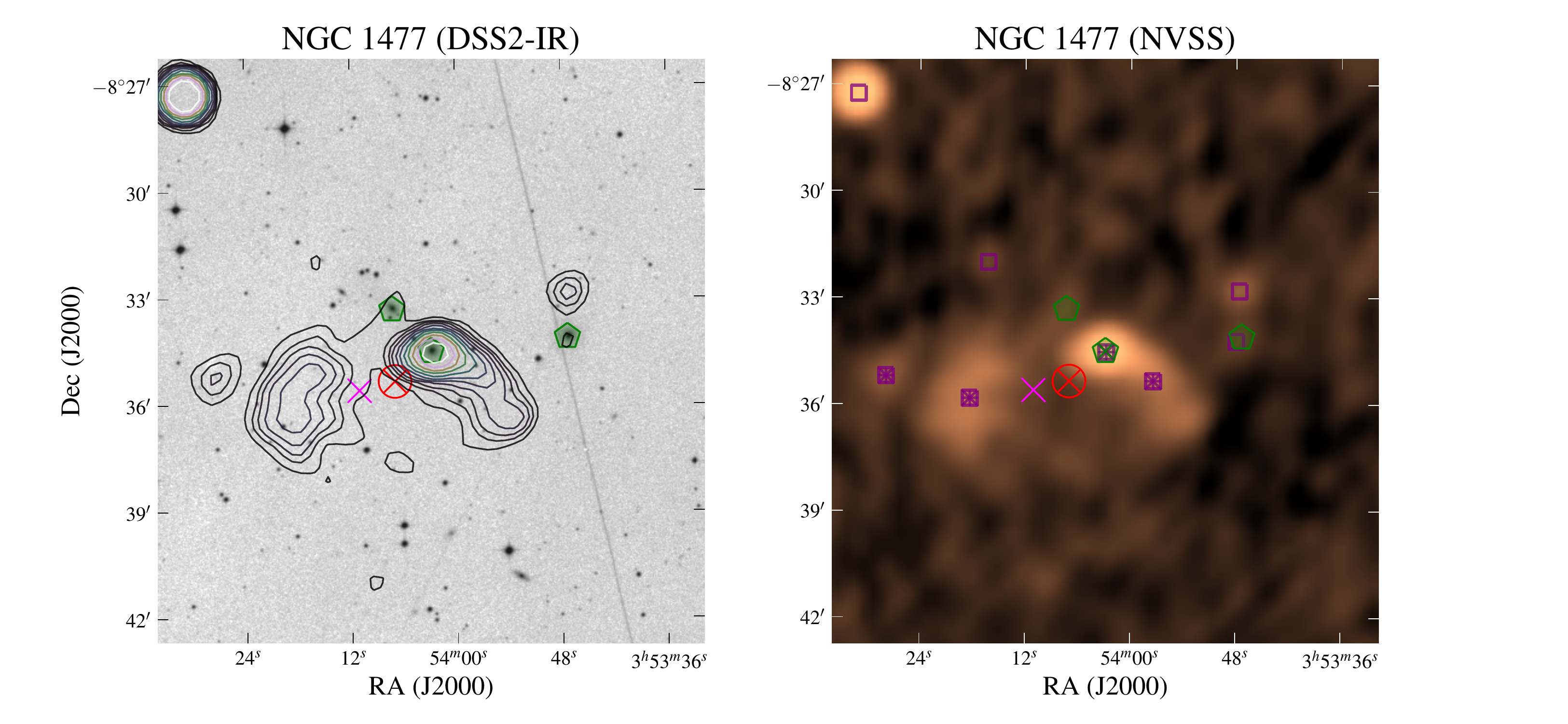}} \qquad
{\includegraphics[trim=10mm 5mm 39mm 5mm, clip, width=.47\textwidth]{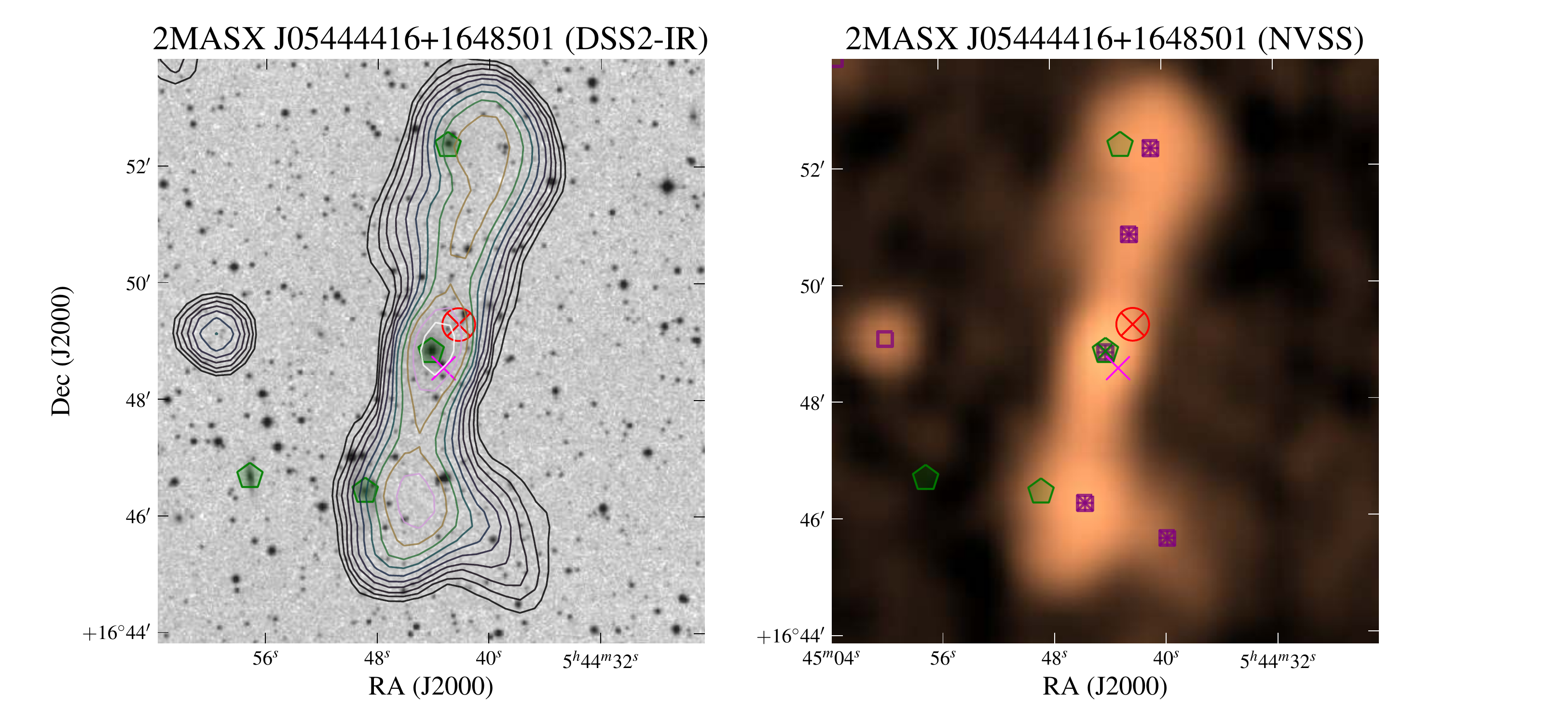}} \\[5pt]
{\includegraphics[trim=10mm 5mm 39mm 5mm, clip, width=.47\textwidth]{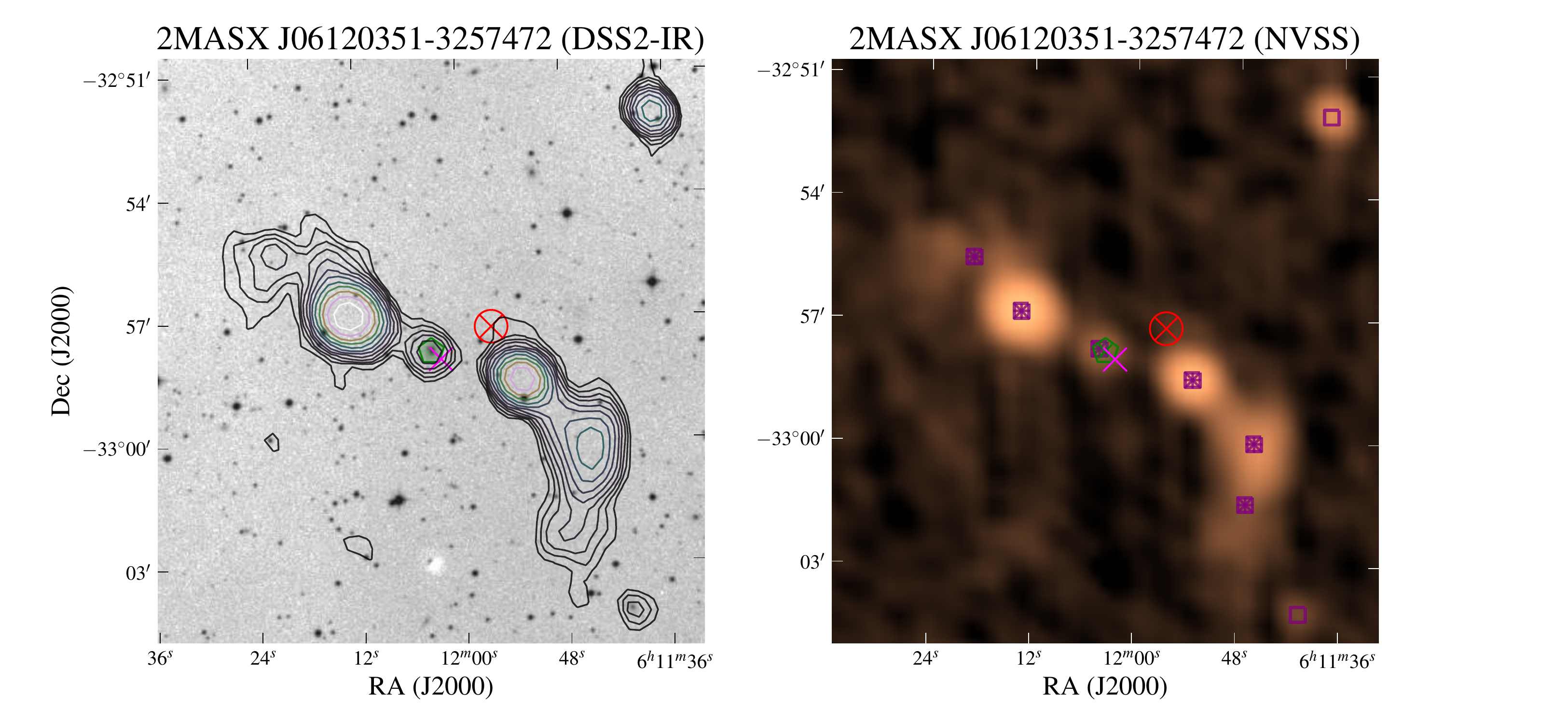}} \qquad
{\includegraphics[trim=10mm 5mm 39mm 5mm, clip, width=.47\textwidth]{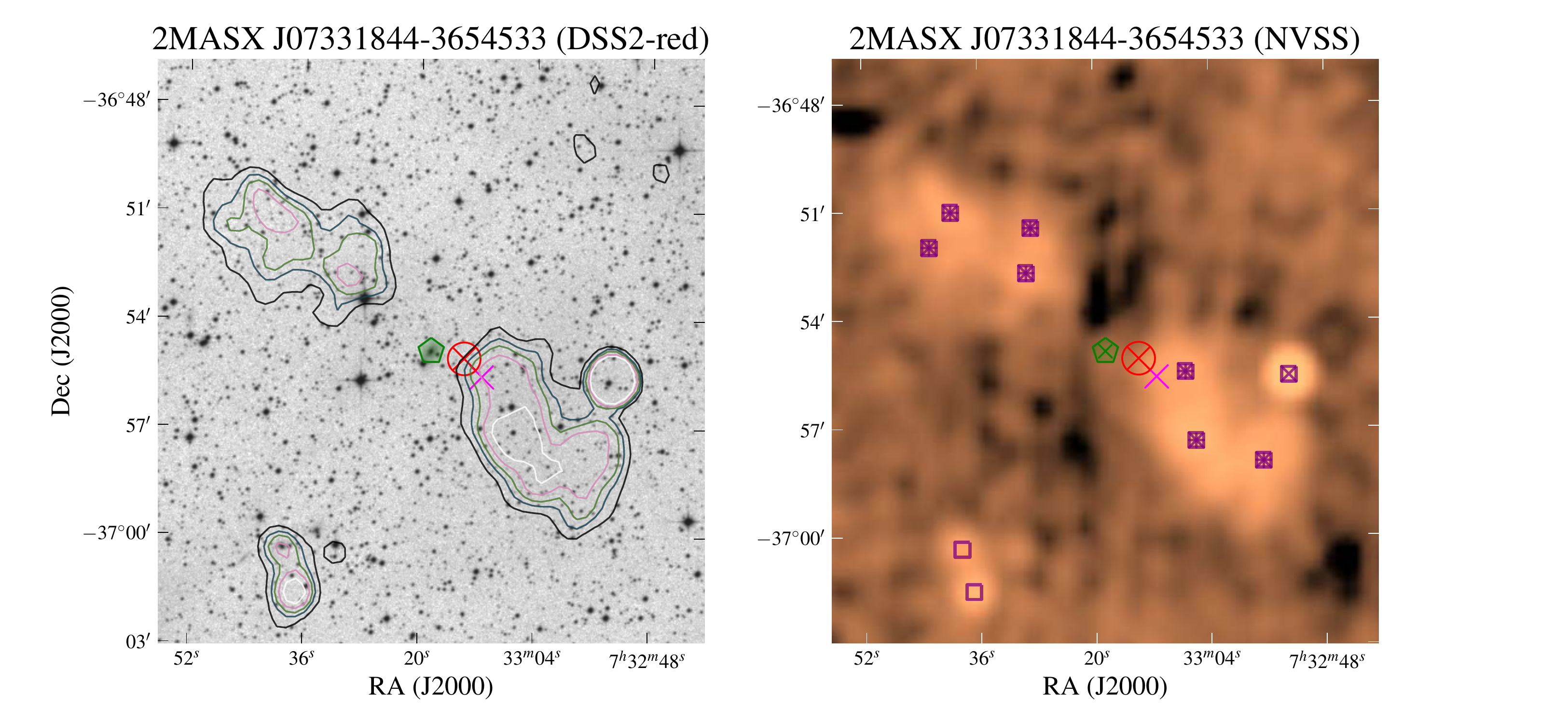}} \\[5pt]
{\includegraphics[trim=10mm 5mm 39mm 5mm, clip, width=.47\textwidth]{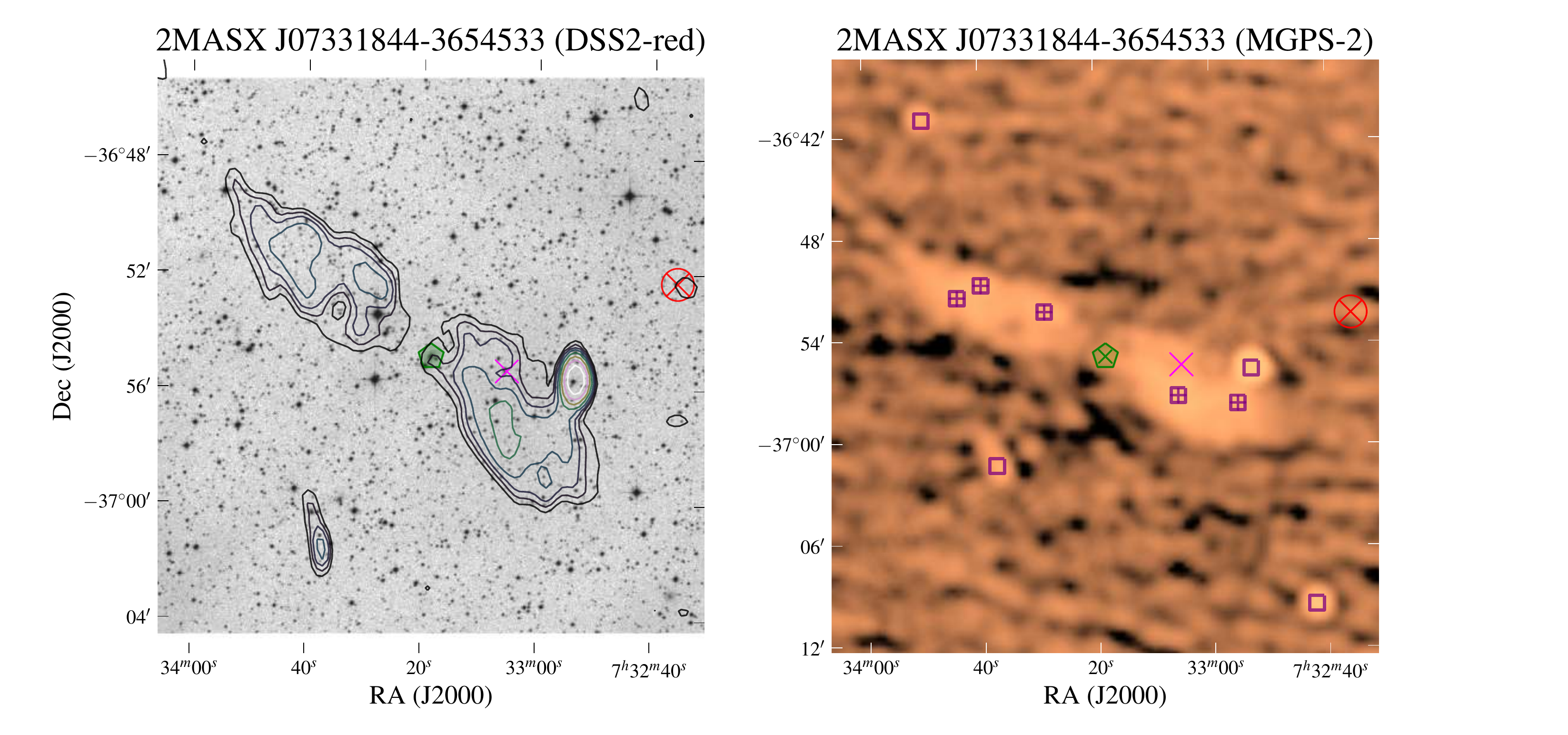}} \qquad
{\includegraphics[trim=10mm 5mm 39mm 5mm, clip, width=.47\textwidth]{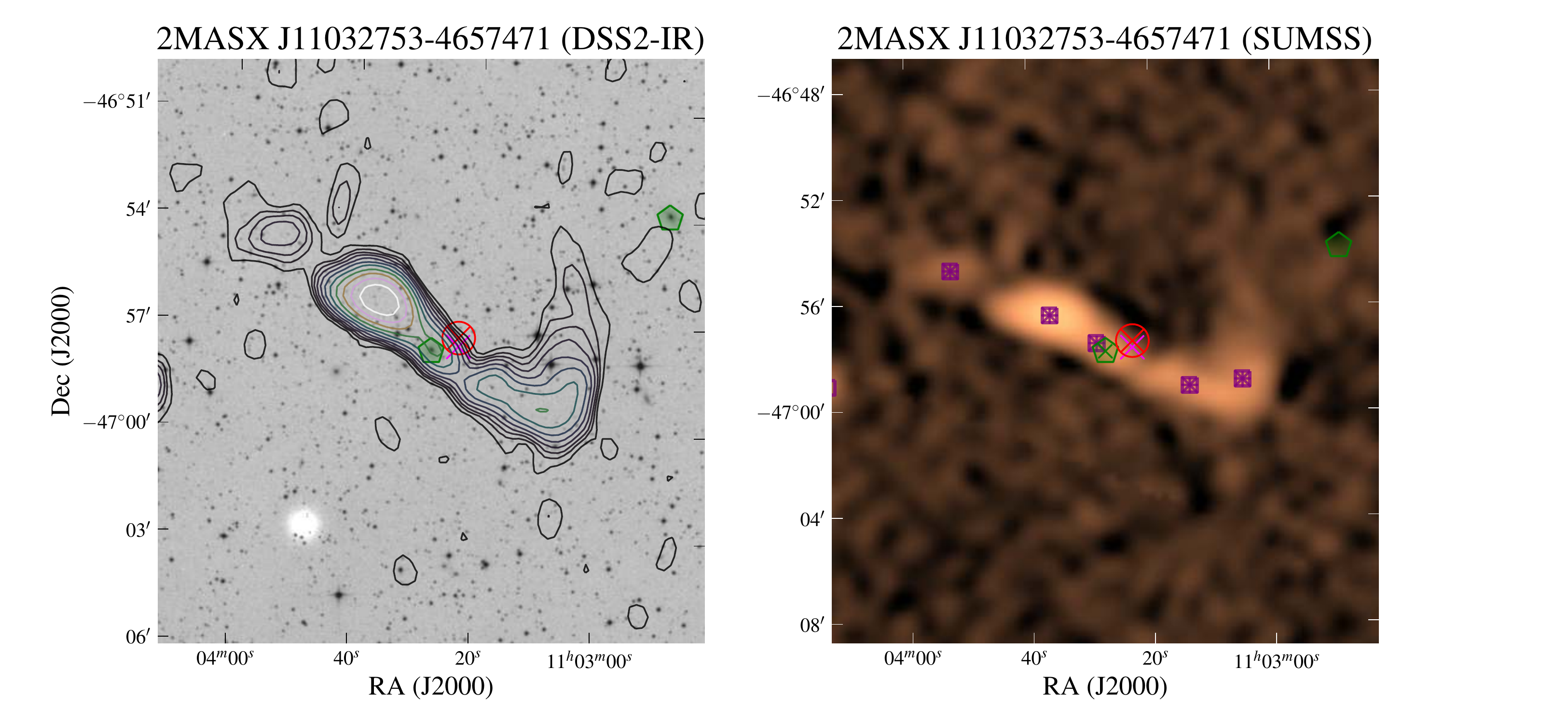}} \\[2pt]
\caption{Examples of newly identified radio-emitting galaxies. } \label{fig:Newly1}
\end{figure*}
\clearpage\begin{figure*}[ht]
{\includegraphics[trim=10mm 5mm 39mm 5mm, clip, width=.47\textwidth]{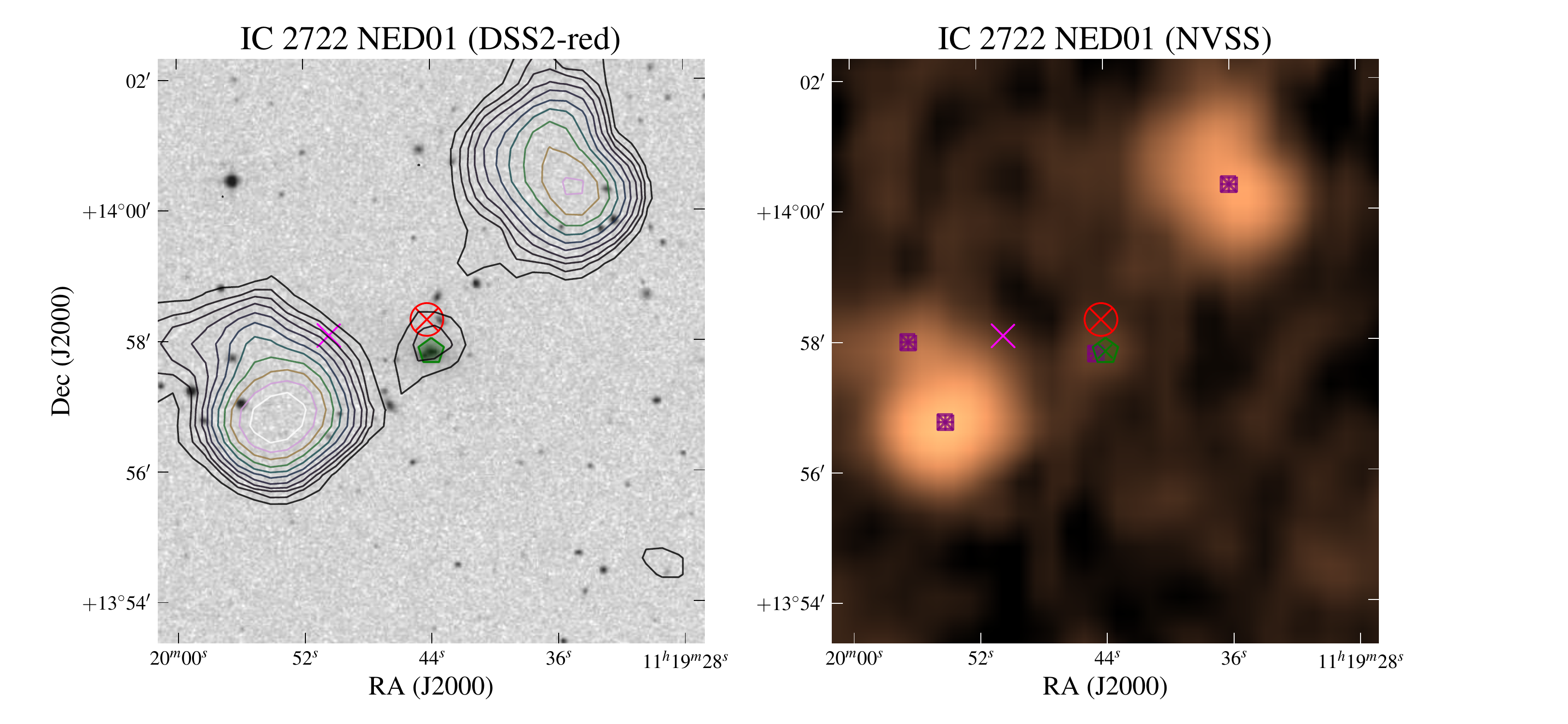}} \qquad
{\includegraphics[trim=10mm 5mm 39mm 5mm, clip, width=.47\textwidth]{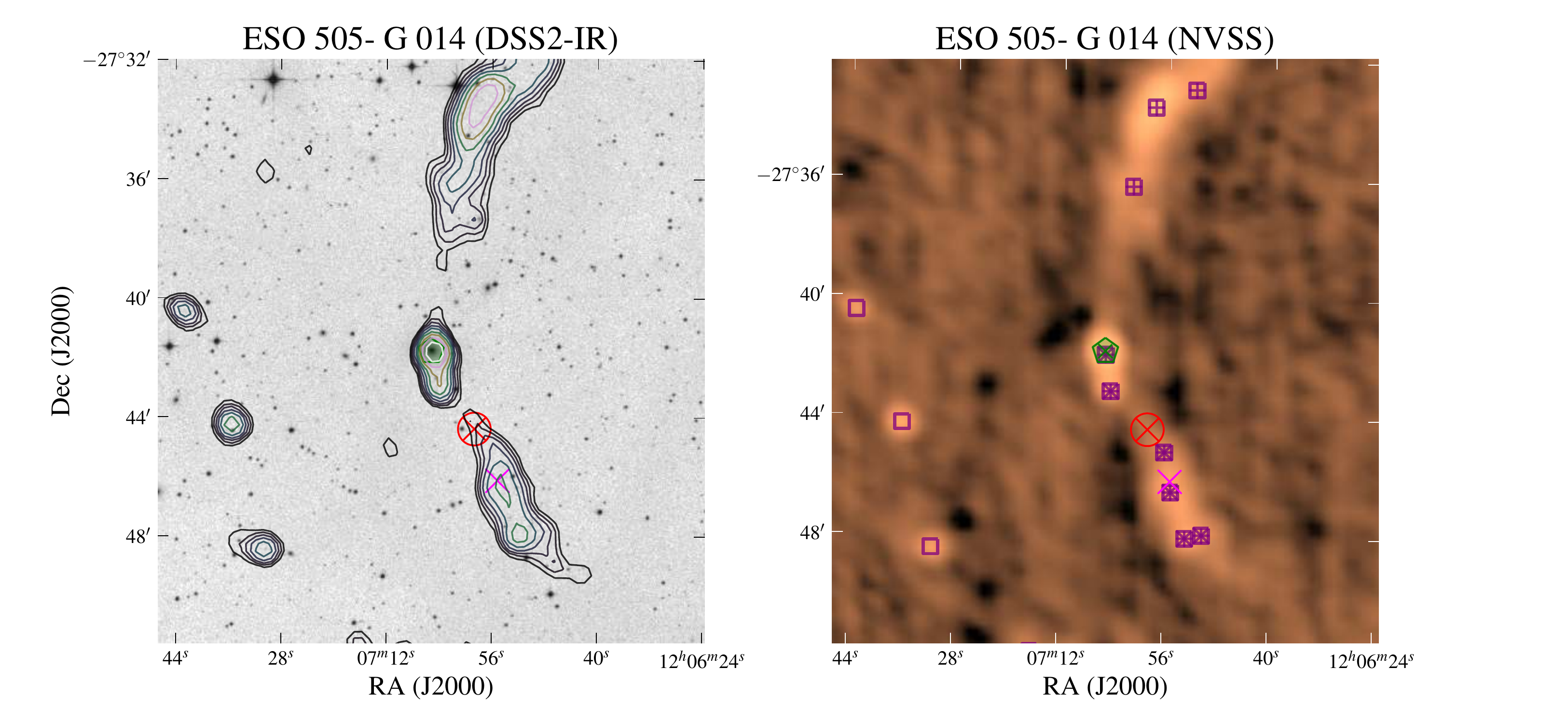}} \\[5pt]
{\includegraphics[trim=10mm 5mm 39mm 5mm, clip, width=.47\textwidth]{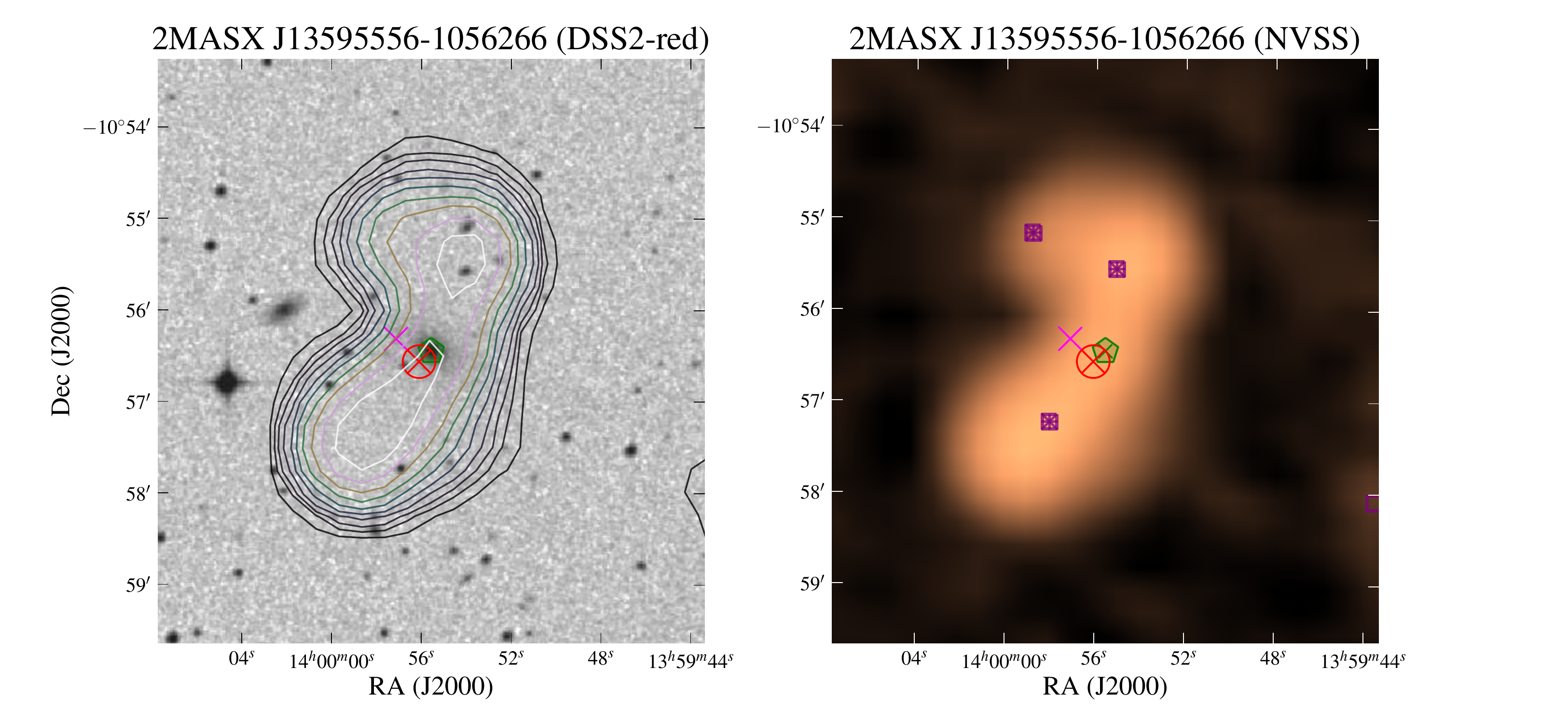}} \qquad
{\includegraphics[trim=10mm 5mm 39mm 5mm, clip, width=.47\textwidth]{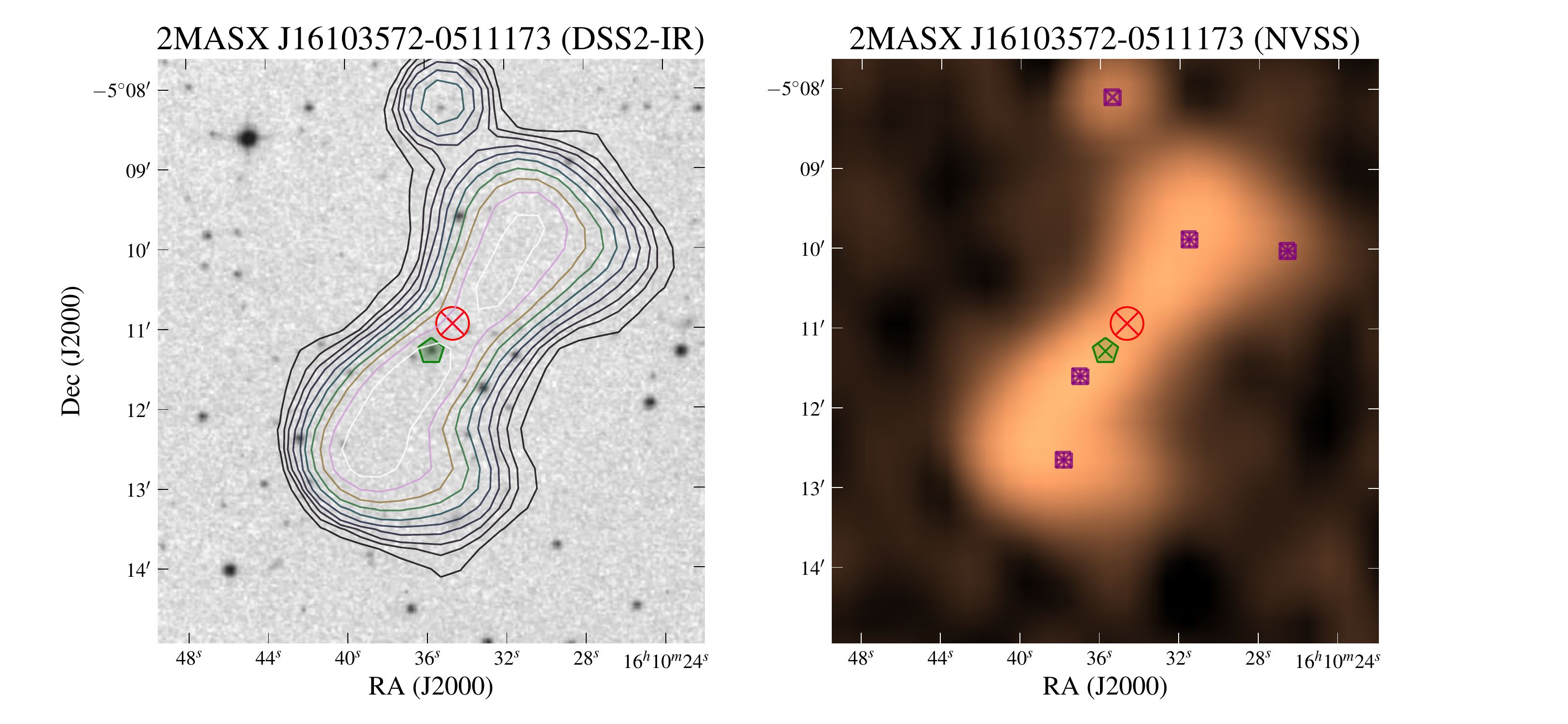}} \\[5pt]
{\includegraphics[trim=10mm 5mm 39mm 5mm, clip, width=.47\textwidth]{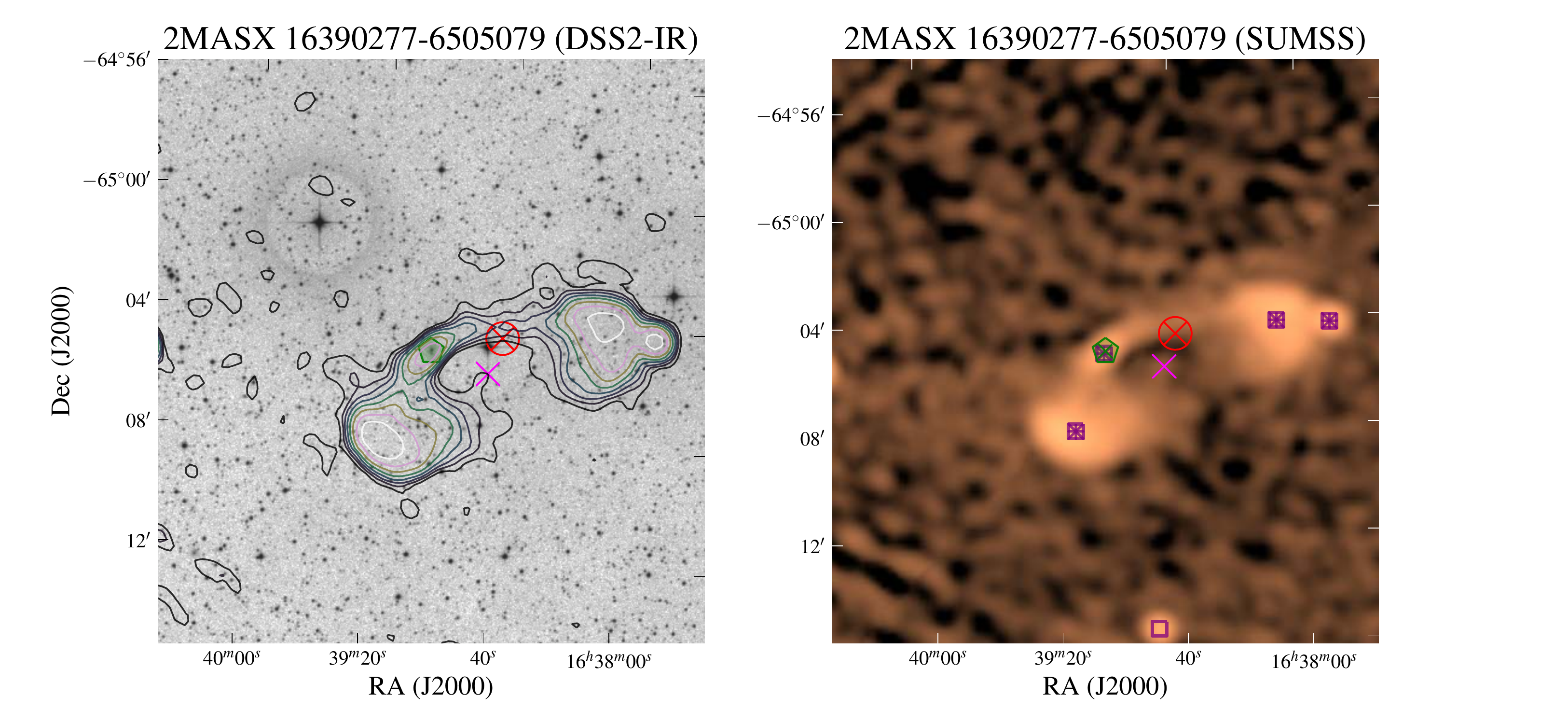}} \qquad
{\includegraphics[trim=10mm 5mm 39mm 5mm, clip, width=.47\textwidth]{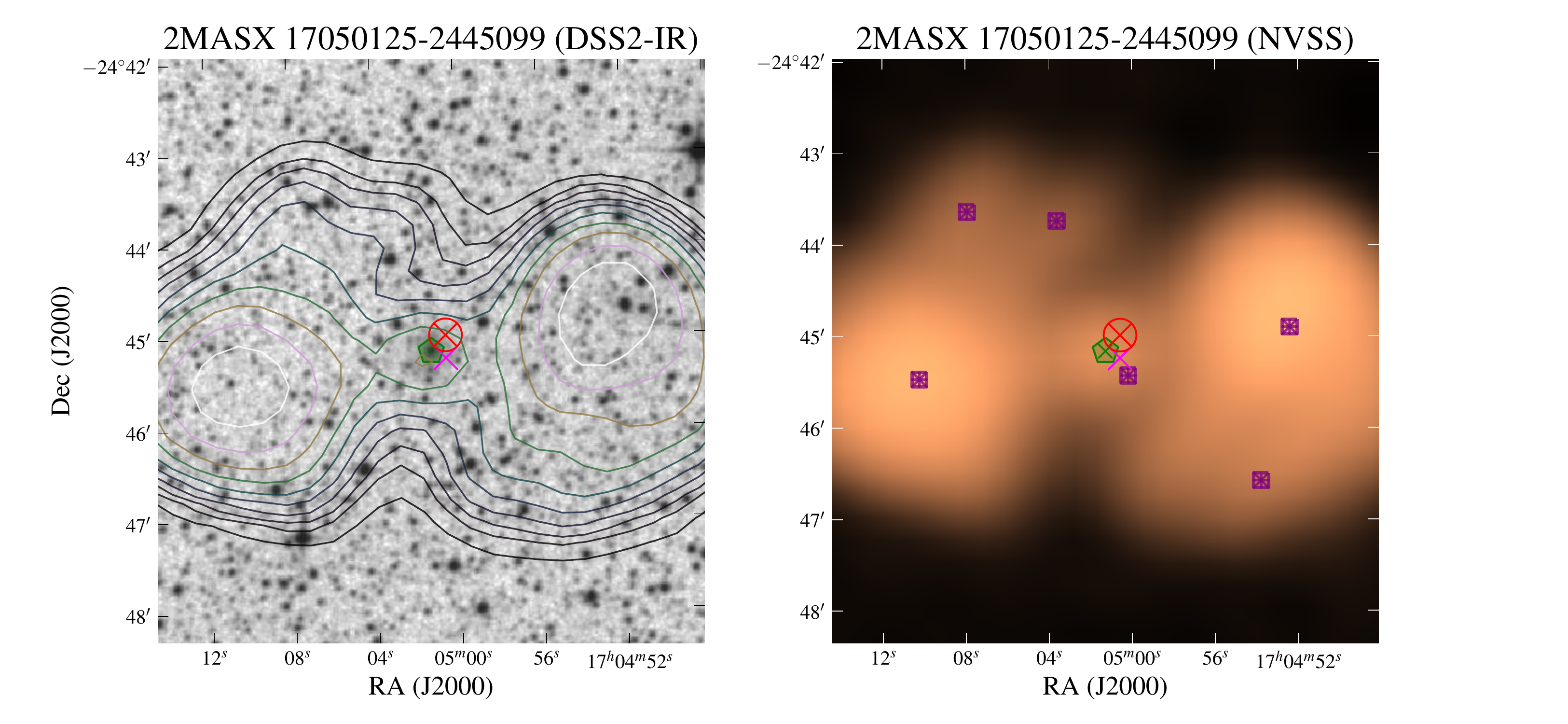}} \\[5pt]
{\includegraphics[trim=10mm 5mm 39mm 5mm, clip, width=.47\textwidth]{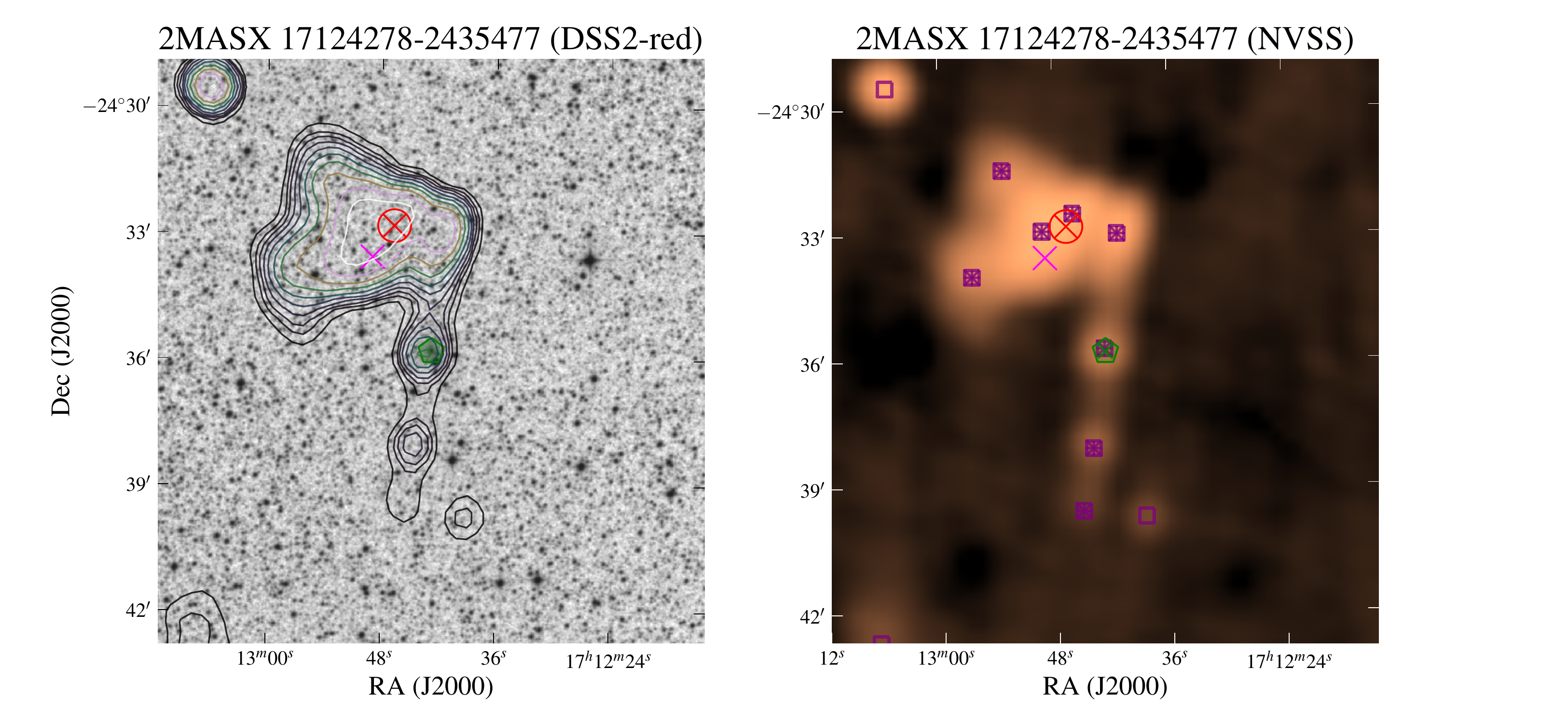}} \qquad
{\includegraphics[trim=10mm 5mm 39mm 5mm, clip, width=.47\textwidth]{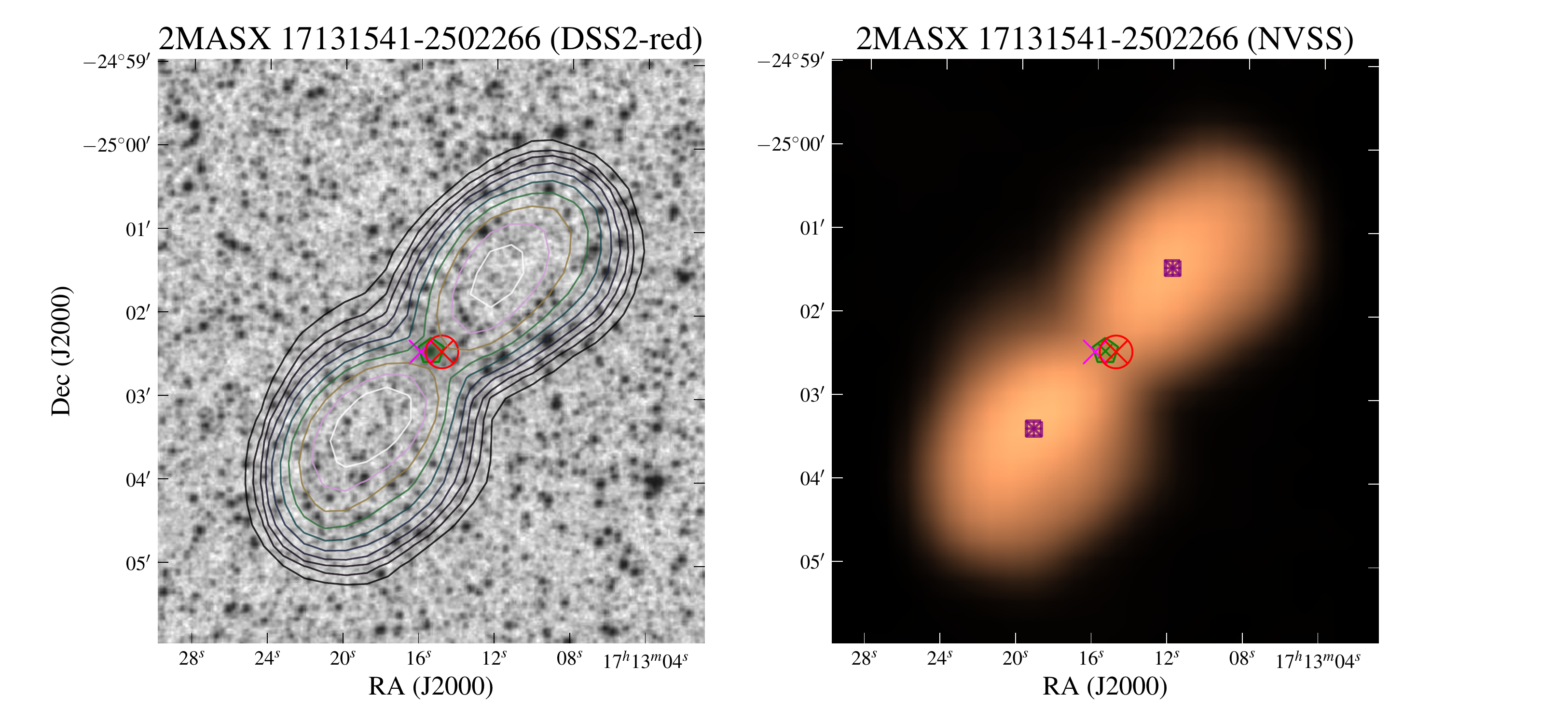}} \\[5pt]
{\includegraphics[trim=10mm 5mm 39mm 5mm, clip, width=.47\textwidth]{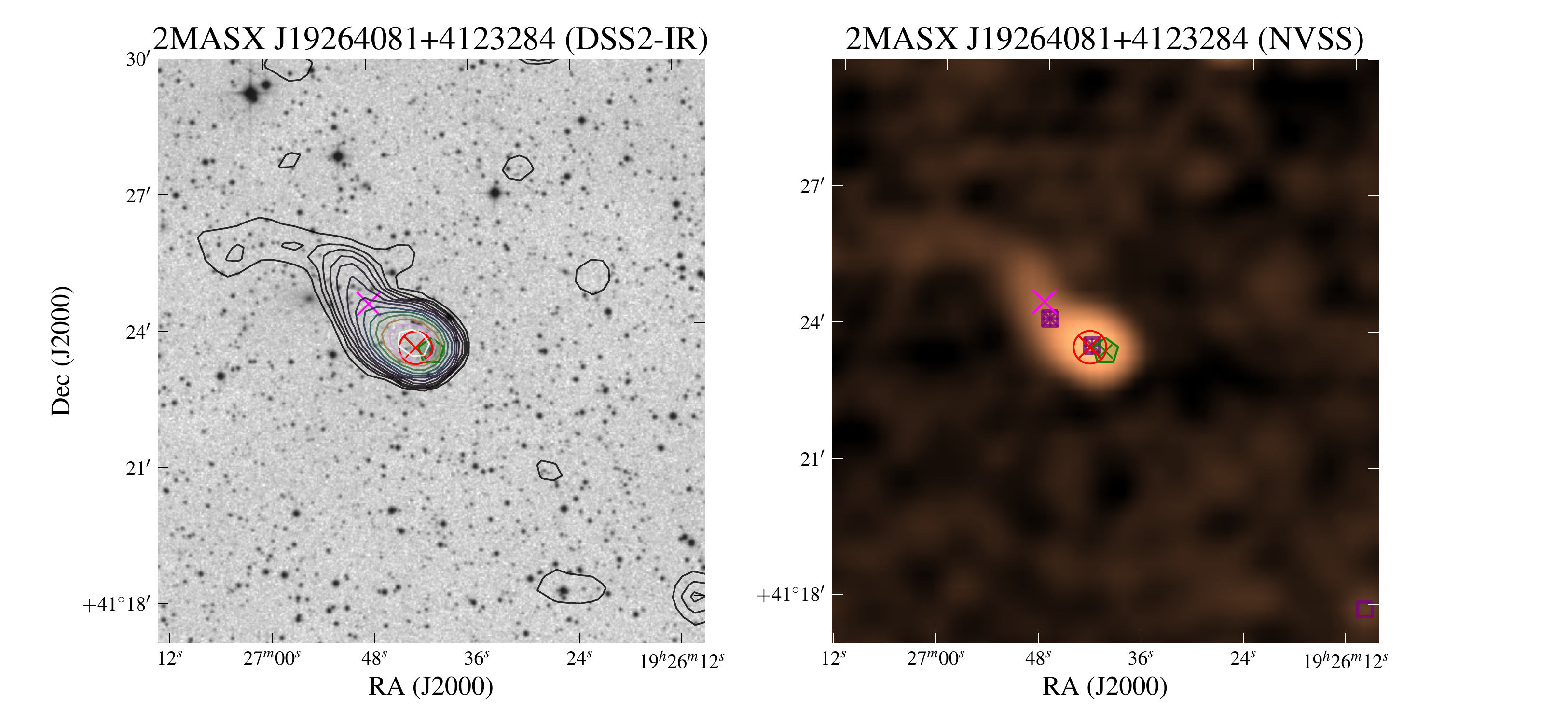}} \qquad
{\includegraphics[trim=10mm 5mm 39mm 5mm, clip, width=.47\textwidth]{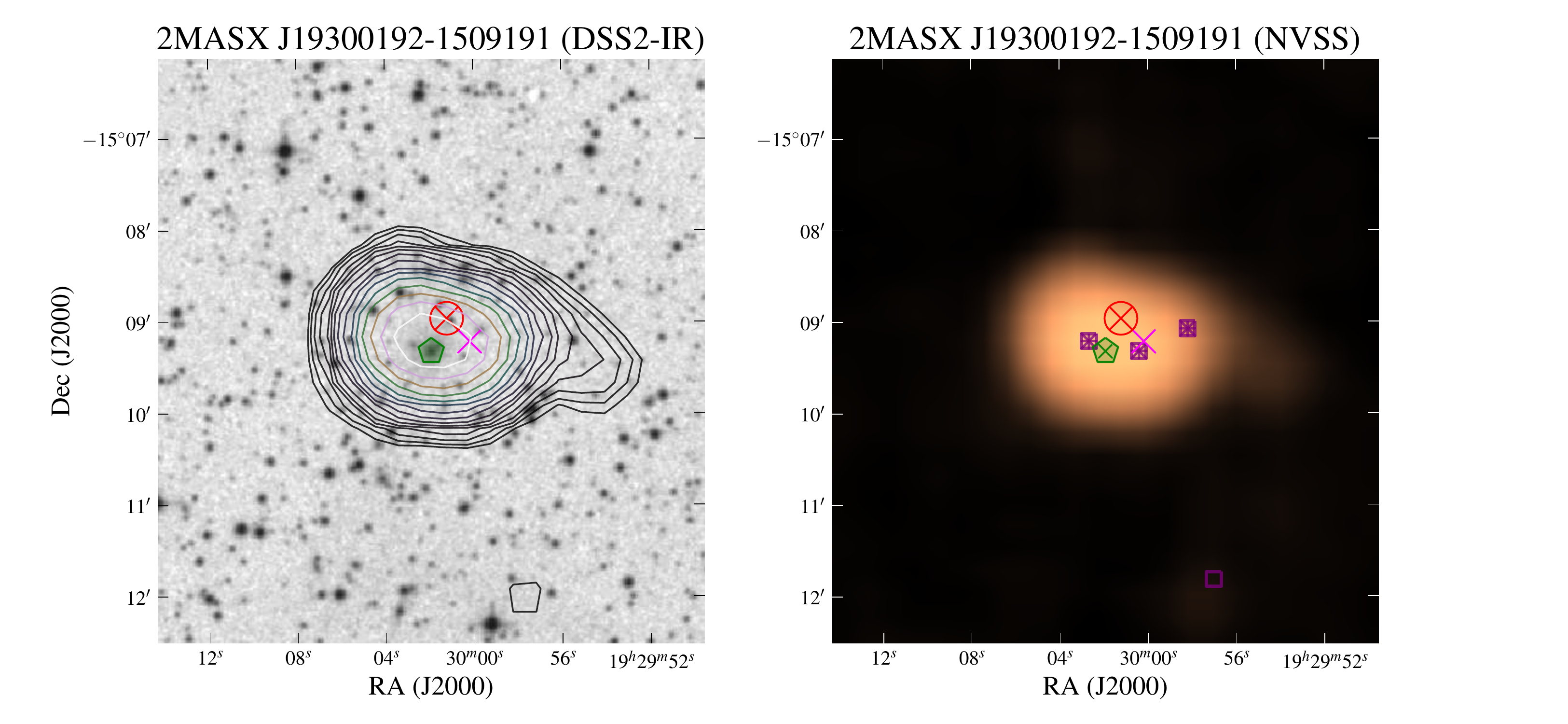}} \\[2pt]
\caption{Examples of newly identified radio-emitting galaxies. } \label{fig:Newly2}
\end{figure*}
\clearpage\begin{figure*}[ht]
{\includegraphics[trim=10mm 5mm 39mm 5mm, clip, width=.47\textwidth]{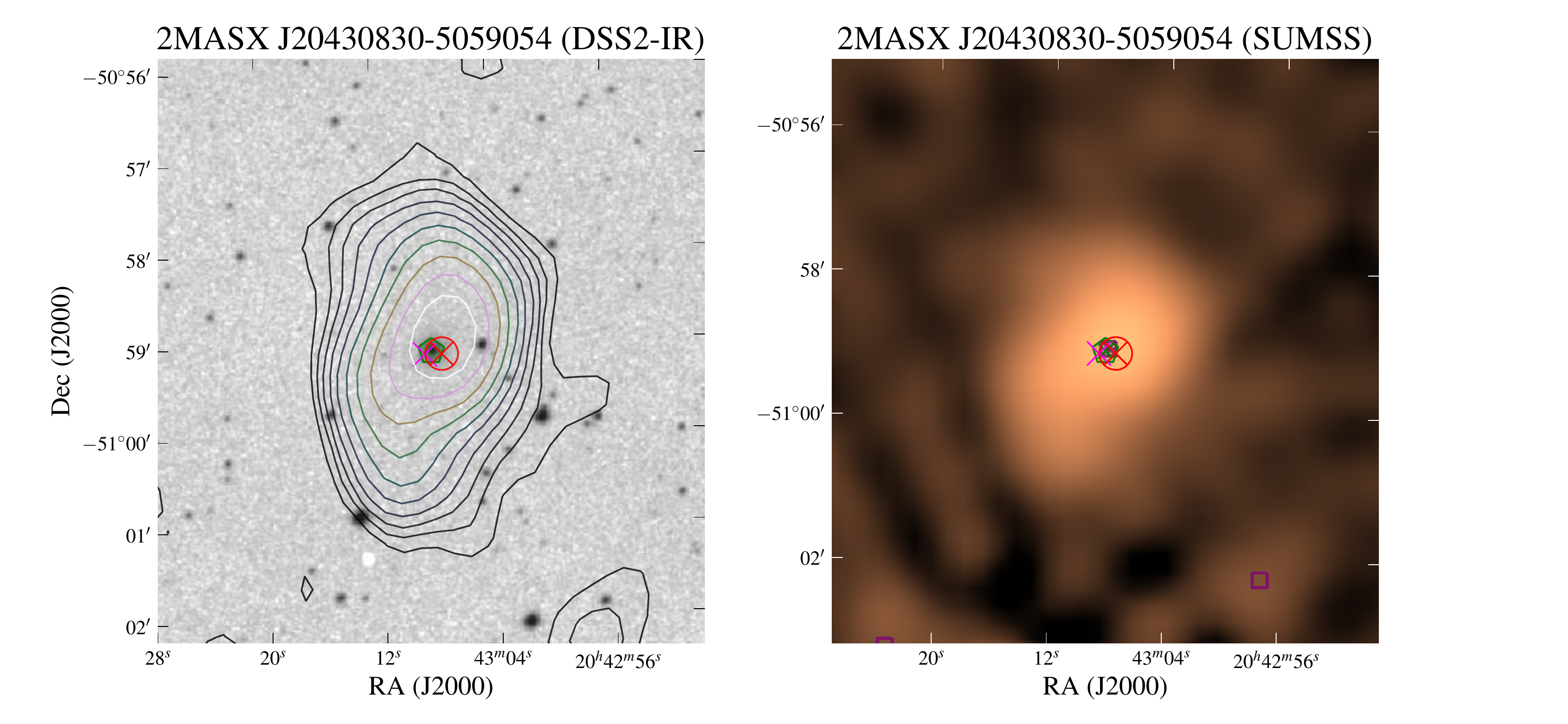}} \qquad
{\includegraphics[trim=10mm 5mm 39mm 5mm, clip, width=.47\textwidth]{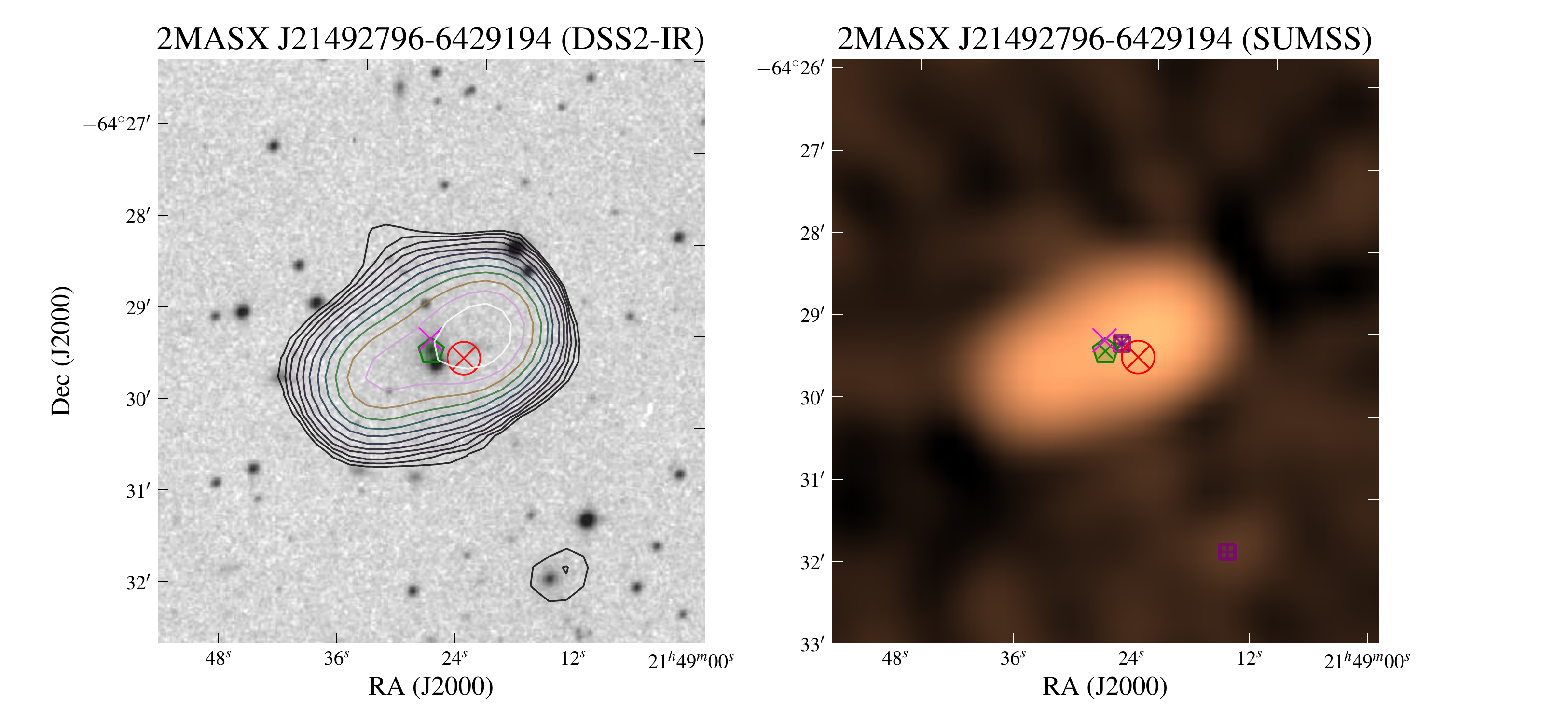}} \\[5pt]
{\includegraphics[trim=10mm 5mm 39mm 5mm, clip, width=.47\textwidth]{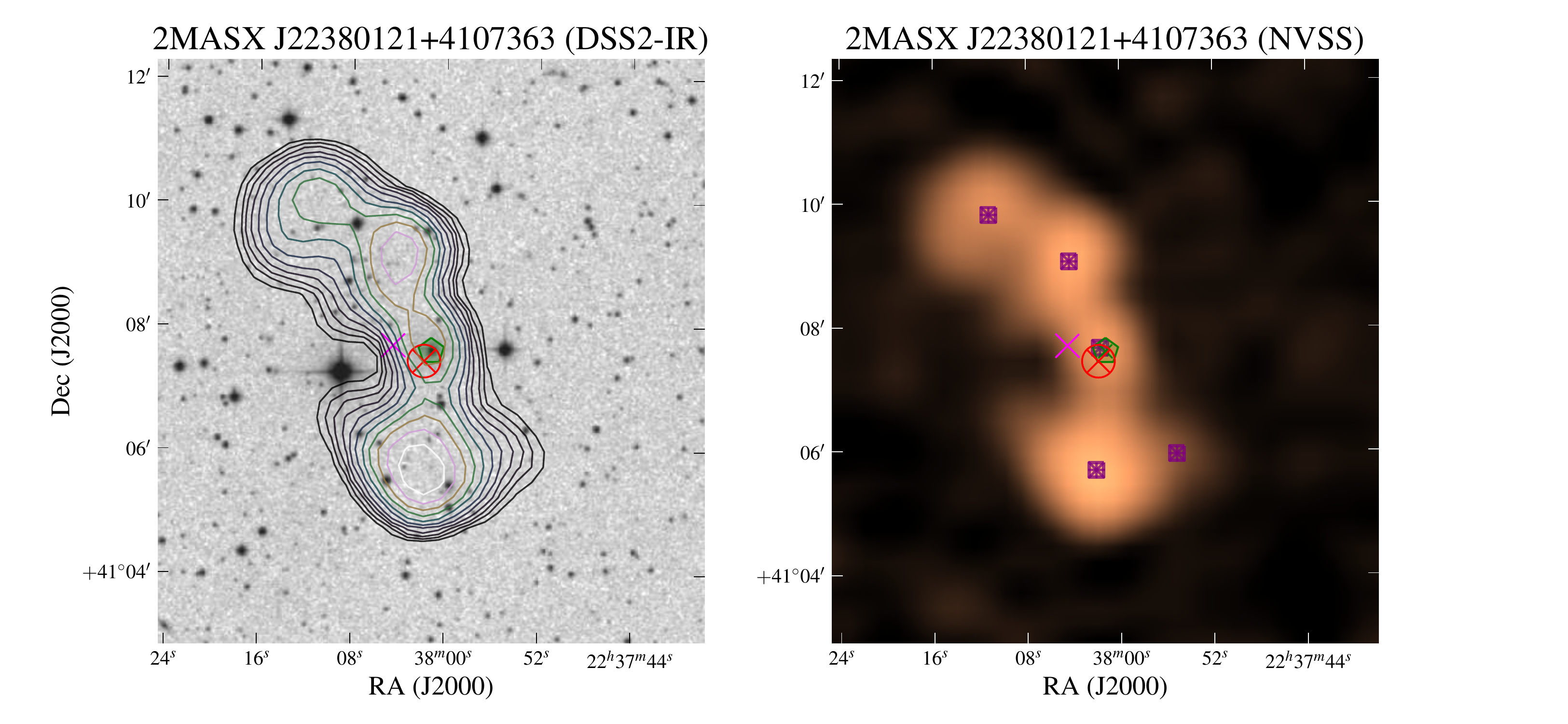}} \qquad
{\includegraphics[trim=10mm 5mm 39mm 5mm, clip, width=.47\textwidth]{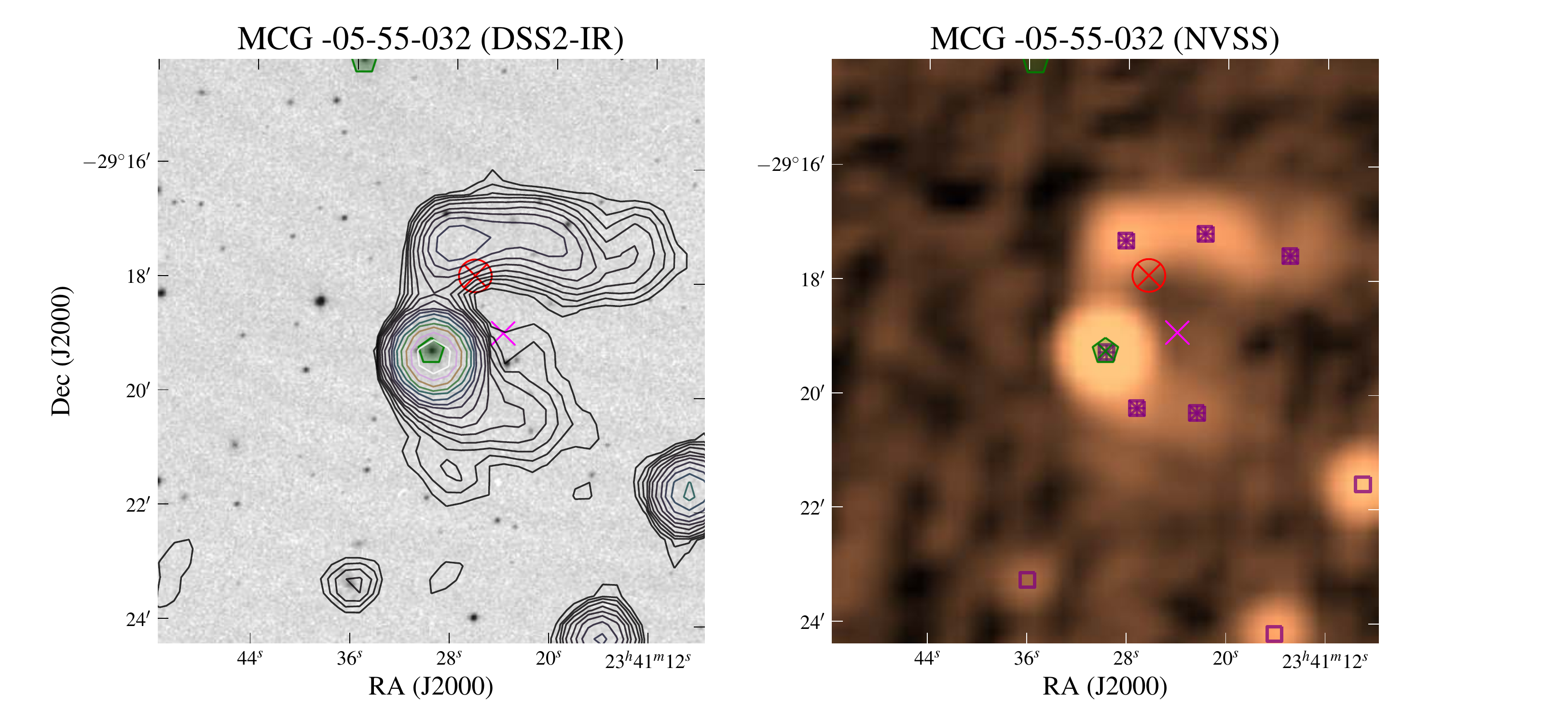}} \\[2pt]
\caption{Examples of newly identified radio-emitting galaxies. } \label{fig:Newly3}
\end{figure*}
\clearpage

\section{Catalog for the volume-limited sample}\label{sec:cats}
Below we present a selection of the columns of the master catalog for the radio galaxies of volume-limited sample (sample~A, defined in sec. \ref{sec:volsample}).  The entire catalog is available in CDS. It can also be obtained at \url{http://ragolu.science.ru.nl}.

\longtab{2}{
\begin{longtable}{crrcrrccccc}
\caption{\label{tab:A} Volume-limited sample of radio galaxies.}\\
\hline\hline
2MASX & ra & dec & $z$ & $D$ & $K$ & $F_{1400}$ & $F_{843}$ & $L_{\rm syn}$ & size & class  \\
  & (deg) & (deg) &  & (Mpc) & (mag) & (Jy)  & (Jy) & $ (\log_{10}\,{\rm erg}\,{\rm s}^{-1}) $ & (arcmin) &  \\
\hline
\endfirsthead
\caption{continued.}\\
\hline \hline
2MASX & ra & dec & $z$ & $D$ & $K$ & $F_{1400}$ & $F_{843}$ & $L_{\rm syn}$ & size & class \\
  & (deg) & (deg) &  & (Mpc) & (mag) & (Jy)  & (Jy) & $ (\log_{10}\,{\rm erg}\,{\rm s}^{-1}) $ & (arcmin) &  \\
\hline
\endhead
\hline
\endfoot
00093552$-$3216365 & $2.39809$ & $-32.27686$ & $0.02546$ & $109.9$ & $9.7$ & $0.43$ & $0.56$ & $39.90$ & $2.4$ & j\\
00391857$+$0319528 & $9.82741$ & $3.33130$ & $0.01472$ & $60.0$ & $9.3$ & $1.3$ & - & $39.84$ & $2.6$ & j\\
00574891$+$3021083 & $14.45378$ & $30.35236$ & $0.01648$ & $69.7$ & $8.0$ & $6.6$ & - & $40.69$ & $8.2$ & j\\
01072493$+$3224452 & $16.85396$ & $32.41256$ & $0.01701$ & $72.5$ & $8.6$ & $4.8$ & - & $40.59$ & $11$ & j\\
01260057$-$0120424 & $21.50248$ & $-1.34514$ & $0.01824$ & $77.0$ & $8.5$ & $3.9$ & - & $40.55$ & $13$ & j\\
01532586$+$7115067 & $28.35774$ & $71.25183$ & $0.02281$ & $101.8$ & $9.8$ & $1.3$ & - & $40.31$ & $8.2$ & j\\
01562095$+$0537437 & $29.08735$ & $5.62887$ & $0.01855$ & $80.2$ & $8.4$ & $1.1$ & - & $40.02$ & $7.0$ & j\\
02231298$+$4259162 & $35.80408$ & $42.98782$ & $0.02200$ & $97.9$ & $9.3$ & $6.2$ & - & $40.96$ & $8.4$ & j\\
02410618$+$0844167 & $40.27572$ & $8.73801$ & $0.02080$ & $91.0$ & $9.7$ & $1.5$ & - & $40.27$ & $11$ & j\\
02574155$+$0601371 & $44.42320$ & $6.02693$ & $0.02215$ & $96.6$ & $9.3$ & $6.0$ & - & $40.93$ & $11$ & j\\
03014235$+$3512203 & $45.42646$ & $35.20563$ & $0.01649$ & $74.4$ & $8.7$ & $1.8$ & - & $40.19$ & $0.29$ & p\\
03082624$+$0406388 & $47.10933$ & $4.11083$ & $0.02865$ & $125.3$ & $9.1$ & $7.6$ & - & $41.26$ & $3.2$ & j\\
03164302$+$4119291 & $49.17918$ & $41.32483$ & $0.01885$ & $84.0$ & $9.2$ & $0.54$ & - & $39.76$ & $4.5$ & j\\
03194823$+$4130420 & $49.95098$ & $41.51168$ & $0.01756$ & $78.7$ & $8.4$ & $23$ & - & $41.33$ & $0.27$ & p\\
03224178$-$3712295 & $50.67412$ & $-37.20820$ & $0.00587$ & $20.9$ & $5.7$ & $0.25$ & $1.7\times 10^{2}$ & $40.92$ & $61$ & j\\
03292389$+$3947318 & $52.34954$ & $39.79221$ & $0.02444$ & $108.0$ & $9.9$ & $1.5$ & - & $40.41$ & $6.3$ & j\\
03341837$+$3921243 & $53.57664$ & $39.35676$ & $0.02059$ & $91.0$ & $9.3$ & $0.96$ & - & $40.08$ & $1.6$ & j\\
04385802$+$0953336 & $69.74182$ & $9.89260$ & $0.02698$ & $117.1$ & $10.3$ & $0.47$ & - & $40.00$ & $8.2$ & j\\
05444416$+$1648501 & $86.18403$ & $16.81393$ & $0.01951$ & $85.0$ & $9.8$ & $0.73$ & - & $39.91$ & $6.6$ & j\\
06153645$+$7102152 & $93.90173$ & $71.03753$ & $0.01351$ & $59.0$ & $9.0$ & $1.1$ & - & $39.77$ & $0.27$ & p\\
07090797$+$4836561 & $107.28328$ & $48.61554$ & $0.01933$ & $84.8$ & $9.4$ & $0.75$ & - & $39.91$ & $4.3$ & j\\
07175757$+$0941218 & $109.48984$ & $9.68934$ & $0.02862$ & $122.7$ & $10.9$ & $0.32$ & - & $39.86$ & $0.29$ & p\\
07204757$-$3407058 & $110.19823$ & $-34.11834$ & $0.02835$ & $119.2$ & $10.1$ & $2.5$ & $3.1$ & $40.73$ & $1.7$ & j\\
08173619$+$5957137 & $124.40091$ & $59.95378$ & $0.02859$ & $125.5$ & $10.3$ & $0.46$ & - & $40.05$ & $2.4$ & j\\
09275281$+$2959085 & $141.97008$ & $29.98573$ & $0.02666$ & $114.5$ & $10.2$ & $0.40$ & - & $39.90$ & $25$ & j\\
10065190$+$1422247 & $151.71620$ & $14.37357$ & $0.02944$ & $125.6$ & $9.7$ & $0.76$ & - & $40.26$ & $5.2$ & j\\
10461032$+$7321108 & $161.54320$ & $73.35307$ & $0.02100$ & $92.4$ & $9.7$ & $0.70$ & - & $39.96$ & $1.1$ & j\\
11082650$-$1015216 & $167.11047$ & $-10.25604$ & $0.02730$ & $114.5$ & $9.7$ & $0.32$ & - & $39.81$ & $0.26$ & p\\
11450498$+$1936229 & $176.27083$ & $19.60635$ & $0.02160$ & $91.2$ & $9.7$ & $5.7$ & - & $40.86$ & $6.1$ & j\\
11564844$-$0405406 & $179.20183$ & $-4.09455$ & $0.02685$ & $111.2$ & $10.4$ & $0.36$ & - & $39.83$ & $7.4$ & j\\
12040140$+$2013559 & $181.00592$ & $20.23227$ & $0.02445$ & $103.9$ & $9.9$ & $0.40$ & - & $39.82$ & $6.8$ & j\\
12080557$+$2514141 & $182.02316$ & $25.23728$ & $0.02253$ & $95.7$ & $10.3$ & $0.75$ & - & $40.02$ & $14$ & j\\
12192326$+$0549289 & $184.84692$ & $5.82471$ & $0.00747$ & $32.1$ & $7.4$ & $10$ & - & $40.22$ & $5.1$ & j\\
12304942$+$1223279 & $187.70593$ & $12.39110$ & $0.00436$ & $18.4$ & $5.9$ & $1.5\times 10^{2}$ & - & $40.88$ & $8.8$ & j\\
12484927$-$4118399 & $192.20532$ & $-41.31109$ & $0.00987$ & $41.7$ & $7.3$ & - & $5.7$ & $40.04$ & $0.0$ & p\\
12530723$-$1029394 & $193.28012$ & $-10.49428$ & $0.01478$ & $67.4$ & $8.7$ & $1.0$ & - & $39.85$ & $15$ & j\\
12543570$-$1234070 & $193.64874$ & $-12.56861$ & $0.01540$ & $69.4$ & $8.2$ & $7.7$ & - & $40.75$ & $4.4$ & j\\
12592333$+$2754418 & $194.84727$ & $27.91161$ & $0.02288$ & $97.7$ & $10.8$ & $0.44$ & - & $39.80$ & $4.6$ & j\\
13010079$-$3226289 & $195.25330$ & $-32.44135$ & $0.01704$ & $73.4$ & $8.6$ & $1.4$ & $1.9$ & $40.07$ & $4.8$ & j\\
13211286$-$4342168 & $200.30356$ & $-43.70469$ & $0.01141$ & $46.9$ & $7.6$ & - & $13$ & $40.50$ & $9.4$ & j\\
13234497$+$3133568 & $200.93747$ & $31.56576$ & $0.01618$ & $70.7$ & $9.6$ & $2.0$ & - & $40.18$ & $12$ & j\\
13245144$+$3622424 & $201.21439$ & $36.37854$ & $0.01749$ & $77.6$ & $9.8$ & $0.88$ & - & $39.90$ & $0.54$ & j\\
13252775$-$4301073 & $201.36565$ & $-43.01871$ & $0.00182$ & $3.6$ & $3.9$ & - & $1.8\times 10^{3}$ & $40.42$ & $73$ & j\\
13360823$-$0829519 & $204.03436$ & $-8.49774$ & $0.02280$ & $95.2$ & $9.3$ & $0.39$ & - & $39.73$ & $4.0$ & j\\
13363905$-$3357572 & $204.16269$ & $-33.96588$ & $0.01247$ & $51.1$ & $7.6$ & $2.4$ & $26$ & $40.89$ & $17$ & j\\
13471216$-$2422224 & $206.80067$ & $-24.37294$ & $0.01953$ & $82.0$ & $10.0$ & $0.59$ & - & $39.78$ & $6.8$ & j\\
14072978$-$2701043 & $211.87405$ & $-27.01787$ & $0.02180$ & $90.3$ & $9.7$ & $0.65$ & - & $39.90$ & $0.24$ & p\\
14095733$+$1732435 & $212.48880$ & $17.54548$ & $0.01619$ & $72.1$ & $9.0$ & $0.81$ & - & $39.80$ & $3.8$ & j\\
14165292$+$1048264 & $214.22060$ & $10.80737$ & $0.02471$ & $105.2$ & $8.9$ & $4.4$ & - & $40.87$ & $7.5$ & j\\
15105610$+$0544416 & $227.73375$ & $5.74490$ & $0.02149$ & $91.2$ & $10.3$ & $0.53$ & - & $39.83$ & $0.91$ & j\\
15573014$+$7041207 & $239.37575$ & $70.68913$ & $0.02572$ & $113.8$ & $9.8$ & $2.0$ & - & $40.59$ & $13$ & j\\
16323175$+$8232165 & $248.13274$ & $82.53791$ & $0.02471$ & $109.7$ & $9.1$ & $2.0$ & - & $40.56$ & $6.9$ & j\\
16525886$+$0224035 & $253.24535$ & $2.40102$ & $0.02448$ & $105.4$ & $9.1$ & $0.43$ & - & $39.86$ & $0.25$ & p\\
17115542$-$2309423 & $257.98090$ & $-23.16181$ & $0.02685$ & $114.6$ & $10.5$ & $0.36$ & - & $39.86$ & $3.4$ & j\\
17124278$-$2435477 & $258.17828$ & $-24.59663$ & $0.02433$ & $103.3$ & $9.7$ & $1.3$ & - & $40.32$ & $8.4$ & j\\
17131541$-$2502266 & $258.31430$ & $-25.04076$ & $0.02857$ & $122.2$ & $10.0$ & $7.2$ & - & $41.21$ & $3.0$ & j\\
17204089$-$0111573 & $260.17044$ & $-1.19920$ & $0.02860$ & $123.9$ & $10.6$ & $0.76$ & - & $40.25$ & $2.2$ & u\\
17234103$-$6500371 & $260.92096$ & $-65.01024$ & $0.01443$ & $59.0$ & $9.3$ & - & $3.7$ & $40.16$ & $0.0$ & p\\
17354374$-$0720527 & $263.93222$ & $-7.34798$ & $0.02426$ & $104.3$ & $9.5$ & $0.57$ & - & $39.98$ & $0.72$ & j\\
17480808$+$5123570 & $267.03369$ & $51.39918$ & $0.02266$ & $100.3$ & $10.8$ & $0.56$ & - & $39.93$ & $1.7$ & j\\
17554844$+$6236435 & $268.95184$ & $62.61216$ & $0.02746$ & $122.0$ & $9.7$ & $0.30$ & - & $39.83$ & $1.1$ & p\\
18363966$+$1943454 & $279.16528$ & $19.72928$ & $0.01614$ & $66.8$ & $9.1$ & $1.6$ & - & $40.02$ & $12$ & j\\
18382625$+$1711496 & $279.60938$ & $17.19715$ & $0.01689$ & $69.5$ & $9.7$ & $7.3$ & - & $40.73$ & $3.4$ & j\\
18403862$-$7709285 & $280.16086$ & $-77.15794$ & $0.01817$ & $73.1$ & $9.5$ & - & $1.2$ & $39.84$ & $0.0$ & p\\
19281700$-$2931442 & $292.07086$ & $-29.52901$ & $0.02442$ & $104.8$ & $9.7$ & $2.1$ & - & $40.55$ & $2.7$ & j\\
19414211$+$5037571 & $295.42551$ & $50.63249$ & $0.02372$ & $105.7$ & $9.4$ & $3.1$ & - & $40.72$ & $5.1$ & j\\
20520232$-$5704076 & $313.00974$ & $-57.06877$ & $0.01175$ & $46.6$ & $8.8$ & - & $2.2$ & $39.72$ & $1.7$ & p\\
21313299$-$3837046 & $322.88742$ & $-38.61790$ & $0.01817$ & $75.0$ & $9.7$ & $0.98$ & $1.1$ & $39.92$ & $2.6$ & j\\
21570595$-$6941236 & $329.27487$ & $-69.68994$ & $0.02827$ & $119.8$ & $10.3$ & - & $44$ & $41.85$ & $5.6$ & j\\
22041766$+$0440021 & $331.07358$ & $4.66722$ & $0.02700$ & $119.4$ & $11.2$ & $0.78$ & - & $40.23$ & $2.7$ & j\\
22144500$+$1350476 & $333.68747$ & $13.84653$ & $0.02616$ & $116.2$ & $10.3$ & $3.3$ & - & $40.83$ & $6.1$ & j\\
22312062$+$3921298 & $337.83585$ & $39.35825$ & $0.01713$ & $73.7$ & $9.9$ & $2.9$ & - & $40.38$ & $17$ & j\\
22382946$+$3519400 & $339.62277$ & $35.32798$ & $0.02759$ & $123.4$ & $9.7$ & $0.36$ & - & $39.92$ & $0.52$ & j\\
22492263$-$3728185 & $342.34439$ & $-37.47178$ & $0.02867$ & $124.0$ & $10.1$ & $0.34$ & $0.47$ & $39.91$ & $0.97$ & j\\
\\
\end{longtable}
}

\end{appendix}

\end{document}